\documentclass[a4paper,10pt]{article}
\usepackage[dvips]{graphicx}
\usepackage{amssymb,amsmath}
\usepackage{cancel}
\numberwithin{equation}{section}

\begin{document}

 \newcommand{\be}[1]{\begin{equation}\label{#1}}
 \newcommand{\ee}{\end{equation}}
 \newcommand{\bea}{\begin{eqnarray}}
 \newcommand{\eea}{\end{eqnarray}}
 \def\disp{\displaystyle}

 \def\gsim{ \lower .75ex \hbox{$\sim$} \llap{\raise .27ex \hbox{$>$}} }
 \def\lsim{ \lower .75ex \hbox{$\sim$} \llap{\raise .27ex \hbox{$<$}} }

 \title{\Large \bf Metric-affine  Myrzakulov  gravity theories with   Gauss-Bonnet and boundary term scalars}

\author{T. Myrzakul$^{1,2}$, K. Yesmakhanova$^{1,2}$, \, N. Myrzakulov$^{1,2}$, \,  S. Myrzakul$^{1,2}$,    \\ K. Myrzakulov$^{1,2}$, \,  K. Yerzhanov$^{1,2}$,  \,  R. Myrzakulov$^{1,2}$, \,  G. Nugmanova$^{1,2}$ \\
$^{1}$\textit{Eurasian  National University, Astana 010008, Kazakhstan}\\
$^{2}$\textit{Ratbay Myrzakulov Eurasian International Centre for Theoretical} \\ \textit{Physics, Astana 010009, Kazakhstan}}
\date{}
 \maketitle

 \begin{abstract}
In this paper, we consider some metric-affine Myrzakulov gravity (MG) theories with Gauss-Bonnet scalars. Also we consider  the MG theories with the boundary term scalars. Note that these MG theories with the Gauss-Bonnet and boundary term scalars were proposed in [arXiv:1205.5266]. Some examples  of Metric-Affine Gravity (MAG) theories are reviewed in the context of the $F(R,T,Q,{\cal T}, {\cal D})$  type models. Then  the generalized  MAG  theory with the curvature, torsion and nonmetricity (the so-called MG-VIII) was studied.   For the FRW spacetime case, in particular, the Lagrangian, Hamilatonian and gravitational equations are obtained. The particular case $F(R,T)=\alpha R+\beta T+\mu Q+\nu{\cal T}$ is investigated  in detail. In quantum case, the corresponding Wheeler-DeWitt equation is obtained. Finally, some gravity theories with the curvature, torsion and nonmetricity   are presented. 
 \end{abstract}
 
\section{Introduction}
At present, General Relativity (GR) is considered the best accepted fundamental theory describing gravity. GR is  described in terms of the Levi-Civita connection, which is  the basis of Riemannian geometry with the Ricci curvature scalar $R$. But GR  can be described in terms
of different geometries from the Riemannian one, for example, $F(R)$ gravity. There are several other alternative gravity theories. For example, one
of the alternative gravity theory  is the so-called teleparallel gravity with the torsion scalar $T$ or its generalization $F(T)$ gravity.  Another possible
alternative gravity theory is the symmetric teleparallel gravity  with  the nonmetricity scalar $Q$ or its generalization $F(Q)$ gravity.  In this paper, we will consider the more general gravity theory, the so-called MG-VIII  with the   action
\begin{equation}
S=\int\sqrt{-g}d^{4}x[F(R,T,Q,{\cal T})+L_{m}].
\end{equation}

This paper is organized as follows. In Sec. 2, we
 briefly review the geometry of the underlying spacetime. In Sec. 3, we present a main information on the MG-VIII gravity. FRW cosmology of the MG-VIII   is studied in Sec. 4. The  specific model $F(R,T)=\alpha R+\beta T+\mu Q+\nu {\cal T}$  is  analyzed in Sec. 5.  The cosmological power-law solution is obtained in Sec. 6. In Sec. 7, the  Wheeler - DeWitt equation is derived.   The relation with the soliton theory is considered in Sec. 8. Some other known gravity theories related with the curvature, torsion and nonmetricity  are presented in Sec. 9.  Final conclusions and remarks are provided in Sec. 10.

\section{Preliminaries}

\subsection{Geometric setup}
Consider a general spacetime with the curvature, torsion and nonmetricity. The corresponding  connection is given by
\begin{equation}
\Gamma^{\rho}_{\,\,\,\,\mu\nu}=\breve{\Gamma}_{\,\,\,\, \mu \nu}^{\rho}+K^{\rho}_{\,\,\,\,\mu\nu}+L^{\rho}_{\,\,\,\,\mu\nu}\,,
\end{equation}
where $\breve{\Gamma}_{\,\,\,\, \mu \nu}^{\rho}$ is the Levi--Civita connection,   $K^{\rho}_{\,\,\,\,\mu\nu}$  is  the  contorsion tensor and 
$L^{\rho}_{\,\,\,\,\mu\nu}$ is  the disformation tensor. These three tensors  have the following forms
\begin{eqnarray}
\breve{\Gamma}^l_{\, \, \, jk} &=& \tfrac{1}{2} g^{lr} \left( \partial _k g_{rj} + \partial _j g_{rk} - \partial _r g_{jk} \right), \\
K^{\rho}_{\,\,\,\,\mu\nu}&=&\frac{1}{2}g^{\rho\lambda}\bigl(T_{\mu\lambda\nu}+T_{\nu\lambda\mu}+T_{\lambda\mu\nu}\bigr)
                                      =-K^{\rho}_{\,\,\,\,\nu\mu} \,, 	\\	
L^{\rho}_{\,\,\,\,\mu\nu}&=&\frac{1}{2}g^{\rho\lambda}\bigl(-Q_{\mu \nu \lambda}-Q_{\nu \mu \lambda} + Q_{\lambda\mu\nu}\bigr)=
                                             L^{\rho}_{\,\,\,\,\nu\mu}.
\end{eqnarray}
Here \begin{equation}
T_{\,\,\,\, \mu \nu}^{\alpha}=2 \Gamma_{\,\,\,\, [\mu \nu]}^{\alpha}\,, \quad Q_{\rho \mu \nu} = \nabla_{\rho} g_{\mu \nu}
\end{equation}
 are the  torsion tensor and the  nonmetricity tensor, respectively. In this generalized spacetime with the curvature, torsion and nonmetricity, let us introduce three scalars as
\begin{eqnarray}
 R&=&g^{\mu\nu}R_{\mu\nu},\label{1.3}\\
 T&=&{S_\rho}^{\mu\nu}\,{T^\rho}_{\mu\nu},\label{1.5}\\
Q&=& -g^{\mu\nu}(L^{\alpha}_{\beta\mu}L^{\beta}_{\nu\alpha}-L^{\alpha}_{\beta\alpha}L^{\beta}_{\mu\nu}),
 \end{eqnarray}  
 where $R$ is the curvature scalar, $T$ is the torsion scalar and $Q$ is the nonmetricity  scalar. Here
\begin{eqnarray}
R_{jk}&=&\partial_i\Gamma_{jk}^i-\partial_j\Gamma_{ik}^i+\Gamma_{ip}^i\Gamma_{jk}^p-\Gamma_{jp}^i\Gamma_{ik}^p, \\
 S^{p\mu\nu}&=&K^{\mu\nu p}-g^{p\nu}T^{\sigma\mu}_{\,\,\sigma}+g^{p\mu}T^{\sigma\nu}_{\,\,\sigma}, \\
K^{\nu}_{p\mu}&=&\frac{1}{2}(T_{p\,\,\,\mu}^{\,\,\nu}+T_{\mu\, \, \, p}^{\,\, \nu}-T^{\nu}_{p\mu}).
 \end{eqnarray}
are the Ricci tensor, the potential and the contorsion tensor, respectively. The key moment of our construction is following: as in our previous  paper \cite{1205.5266}, here we assume that these three scalars have the following forms 
 \begin{eqnarray}
 R&=&u+R_{s},\\
 T&=&v+T_{s},\\
Q&=&w+Q_{s},
 \end{eqnarray}
 where $u=u(\Gamma^{\rho}_{\,\,\,\,\mu\nu}; x_i; g_{ij}, \dot{g_{ij}},\ddot{g_{ij}}, ... ; f_j), \quad v=v(\Gamma^{\rho}_{\,\,\,\,\mu\nu}; x_i; g_{ij}, \dot{g_{ij}},\ddot{g_{ij}}, ... ; g_j)$  and $w=w(\Gamma^{\rho}_{\,\,\,\,\mu\nu}; x_i; \\  g_{ij},   \dot{g_{ij}},\ddot{g_{ij}}, ... ; h_j)$ are some real functions. Here: i) $R_{s}=R^{(LC)}$ is the curvature scalar corresponding to the Levi-Civita connection with the vanishing torsion and nonmetricity ($T=Q=0$); ii) $T_{s}=T^{(WC)}$ is the torsion scalar for the purely Weitzenb\"{o}ck connection with the vanishing curvature and nonmetricity ($R=Q=0$); iii) $Q_{s}=Q^{(NM)}$ is the nonmetricity scalar with the vanishing torsion and curvature ($R=T=0$).

Consider the  Friedmann-Robertson-Walker
 (FRW) spacetime.  The flat FRW spacetime is described by the metric
 \begin{equation}
 ds^2=-N^{2} (t)dt^2+a^2(t)(dx^2+dy^2+dz^2),
 \end{equation}
 where $a=a(t)$ is the scale factor, $N(t)$ is the lapse function. The orthonormal tetrad
 components $e_i(x^\mu)$ are related to the metric through
 \begin{equation}
 g_{\mu\nu}=\eta_{ij}e_\mu^i e_\nu^j\,,
 \end{equation}
 where the Latin indices $i$, $j$ run over $0,..., 3$ for
 the tangent space of the manifold, while the Greek letters $\mu$,~$\nu$ are
 the coordinate indices on the manifold, also running over $0, ..., 3$. 
  With the FRW metric ansatz the three variables $R_{s}, T_{s}, Q_{s}$  look like (we assume that $N=1$)
 \begin{eqnarray}
 R_s&=&R^{LC}=6(\dot{H}+2H^2),\\
   T_s&=&T^{WC}=-6H^2,\\
	Q_s&=&Q^{NM}= 6H^2,
 \end{eqnarray}  
 where $H=(\ln a)_{t}$ is the Hubble parameter. Therefore, these  three scalars ($R, T, Q)$ of the metric - affine spacetime (in the FRW case) take the following forms 
 \begin{eqnarray}
 R&=&u+6(\dot{H}+2H^2),\\
   T&=&v-6H^2,\\
	Q&=& w+6H^2,
 \end{eqnarray}
 where $u, v, w$ are some real functions of $\Gamma^{\rho}_{\,\,\,\,\mu\nu}; t, a, \dot{a}, \ddot{a}, R_{s}, T_{s}, Q_{s},  ...$ and so on.

The concepts of torsion and non-metricity allow to classify different geometries  on which we will make observations throughout this work.
\subsection{Variations}
\subsubsection{Variations of the torsion variables}
Let us now derive the variations for the torsion tensor $(S_{\mu\nu}^{\;\;\;\;\alpha})$ and torsion vector $(S_{\mu}\equiv S_{\mu\alpha}^{\;\;\;\;\alpha})$ since we will be using them in the various theories we are going to study. Firstly, note that since the torsion does not depend on the metric, the $\delta g^{\mu\nu}$ variation is identically zero, namely
	\begin{equation}
	\delta_{g}S_{\mu\nu}^{\;\;\;\;\alpha}=\frac{\delta S_{\mu\nu}^{\;\;\;\;\alpha}}{\delta g^{\kappa\lambda}}\delta g^{\kappa\lambda}=0
	\end{equation}
	as well as\footnote{This is so because in order to form the torsion vector $S_{\mu}$ we need only contract an upper with a lower index without the use of any metric. Notice also that if we were to form another vector by contracting the first two indices of the torsion with the metric tensor, the result would yield zero due to the fact that the torsion is antisymmetric in its first two indices while the metric tensor is symmetric. In words, $\tilde{S}^{\mu}\equiv g^{\alpha\beta}S_{\alpha\beta}^{\;\;\;\;\mu}=0$. }
	\begin{equation}
	\delta_{g}S_{\mu}=0
	\end{equation}
	Now to proceed with the $\Gamma$-variation we recall that we want to have a common factor $\delta \Gamma^{\lambda}_{\;\;\;\mu\nu}$ appearing in the variation. Thus, we express the torsion tensor as
	\begin{gather}
	S_{\alpha\beta}^{\;\;\;\;\lambda}=\frac{1}{2}(\Gamma^{\lambda}_{\;\;\;\alpha\beta}-\Gamma^{\lambda}_{\;\;\;\beta\alpha})=\frac{1}{2}(\delta_{\alpha}^{\mu}\delta_{\beta}^{\nu}\Gamma^{\lambda}_{\;\;\;\mu\nu}-\delta_{\alpha}^{\nu}\delta_{\beta}^{\mu}\Gamma^{\lambda}_{\;\;\;\mu\nu})= \nonumber \\
	=\frac{1}{2}(\delta_{\alpha}^{\mu}\delta_{\beta}^{\nu}-\delta_{\alpha}^{\nu}\delta_{\beta}^{\mu})\Gamma^{\lambda}_{\;\;\;\mu\nu}=\delta_{\alpha}^{[\mu}\delta_{\beta}^{\nu]}\Gamma^{\lambda}_{\;\;\;\mu\nu} \Rightarrow  \nonumber
	\end{gather}
	\begin{equation}
	S_{\alpha\beta}^{\;\;\;\;\lambda}=\delta_{\alpha}^{[\mu}\delta_{\beta}^{\nu]}\Gamma^{\lambda}_{\;\;\;\mu\nu}
	\end{equation}
	such that
	\begin{equation}
	\delta_{\Gamma}S_{\alpha\beta}^{\;\;\;\;\lambda}=\delta_{\alpha}^{[\mu}\delta_{\beta}^{\nu]}\delta\Gamma^{\lambda}_{\;\;\;\mu\nu}
	\end{equation}
	So long as the torsion vector is concerned we contract the above in $\beta,\lambda$ to obtain
	\begin{equation}
	\delta_{\Gamma}S_{\alpha}=\delta_{\Gamma}S_{\alpha\lambda}^{\;\;\;\;\lambda}=\delta_{\alpha}^{[\mu}\delta_{\lambda}^{\nu]}\delta\Gamma^{\lambda}_{\;\;\;\mu\nu}
	\end{equation}
	and for the torsion pseudo-vector (in $4$-dim)
	\begin{equation}
	\delta_{\Gamma}\tilde{S}^{\alpha}=\epsilon^{\alpha\mu\nu}_{\;\;\;\;\;\; \lambda}\delta \Gamma^{\lambda}_{\;\;\;\mu\nu}
	\end{equation}
	Having performed the variations of the torsion, we know proceed to derive the variations of the non-metricity tensor with respect to both the metric tensor and the connection.

\subsubsection{Variations of the nonmetricity  variables}
Let us firstly obtain the variation of the non-metricity tensor with respect to the connection. To do so we single out a common $\Gamma^{\lambda}_{\;\;\;\mu\nu}$-factor in the expression of the non-metricity as we did with the torsion. We have
	\begin{gather}
	Q_{\rho\alpha\beta}=-\nabla_{\rho}g_{\alpha\beta}=-\partial_{\rho}g_{\alpha\beta}+\Gamma^{\lambda}_{\;\;\;\alpha\rho}g_{\lambda\beta}+\Gamma^{\lambda}_{\;\;\;\beta\rho}g_{\lambda\alpha}= \nonumber \\
	=-\partial_{\rho}g_{\alpha\beta}+\delta_{\alpha}^{\mu}\delta_{\rho}^{\nu}\Gamma^{\lambda}_{\;\;\;\mu\nu}g_{\lambda\beta}+\delta_{\beta}^{\mu}\delta_{\rho}^{\nu}\Gamma^{\lambda}_{\;\;\;\mu\nu}g_{\lambda\alpha}= \nonumber \\
	=-\partial_{\rho}g_{\alpha\beta}+\delta_{\rho}^{\nu}(\delta_{\alpha}^{\mu}g_{\lambda\beta}+\delta_{\beta}^{\mu}g_{\lambda\alpha})\Gamma^{\lambda}_{\;\;\;\mu\nu}\Rightarrow  \nonumber
	\end{gather}
	\begin{equation}
	Q_{\rho\alpha\beta}=-\partial_{\rho}g_{\alpha\beta}+\delta_{\rho}^{\nu}2\delta_{(\alpha}^{\mu} g_{\beta )\lambda}\Gamma^{\lambda}_{\;\;\;\mu\nu}
	\end{equation}
	Therefore, variation with respect to the connection, immediately gives
	\begin{equation}
	\delta_{\Gamma}Q_{\rho\alpha\beta}=\delta_{\rho}^{\nu}2\delta_{(\alpha}^{\mu} g_{\beta )\lambda}\delta\Gamma^{\lambda}_{\;\;\;\mu\nu} \label{nonqt}
	\end{equation}
	Let us now vary with respect to the metric tensor. Using the above definition of non-metricity along with the identity
	\begin{equation}
	\delta g_{\alpha\beta}=-g_{\mu\alpha}g_{\nu\beta}\delta g^{\mu\nu}
	\end{equation}
	it follows that
	\begin{gather}
	\delta_{g} Q_{\rho\alpha\beta}=-\partial_{\rho}\delta g_{\alpha\beta}+\Gamma^{\lambda}_{\;\;\;\alpha\rho}\delta g_{\lambda\beta}+\Gamma^{\lambda}_{\;\;\;\beta\rho}\delta g_{\lambda\alpha} = \nonumber \\
	=\partial_{\rho}(g_{\mu\alpha}g_{\nu\beta}\delta g^{\mu\nu})-\Gamma^{\lambda}_{\;\;\;\alpha\rho}g_{\lambda\mu}g_{\nu\beta}\delta g^{\mu\nu}-\Gamma^{\lambda}_{\;\;\;\beta\rho}g_{\lambda\mu}g_{\nu\alpha}\delta g^{\mu\nu}= \nonumber \\
	=\partial_{\rho}(g_{\mu\alpha}g_{\nu\beta}\delta g^{\mu\nu})-(\delta g^{\mu\nu})g_{\lambda\mu}2 g_{\nu (\alpha}\Gamma^{\lambda}_{\;\;\;\beta)\rho} \nonumber
	\end{gather}
	Thus, one has
	\begin{equation}
	\delta_{g} Q_{\rho\alpha\beta}=\partial_{\rho}(g_{\mu\alpha}g_{\nu\beta}\delta g^{\mu\nu})-(\delta g^{\mu\nu})2 g_{\lambda\mu} g_{\nu (\alpha}\Gamma^{\lambda}_{\;\;\;\beta)\rho} 
	\end{equation}
	We continue by varying the Weyl vector
	\begin{equation}
	Q_{\nu}\equiv -g^{\alpha\beta}\nabla_{\nu}g_{\alpha\beta}=-g^{\alpha\beta}\partial_{\nu}g_{\alpha\beta}+2 \Gamma^{\lambda}_{\;\;\;\lambda\nu}
	\end{equation}
	Variation with respect to the connection yields\footnote{This may also be obtained by contracting ($\ref{nonqt}$) with $g^{\alpha\beta}$. Of course, this can be done because the $\Gamma$-variation commutes with the metric tensor. However, this is not true for the $g$-variation.}
	\begin{equation}
	\delta_{\Gamma}Q_{\nu}=2\delta \Gamma^{\lambda}_{\;\;\;\lambda\nu}=\delta \Gamma^{\lambda}_{\;\;\;\mu\nu}2\delta_{\lambda}^{\mu} \Rightarrow\nonumber
	\end{equation}
	\begin{equation}
	\delta_{\Gamma}Q_{\rho}=2 \delta_{\rho}^{\nu}\delta_{\lambda}^{\mu}\delta \Gamma^{\lambda}_{\;\;\;\mu\nu}
	\end{equation}
	While variation with respect to the metric tensor gives
	\begin{equation}
	\delta_{g}Q_{\rho}=-(\delta g^{\mu\nu})\partial_{\rho}g_{\mu\nu}-g^{\alpha\beta}\partial_{\rho}\delta g_{\alpha\beta}
	\end{equation}
	Now, expanding the second term, we have
	\begin{gather}
	g^{\alpha\beta}\partial_{\rho}\delta g_{\alpha\beta}=-g^{\alpha\beta}\partial_{\rho}(g_{\mu\alpha}g_{\nu\beta}\delta g^{\mu\nu})= \nonumber \\
	=-g_{\mu\nu}\partial_{\rho}\delta g^{\mu\nu}-2(\delta g^{\mu\nu})\partial_{\rho}g_{\mu\nu}
	\end{gather}
	such that
	\begin{gather}
	\delta_{g}Q_{\rho}=-(\delta g^{\mu\nu})\partial_{\rho}g_{\mu\nu}+g_{\mu\nu}\partial_{\rho}\delta g^{\mu\nu}+2(\delta g^{\mu\nu})\partial_{\rho}g_{\mu\nu}= \nonumber \\
	=g_{\mu\nu}\partial_{\rho}\delta g^{\mu\nu}+(\delta g^{\mu\nu})\partial_{\rho}g_{\mu\nu}=\partial_{\rho}(g_{\mu\nu}\delta g^{\mu\nu})   \nonumber 
	\end{gather}
	Thus, the $g$-variation of the Weyl vector has the handy form
	\begin{equation}
	\delta_{g}Q_{\rho}=\partial_{\rho}(g_{\mu\nu}\delta g^{\mu\nu})  
	\end{equation}
	Let us now proceed by varying the second non-metricity vector $2nmv$. Recall that the latter is given by
	\begin{equation}
	\tilde{Q}_{\beta}=g^{\rho\alpha}Q_{\rho\alpha\beta=}=-g^{\rho\alpha}\partial_{\rho}g_{\alpha\beta}+(g^{\mu\nu}g_{\beta\lambda}+\delta_{\beta}^{\mu}\delta_{\lambda}^{\nu})\Gamma^{\lambda}_{\;\;\;\mu\nu}
	\end{equation}
	Variation with respect to the connection immediately gives
	\begin{equation}
	\delta_{\Gamma}\tilde{Q}_{\beta}=(g^{\mu\nu}g_{\beta\lambda}+\delta_{\beta}^{\mu}\delta_{\lambda}^{\nu})\delta \Gamma^{\lambda}_{\;\;\;\mu\nu}
	\end{equation}
	while variation with respect to the metric tensor reads
	\begin{gather}
	\delta_{g}\tilde{Q}_{\beta}=-(\delta g^{\mu\nu})\partial_{\mu}g_{\nu\beta}-g^{\rho\alpha}\partial_{\rho}\delta g_{\alpha\beta}+(\delta g^{\mu\nu})g_{\beta\lambda}\Gamma^{\lambda}_{\;\;\;\mu\nu}+g^{\mu\nu}\Gamma^{\lambda}_{\;\;\;\mu\nu}\delta g_{\beta\lambda}= \nonumber \\
	=\delta g^{\mu\nu}\Big[ -\partial_{\mu}g_{\nu\beta} +g_{\lambda\beta}\Gamma^{\lambda}_{\;\;\;\mu\nu}\Big] -g^{\rho\alpha}\partial_{\rho}\delta g_{\alpha\beta}+g^{\mu\nu}\Gamma^{\lambda}_{\;\;\;\mu\nu}\delta g_{\beta\lambda}
	\end{gather}
	Now using
	\begin{equation}
	\delta g_{\alpha\beta}=-g_{\alpha\mu}g_{\beta\nu}\delta g^{\mu\nu}
	\end{equation}
	it can easily be shown that
	\begin{equation}
	g^{\rho\alpha}\partial_{\rho}\delta g_{\alpha\beta}=-g_{\beta\nu}g^{\rho\alpha}(\partial_{\rho}g_{\mu\alpha})\delta g^{\mu\nu}-\partial_{\mu}(g_{\nu\beta}\delta g^{\mu\nu})
	\end{equation}
	as well as
	\begin{equation}
	g^{\mu\nu}\Gamma^{\lambda}_{\;\;\;\mu\nu}\delta g_{\beta\lambda}=-g^{\rho\sigma}\Gamma^{\alpha}_{\;\;\;\rho\sigma}g_{\mu\alpha}g_{\nu\beta}\delta g^{\mu\nu}
	\end{equation}
	and upon using these, the g-variation of $\tilde{Q}_{\beta}$ reads
	\begin{equation}
	\delta_{g} \tilde{Q}_{\beta}=\delta g^{\mu\nu}\Big[ g_{\nu\beta}g^{\rho\alpha}(\partial_{\rho}g_{\mu\alpha})+\Gamma^{\lambda}_{\;\;\;\mu\nu}g_{\lambda\beta}-g^{\rho\sigma}\Gamma^{\alpha}_{\;\;\;\rho\sigma}g_{\mu\alpha}g_{\nu\beta}\Big]+g_{\nu\beta}(\partial_{\mu}\delta g^{\mu\nu})
	\end{equation}
	Notice that there is a quicker and more elegant way to derive the $g-$variation of non-metricity. This comes about by first recalling that the general covariant derivative $\nabla_{\alpha}$ does not depend on the metric tensor. Then, using the definition of the variation, one has
	\begin{equation}
	\delta_{g}Q_{\alpha\mu\nu}=-\nabla_{\alpha}(g_{\mu\nu}+\delta g_{\mu\nu})+\nabla_{\alpha}g_{\mu\nu}=-\nabla_{\alpha}\delta g_{\mu\nu}
\end{equation}
	and also
	\begin{equation}
	\delta_{g}Q_{\alpha}^{\;\;\mu\nu}=\nabla_{\alpha}(g^{\mu\nu}+\delta g^{\mu\nu})-\nabla_{\alpha}g_{\mu\nu}=+\nabla_{\alpha}\delta g^{\mu\nu}
	\end{equation}
	So, when coupled to a tensor filed (or a tensor density) $T^{\alpha}_{\;\;\mu\nu}$ we have
	\begin{equation}
	T^{\alpha}_{\;\;\mu\nu}\delta_{g}Q_{\alpha}^{\;\;\mu\nu}=\nabla_{\alpha}(T^{\alpha}_{\;\;\mu\nu}\delta g^{\mu\nu})-(\delta g^{\mu\nu})\nabla_{\alpha}T^{\alpha}_{\;\;\mu\nu}
	\end{equation}
	where we have employed Leibniz's rule for the covariant derivatives. Next we derive the variations of the Riemann tensor.

\subsubsection{Variations of the Riemann  variables}
For the sake of completeness we also give here the variations of the Riemann tensor (and its related contractions) with respect to the independent connection and the metric. First notice that the prototype of the Riemann tensor
	\begin{equation}
	R^{\mu}_{\;\;\;\nu\alpha\beta}:=2\partial_{[\alpha}\Gamma^{\mu}_{\;\;\;|\nu|\beta]}+2\Gamma^{\mu}_{\;\;\;\rho[\alpha}\Gamma^{\rho}_{\;\;\;|\nu|\beta]} \label{defriem}
	\end{equation}
	does not depend on the metric and therefore
	\begin{equation}
	\delta_{g} R^{\mu}_{\;\;\;\nu\alpha\beta}=0
	\end{equation}
	When the first index is brought down however we have a metric tensor dependence since
	\begin{equation}
	R_{\rho\nu\alpha\beta}=g_{\mu\rho}R^{\mu}_{\;\;\;\nu\alpha\beta}
	\end{equation}
	and thus
	\begin{equation}
	\delta_{g}R_{\rho\nu\alpha\beta}=(\delta g_{\mu\rho})R^{\mu}_{\;\;\;\nu\alpha\beta}=-(\delta g^{\kappa\lambda})g_{\mu\kappa}g_{\rho\lambda}R^{\mu}_{\;\;\;\nu\alpha\beta}=-(\delta g^{\kappa\lambda})g_{\rho\lambda}R_{\kappa\nu\alpha\beta}
	\end{equation}
	Now, to derive the variation with respect to the connection we start by $(\ref{defriem})$ and compute
	\begin{equation}
	\delta_{\Gamma}R^{\mu}_{\;\;\;\nu\alpha\beta}=R^{\mu}_{\;\;\;\nu\alpha\beta}[ \Gamma +\delta \Gamma]-R^{\mu}_{\;\;\;\nu\alpha\beta}[ \Gamma ]
	\end{equation}
	and expanding $R^{\mu}_{\;\;\;\nu\alpha\beta}[ \Gamma +\delta \Gamma]$ to linear order in $\delta \Gamma$ we finally arrive at
	\begin{equation}
	\delta_{\Gamma}R^{\mu}_{\;\;\;\nu\alpha\beta}=\nabla_{\alpha}(\delta \Gamma^{\mu}_{\;\;\;\nu\beta})-\nabla_{\beta}(\delta \Gamma^{\mu}_{\;\;\;\nu\alpha})-2 S_{\alpha\beta}^{\;\;\;\;\lambda}\delta\Gamma^{\mu}_{\;\;\;\nu\lambda}
	\end{equation}
	Having obtained all he necessary setup we are now in a position to study Metric-Affine Theories of Gravity. We do so in what follows.

\subsubsection{Variations of the geometrical scalars}
\subsubsection{Variations of the energy-momentum  variables}
\section{Harko paper appendices}
\subsection{Calculation of $Q=-Q_{\alpha\mu\nu}P^{\alpha\mu\nu}$}\label{app1}
According to Eq.~(\ref{eq:Q} )and Eq.~(\ref{eq:disformation}), we have
\begin{eqnarray}
 &&Q\equiv -g^{\mu \nu}\left( L^{\alpha}_{\ \ \beta\mu}L^{\beta}_{\ \ \nu\alpha} - L^{\alpha}_{\ \ \beta\alpha} L^{\beta}_{\ \ \mu \nu} \right) , \\
&& L^{\alpha}_{\ \ \beta \mu}= -\frac{1}{2} g^{\alpha \lambda}\left( Q_{\mu\beta\lambda} +Q_{\beta\lambda\mu} - Q_{\lambda\mu\beta} \right) , \\
&& L^{\beta}_{\ \ \nu \alpha}= -\frac{1}{2} g^{\beta \rho}\left(Q_{\alpha\nu\rho} +Q_{\nu\rho\alpha} - Q_{\rho\alpha\nu}\right) , \\
&&  L^{\alpha}_{\ \ \beta \alpha}= -\frac{1}{2} g^{\alpha \lambda}\left( Q_{\alpha\beta\lambda} +Q_{\beta\lambda\alpha} - Q_{\lambda\alpha\beta}\right) \nonumber \\
&& = -\frac{1}{2}\left( \tilde{Q}_{\beta} + Q_{\beta} - \tilde{Q}_{\beta} \right) = -\frac{1}{2} Q_{\beta}, \\
&& L^{\beta}_{\ \ \mu\nu}= -\frac{1}{2} g^{\beta \rho}\left( Q_{\nu\mu\rho} +Q_{\mu\rho\nu} - Q_{\rho\nu\mu} \right) .
\end{eqnarray}
Thus, we obtain
\begin{eqnarray}
&&-g^{\mu \nu} L^{\alpha}_{\ \ \beta\mu}L^{\beta}_{\ \ \nu\alpha}= -\frac{1}{4}g^{\mu\nu}g^{\alpha\lambda}g^{\beta\rho}\left( Q_{\mu\beta\lambda} +Q_{\beta\lambda\mu} - Q_{\lambda\mu\beta} \right) \nonumber \\
&& \times \left( Q_{\alpha\nu\rho} +Q_{\nu\rho\alpha} - Q_{\rho\alpha\nu} \right) = -\frac{1}{4} \left( Q^{\nu\rho\alpha} +Q^{\rho\alpha\nu}-Q^{\alpha\nu\rho} \right) \nonumber \\
&&\times \left( Q_{\alpha\nu\rho} +Q_{\nu\rho\alpha} - Q_{\rho\alpha\nu} \right) = -\frac{1}{4} (  \cancel{Q^{\nu\rho\alpha}Q_{\alpha\nu\rho}} +Q^{\nu\rho\alpha}Q_{\nu\rho\alpha} \nonumber \\
&& \bcancel{- Q^{\nu\rho\alpha}Q_{\rho\alpha\nu}}  + Q^{\rho\alpha\nu}Q_{\alpha\nu\rho} +\bcancel{Q^{\rho\alpha\nu}Q_{\nu\rho\alpha}} - Q^{\rho\alpha\nu}Q_{\rho\alpha\nu} \nonumber \\
&& -Q^{\alpha\nu\rho}Q_{\alpha\nu\rho} \cancel{-Q^{\alpha\nu\rho}Q_{\nu\rho\alpha}} +Q^{\alpha\nu\rho} Q_{\rho\alpha\nu})  \nonumber \\
&& =  -\frac{1}{4}\left( 2 Q^{\alpha\nu\rho} Q_{\rho\alpha\nu}-Q^{\alpha\nu\rho}Q_{\alpha\nu\rho} \right) , \\
&&g^{\mu \nu}L^{\alpha}_{\ \ \beta\alpha} L^{\beta}_{\ \ \mu \nu}= \frac{1}{4}g^{\mu\nu}g^{\beta\rho}Q_{\beta}\left( Q_{\nu\mu\rho} +Q_{\mu\rho\nu} - Q_{\rho\nu\mu} \right) \nonumber \\
&& = \frac{1}{4}Q^{\rho}\left( 2\tilde{Q}_{\rho}-Q_{\rho}  \right), \\
&& Q =- \frac{1}{4}\left( -Q^{\alpha\nu\rho}Q_{\alpha\nu\rho}+ 2 Q^{\alpha\nu\rho} Q_{\rho\alpha\nu} - 2Q^{\rho}\tilde{Q}_{\rho} + Q^{\rho}Q_{\rho} \right). \nonumber \\
\label{eq:deltaQ}
\end{eqnarray}
Then, according to Eq.~(\ref{eq:superpotential}), we have
\begin{eqnarray}
&&P^{\alpha\mu\nu}=  \frac{1}{4}\bigg[  -Q^{\alpha \mu \nu}+ Q^{\mu \alpha  \nu} +Q^{\nu \alpha  \mu} + Q^{\alpha}g^{\mu \nu} - \tilde{Q}^{\alpha}g^{\mu\nu}\nonumber \\
&&
-\frac{1}{2}\left(  g^{\alpha\mu}Q^{\nu} + g^{\alpha\nu}Q^{\mu}  \right) \bigg ], \\
&& -Q_{\alpha\mu\nu}P^{\alpha\mu\nu}= -\frac{1}{4}\bigg[  -Q_{\alpha\mu\nu}Q^{\alpha \mu \nu}+ Q_{\alpha\mu\nu}Q^{\mu \alpha  \nu} \nonumber \\
&&+Q_{\alpha\mu\nu}Q^{\nu \alpha  \mu} + Q_{\alpha\mu\nu}Q^{\alpha}g^{\mu \nu} - Q_{\alpha\mu\nu}\tilde{Q}^{\alpha}g^{\mu\nu} \nonumber \\
&&-\frac{1}{2}Q_{\alpha\mu\nu}\left(  g^{\alpha\mu}Q^{\nu} + g^{\alpha\nu}Q^{\mu} \right) \bigg ] = -\frac{1}{4}(-Q_{\alpha\mu\nu}Q^{\alpha \mu \nu}\nonumber \\
&&+2 Q_{\alpha\mu\nu}Q^{\mu \alpha  \nu} + Q_{\alpha}Q^{\alpha}-2Q_{\alpha}\tilde{Q}^{\alpha}) =Q.\label{eq:Qproof}
\end{eqnarray}

To obtain the above result we have used the relations  $Q_{\alpha\mu\nu}Q^{\mu \alpha  \nu}=Q_{\alpha\mu\nu}Q^{\nu \alpha  \mu}$, which is valid since $Q_{\alpha\mu\nu}Q^{\mu \alpha  \nu}=Q_{\alpha\nu\mu}Q^{\mu \alpha  \nu}=Q^{\alpha\nu\mu}Q_{\mu \alpha  \nu}=Q^{\nu\mu\alpha}Q_{\alpha  \nu\mu}=Q_{\alpha\mu\nu}Q^{\nu \alpha  \mu}$. Hence, we have proved that $Q=-Q_{\alpha\mu\nu}P^{\alpha\mu\nu}$, a relation which is very useful in later calculations.

\subsection{Calculation of the variation of $\delta Q$}\label{app2}

Before the presentation of the detailed variation of $\delta Q$, we write down all the nonmetricity tensors for later applications. They are obtained as
\begin{eqnarray}
\hspace{-0.8cm}&&Q_{\alpha\mu\nu}=\nabla_{\alpha}g_{\mu\nu}, \\
\hspace{-0.8cm}&&Q^{\alpha}_{\ \ \mu\nu}=g^{\alpha\beta}Q_{\beta\mu\nu}=g^{\alpha\beta}\nabla_{\beta} g_{\mu\nu}=\nabla^{\alpha}g_{\mu\nu}, \\
\hspace{-0.8cm}&& Q_{\alpha \ \ \nu}^{\ \ \mu}=g^{\mu\rho}Q_{\alpha\rho\nu}=g^{\mu\rho}\nabla_{\alpha}g_{\rho\nu} = - g_{\rho\nu} \nabla_{\alpha}g^{\mu\rho},\\
\hspace{-0.8cm}&& Q_{\alpha\mu}^{\ \ \ \ \nu}=g^{\nu\rho}Q_{\alpha\mu\rho}=g^{\nu\rho}\nabla_{\alpha}g_{\mu\rho}=-g_{\mu\rho}\nabla_{\alpha}g^{\nu\rho}, \\
\hspace{-0.8cm}&& Q^{\alpha\mu}_{\ \ \ \ \nu}=g^{\alpha\beta}g^{\mu\rho}\nabla_{\beta}g_{\rho\nu}=g^{\mu\rho}\nabla^{\alpha}g_{\rho\nu}=-g_{\rho\nu}\nabla^{\alpha}g^{\mu\rho},\nonumber \\ \\
\hspace{-0.8cm}&& Q^{\alpha \ \ \nu}_{\ \ \mu}= g^{\alpha\beta}g^{\nu\rho}\nabla_{\beta}g_{\mu\rho}=g^{\nu\rho}\nabla^{\alpha}g_{\mu\rho}=-g_{\mu\rho}\nabla^{\alpha}g^{\nu\rho} ,\nonumber \\ \\
\hspace{-0.8cm}&&Q_{\alpha}^{\ \ \mu\nu}=g^{\mu\rho}g^{\nu\sigma}\nabla_{\alpha}g_{\rho\sigma}=-g^{\mu\rho}g_{\rho\sigma}\nabla_{\alpha}g^{\nu\sigma}=-\nabla_{\alpha}g^{\mu\nu}, \nonumber \\ \\
\hspace{-0.8cm}&& Q^{\alpha\mu\nu}=-\nabla^{\alpha}g^{\mu\nu}
\end{eqnarray}

Let us find the variation of $Q$ by using Eq.~(\ref{eq:deltaQ}),
\begin{eqnarray}\label{eq:deltaQre}
&&\delta Q \nonumber \\
&&= - \frac{1}{4}\delta\left( -Q^{\alpha\nu\rho}Q_{\alpha\nu\rho}+ 2 Q^{\alpha\nu\rho} Q_{\rho\alpha\nu} - 2Q^{\rho}\tilde{Q}_{\rho} + Q^{\rho}Q_{\rho}  \right) \nonumber \\
&& =-\frac{1}{4}( -\delta Q^{\alpha\nu\rho}Q_{\alpha\nu\rho} -Q^{\alpha\nu\rho}\delta Q_{\alpha\nu\rho} + 2 \delta Q^{\alpha\nu\rho} Q_{\rho\alpha\nu} \nonumber \\
&& + 2 Q^{\alpha\nu\rho} \delta Q_{\rho\alpha\nu}- 2 \delta Q^{\rho}\tilde{Q}_{\rho}-2Q^{\rho}\delta \tilde{Q}_{\rho} + \delta Q^{\rho}Q_{\rho} \nonumber\\
&&+ Q^{\rho}\delta Q_{\rho})  \nonumber \\
&&= -\frac{1}{4}\Big[ Q_{\alpha\nu\rho}  \nabla^{\alpha}\delta g^{\nu\rho}-Q^{\alpha\nu\rho} \nabla_{\alpha}\delta g_{\nu\rho} - 2 Q_{\rho\alpha\nu} \nabla^{\alpha}\delta g^{\nu\rho} \nonumber \\
&&+ 2 Q^{\alpha\nu\rho} \nabla_{\rho}\delta g_{\alpha\nu}- 2\tilde{Q}_{\rho} \delta (-g_{\mu\nu}\nabla^{\rho}g^{\mu\nu})-2Q^{\rho}\delta (\nabla^{\lambda} g_{\rho\lambda}) \nonumber \\
&&+Q_{\rho} \delta (-g_{\mu\nu}\nabla^{\rho}g^{\mu\nu})+ Q^{\rho}\delta (-g_{\mu\nu}\nabla_{\rho}g^{\mu\nu})\Big] \nonumber \\
&&= -\frac{1}{4}\Big[ Q_{\alpha\nu\rho}  \nabla^{\alpha}\delta g^{\nu\rho}-Q^{\alpha\nu\rho} \nabla_{\alpha}\delta g_{\nu\rho} - 2 Q_{\rho\alpha\nu} \nabla^{\alpha}\delta g^{\nu\rho} \nonumber \\
&&+ 2 Q^{\alpha\nu\rho} \nabla_{\rho}\delta g_{\alpha\nu}+ 2\tilde{Q}_{\rho}\nabla^{\rho}g^{\mu\nu} \delta g_{\mu\nu}+ 2\tilde{Q}_{\rho}  g_{\mu\nu}\nabla^{\rho}\delta g^{\mu\nu}\nonumber \\
&&-2Q^{\rho} \nabla^{\lambda}\delta g_{\rho\lambda} -Q_{\rho}\nabla^{\rho}g^{\mu\nu} \delta g_{\mu\nu} -  Q_{\rho}  g_{\mu\nu}\nabla^{\rho}\delta g^{\mu\nu} \nonumber \\
&&- Q^{\rho}\nabla_{\rho}g^{\mu\nu}\delta g_{\mu\nu}   - Q^{\rho} g_{\mu\nu}\nabla_{\rho}\delta g^{\mu\nu} \Big] \nonumber \\
&&= -\frac{1}{4}\Big[ Q_{\alpha\nu\rho}  \nabla^{\alpha}\delta g^{\nu\rho}-Q^{\alpha\nu\rho} \nabla_{\alpha}\delta g_{\nu\rho} - 2 Q_{\rho\alpha\nu} \nabla^{\alpha}\delta g^{\nu\rho} \nonumber \\
&&+ 2 Q^{\alpha\nu\rho} \nabla_{\rho}\delta g_{\alpha\nu}+ 2\tilde{Q}_{\rho}\nabla^{\rho}g^{\mu\nu} \delta g_{\mu\nu}+ 2\tilde{Q}_{\rho}  g_{\mu\nu}\nabla^{\rho}\delta g^{\mu\nu}\nonumber \\
&&-2Q^{\rho} \nabla^{\lambda}\delta g_{\rho\lambda} -Q_{\rho}\nabla^{\rho}g^{\mu\nu} \delta g_{\mu\nu} -  Q_{\rho}  g_{\mu\nu}\nabla^{\rho}\delta g^{\mu\nu} \nonumber \\
&&- Q^{\rho}\nabla_{\rho}g^{\mu\nu}\delta g_{\mu\nu}   - Q^{\rho} g_{\mu\nu}\nabla_{\rho}\delta g^{\mu\nu} \Big].
\end{eqnarray}

In order to simplify the above equation we can use several useful equations, which are given below as
\begin{eqnarray}
&&\delta g_{\mu\nu}=-g_{\mu\alpha}\delta g^{\alpha\beta} g_{\beta\nu}, \\
&&-Q^{\alpha\nu\rho}\nabla_{\alpha}\delta g_{\nu\rho}=-Q^{\alpha\nu\rho}\nabla_{\alpha} \left( -g_{\nu\lambda}\delta g^{\lambda\theta} g_{\theta\rho}   \right) \nonumber \\
&& = 2 Q^{\alpha\nu}_{\ \ \ \ \theta}Q_{\alpha\nu\lambda} \delta g^{\lambda\theta}+Q_{\alpha\lambda\theta}\nabla^{\alpha}g^{\lambda\theta}\nonumber \\
&&=2 Q^{\alpha\sigma}_{\ \ \ \ \nu}Q_{\alpha\sigma\mu} \delta g^{\mu\nu}+Q_{\alpha\nu\rho}\nabla^{\alpha}g^{\nu\rho}, \\
&&  2 Q^{\alpha\nu\rho} \nabla_{\rho}\delta g_{\alpha\nu}=-4 Q_{\mu}^{\ \ \sigma\rho}Q_{\rho\sigma\nu}\delta g^{\mu\nu} -2 Q_{\nu\rho\alpha}\nabla^{\alpha}\delta g^{\nu\rho}, \nonumber \\ \\
&&-2Q^{\rho} \nabla^{\lambda}\delta g_{\rho\lambda}=2 Q^{\alpha}Q_{\nu\alpha\mu}\delta g^{\mu\nu} + 2 Q_{\mu}\tilde{Q}_{\nu}\delta g^{\mu\nu}  \nonumber \\
&&+ 2 Q_{\nu}g_{\alpha\rho}\nabla^{\alpha}g^{\nu\rho}.
\end{eqnarray}

Thus, Eq.~(\ref{eq:deltaQre}) takes the form
\begin{eqnarray}
&&\delta Q  \nonumber \\
&&= -\frac{1}{4}\Big[ Q_{\alpha\nu\rho}  \nabla^{\alpha}\delta g^{\nu\rho}+2 Q^{\alpha\sigma}_{\ \ \ \ \nu}Q_{\alpha\sigma\mu} \delta g^{\mu\nu}+Q_{\alpha\nu\rho}\nabla^{\alpha}g^{\nu\rho} \nonumber \\
&&- 2 Q_{\rho\alpha\nu} \nabla^{\alpha}\delta g^{\nu\rho} -4 Q_{\mu}^{\ \ \sigma\rho}Q_{\rho\sigma\nu}\delta g^{\mu\nu} -2 Q_{\nu\rho\alpha}\nabla^{\alpha}\delta g^{\nu\rho}  \nonumber \\
&&+ 2\tilde{Q}^{\rho}Q_{\rho\mu\nu} \delta g^{\mu\nu}+ 2\tilde{Q}_{\alpha}  g_{\nu\rho}\nabla^{\alpha}\delta g^{\nu\rho}+2 Q^{\alpha}Q_{\nu\alpha\mu}\delta g^{\mu\nu} \nonumber \\
&& + 2 Q_{\mu}\tilde{Q}_{\nu}\delta g^{\mu\nu} + 2 Q_{\nu}g_{\alpha\rho}\nabla^{\alpha}g^{\nu\rho}  -Q^{\rho}Q_{\rho\mu\nu} \delta g^{\mu\nu}  \nonumber \\
&&-  Q_{\alpha}  g_{\nu\rho}\nabla^{\alpha}\delta g^{\nu\rho}- Q^{\rho}Q_{\rho\mu\nu}\delta g^{\mu\nu}   - Q_{\alpha} g_{\nu\rho}\nabla^{\alpha}\delta g^{\nu\rho} \Big] \nonumber \\
&& = 2 P_{\alpha\nu\rho} \nabla^{\alpha}\delta g^{\nu\rho} - \left( P_{\mu\alpha\beta}Q_{\nu}^{\ \ \alpha\beta} -2 Q^{\alpha\beta}_{\ \ \ \ \mu}P_{\alpha\beta\nu} \right) \delta g^{\mu\nu}, \nonumber \\
\end{eqnarray}
where we have used the relations
\begin{eqnarray}
&&2 P_{\alpha\nu\rho}= -\frac{1}{4}\big [ 2 Q_{\alpha\nu\rho} -2 Q_{\rho\alpha\nu}-2 Q_{\nu\rho\alpha}  \nonumber \\
&&+ 2 (\tilde{Q}_{\alpha} - Q_{\alpha}) g_{\nu \rho}+ 2 Q_{\nu}g_{\alpha\rho}  \big ], \\
&&4\left( P_{\mu\alpha\beta}Q_{\nu}^{\ \ \alpha\beta} -2 Q^{\alpha\beta}_{\ \ \ \ \mu}P_{\alpha\beta\nu}\right) = 2 Q^{\alpha\beta}_{\ \ \ \ \nu}Q_{\alpha\beta\mu}\nonumber \\
&& -4 Q_{\mu}^{\ \ \alpha\beta}Q_{\beta\alpha\nu} +2 \tilde{Q}^{\alpha}Q_{\alpha\mu\nu} + 2 Q^{\alpha}Q_{\nu\alpha\mu} \nonumber \\
&&+2Q_{\mu}\tilde{Q}_{v}-Q^{\alpha}Q_{\alpha\mu\nu}.
\end{eqnarray}

\subsection{Variation of the gravitational action with respect to the connection}\label{app3}

The full action of the $f(Q,T)$ theory supplemented  with the Lagrangian multipliers is
\begin{eqnarray}
S= \int \mathrm{d} ^4 x \bigg[ \frac{ \sqrt{-g}}{16 \pi} f(Q, T) + \mathcal{L}_{M} \sqrt{-g}\nonumber \\
+ \lambda_{\alpha}^{\ \ \beta\gamma} T^{\alpha}_{\ \ \beta \gamma}  +\xi_{\alpha}^{\ \ \beta\mu\nu} R^{\alpha}_{\ \ \beta \mu\nu}  \bigg].
\end{eqnarray}
We can vary the action separately, thus obtaining
\begin{eqnarray}
&&\delta \bigg[  \frac{ \sqrt{-g}}{16 \pi} f(Q, T) + \mathcal{L}_{M} \sqrt{-g} \bigg]\nonumber \\
&&=  \bigg( \frac{4 \sqrt{-g}}{16 \pi} f_{Q} P^{\mu\nu}_{\ \ \ \ \alpha} +H_{\alpha}^{\ \ \mu \nu} \bigg)\delta \hat{\Gamma}^{\alpha}_{\ \ \mu\nu}\;, \\
&& \delta \big(\lambda_{\alpha}^{\ \ \mu\nu} T^{\alpha}_{\ \ \mu \nu} \big) = 2 \lambda_{\alpha}^{\ \ \mu\nu} \delta \hat{\Gamma}^{\alpha}_{\ \ \mu \nu}, \\
&& \delta \big( \xi_{\alpha}^{\ \ \beta\mu\nu} R^{\alpha}_{\ \ \beta \mu\nu} \big)=\xi_{\alpha}^{\ \ \beta\mu\nu} \Big[  \nabla_{\mu} \big( \delta \hat{\Gamma}^{\alpha}_{\ \ \nu\beta}\big)- \nabla_{\nu} \big( \delta \hat{\Gamma}^{\alpha}_{\ \ \mu\beta}\big)       \Big]\nonumber \\
&&=2\xi_{\alpha}^{\ \ \nu\beta\mu}\nabla_{\beta}\big(  \delta  \hat{\Gamma}^{\alpha}_{\ \ \mu\nu}   \big) \simeq  2\big( \nabla_{\beta} \xi_{\alpha}^{\ \ \nu\beta\mu}\big)  \delta  \hat{\Gamma}^{\alpha}_{\ \ \mu\nu} .
\end{eqnarray}

Thus,
\begin{eqnarray}
\delta S= \int \mathrm{d} ^4 x \bigg( \frac{4 \sqrt{-g}}{16 \pi} f_{Q} P^{\mu\nu}_{\ \ \ \ \alpha} +H_{\alpha}^{\ \ \mu \nu} +  2\lambda_{\alpha}^{\ \ \mu\nu} \nonumber \\
 + 2 \nabla_{\beta} \xi_{\alpha}^{\ \ \nu\beta\mu}    \bigg)    \delta  \hat{\Gamma}^{\alpha}_{\ \ \mu\nu} .
\end{eqnarray}

To eliminate the Lagrange multipliers, we take two covariant derivatives $\nabla_{\mu}\nabla_{\nu}$ or $\nabla_{\nu}\nabla_{\mu}$ (considering vanishing curvature tensor) of the integrand, and thus we finally arrive to Eq.~(\ref{eq:feqconnection}).

\subsection{Metric divergence of (1,1)-form field equations}\label{app4}

The metric divergence of the gravitational field equation Eq.~(\ref{eq:metricdivfeq}) of the $f(Q,T)$ theory is
\begin{eqnarray}
&&{\cal D}_{\mu}\Big[ f_{T}\big(T^{\mu}_{\ \ \nu}+\Theta^{\mu}_{\ \ \nu}\big)-8\pi T^{\mu}_{\ \ \nu}\Big]=\frac{1}{2}\partial_{\nu}f \nonumber \\
&&+{\cal D}_{\mu}\Big(f_{Q}Q_{\nu}^{\ \ \alpha\beta}P^{\mu}_{\ \ \alpha\beta}\Big)+{\cal D}_{\mu}\bigg[\frac{2}{\sqrt{-g}}\nabla_{\alpha}\big(f_{Q}\sqrt{-g} P^{\alpha\mu}_{\ \ \ \ \nu}\big) \bigg], \nonumber \\
\end{eqnarray}
where we have
\begin{eqnarray}
\hspace{-0.5cm}&&{\cal D}_{\mu}\Big(f_{Q}Q_{\nu}^{\ \ \alpha\beta}P^{\mu}_{\ \ \alpha\beta}\Big)=\nabla_{\mu}\Big(f_{Q}Q_{\nu}^{\ \ \alpha\beta}P^{\mu}_{\ \ \alpha\beta}\Big) \nonumber \\
\hspace{-0.5cm}&&+ \frac{1}{2}Q_{\mu}\Big(f_{Q}Q_{\nu}^{\ \ \alpha\beta}P^{\mu}_{\ \ \alpha\beta}\Big) + L^{\rho}_{\ \ \mu\nu}\Big(f_{Q}Q_{\rho}^{\ \ \alpha\beta}P^{\mu}_{\ \ \alpha\beta}\Big) ,\\
\hspace{-0.5cm}&& {\cal D}_{\mu}\bigg[\frac{2}{\sqrt{-g}}\nabla_{\alpha}\big(f_{Q}\sqrt{-g} P^{\alpha\mu}_{\ \ \ \ \nu}\big) \bigg] \nonumber \\
\hspace{-0.5cm}&&=\frac{2}{\sqrt{-g}}{\cal D}_{\mu}\bigg[\nabla_{\alpha}\big(f_{Q}\sqrt{-g} P^{\alpha\mu}_{\ \ \ \ \nu}\big) \bigg] \nonumber \\
\hspace{-0.5cm}&& = \frac{2}{\sqrt{-g}} \nabla_{\mu}\nabla_{\alpha}\big(f_{Q}\sqrt{-g} P^{\alpha\mu}_{\ \ \ \ \nu}\big) \nonumber \\
\hspace{-0.5cm}&&+\frac{1}{\sqrt{-g}} Q_{\mu}\nabla_{\alpha}\big(f_{Q}\sqrt{-g} P^{\alpha\mu}_{\ \ \ \ \nu}\big) \nonumber \\
\hspace{-0.5cm}&& +\frac{2}{\sqrt{-g}} L^{\rho}_{\ \ \mu\nu}\nabla_{\alpha}\big(f_{Q}\sqrt{-g} P^{\alpha\mu}_{\ \ \ \ \rho}\big),
\end{eqnarray}
which gives
\begin{eqnarray}
&&{\cal D}_{\mu}\Big[ f_{T}\big(T^{\mu}_{\ \ \nu}+\Theta^{\mu}_{\ \ \nu}\big)-8\pi T^{\mu}_{\ \ \nu}\Big]+\frac{8\pi}{\sqrt{-g}}\nabla_{\alpha}\nabla_{\mu}H_{\nu}^{\ \ \alpha\mu} \nonumber \\
&&=\frac{1}{2}\partial_{\nu}f+\nabla_{\mu}\Big(f_{Q}Q_{\nu}^{\ \ \alpha\beta}P^{\mu}_{\ \ \alpha\beta}\Big)+\frac{1}{2}Q_{\mu}\Big(f_{Q}Q_{\nu}^{\ \ \alpha\beta}P^{\mu}_{\ \ \alpha\beta}\Big) \nonumber \\
&&+ L^{\rho}_{\ \ \mu\nu}\Big(f_{Q}Q_{\rho}^{\ \ \alpha\beta}P^{\mu}_{\ \ \alpha\beta}\Big) +\frac{2}{\sqrt{-g}} L^{\rho}_{\ \ \mu\nu}\nabla_{\alpha}\Big(f_{Q}\sqrt{-g} P^{\alpha\mu}_{\ \ \ \ \rho}\Big)\nonumber \\
&& +\frac{1}{\sqrt{-g}} Q_{\mu}\nabla_{\alpha}\Big(f_{Q}\sqrt{-g} P^{\alpha\mu}_{\ \ \ \ \nu}\Big)=\sum_{i=1}^{10} E_i.
\end{eqnarray}

For the sake of clarity, in the above equation we have defined
\begin{eqnarray}
&&E_1=\frac{1}{2}\partial_{\nu}f ,\\
&&E_2=\Big(\nabla_{\mu}f_{Q}\Big)Q_{\nu\alpha\beta}P^{\mu\alpha\beta},\\
&&E_3= f_{Q}\Big( \nabla_{\mu}Q_{\nu\alpha\beta}\Big)P^{\mu\alpha\beta}, \\
&& E_4=f_{Q}Q_{\nu\alpha\beta}\Big(\nabla_{\mu}P^{\mu\alpha\beta}\Big),\\
&& E_5=\frac{1}{2}f_{Q} Q_{\mu}Q_{\nu\alpha\beta}P^{\mu\alpha\beta}, \\
&& E_6=f_{Q} L^{\rho}_{\ \ \mu\nu}Q_{\rho\alpha\beta}P^{\mu\alpha\beta},\\
&& E_7= 2\Big(\nabla_{\alpha}f_{Q}\Big) L^{\rho}_{\ \ \mu\nu}P^{\alpha\mu}_{\ \ \ \ \rho},\\
&& E_8=f_{Q}Q_{\alpha}L^{\rho}_{\ \ \mu\nu}P^{\alpha\mu}_{\ \ \ \ \rho}, \\
&& E_9=2 f_{Q}L^{\rho}_{\ \ \mu\nu}\nabla_{\alpha}P^{\alpha\mu}_{\ \ \ \ \rho},\\
&& E_{10}=\frac{1}{\sqrt{-g}} Q_{\mu}\nabla_{\alpha}\Big(f_{Q}\sqrt{-g} P^{\alpha\mu}_{\ \ \ \ \nu}\Big).
\end{eqnarray}

Then, we can find the following relations
\begin{eqnarray}
&&E_2+E_7=\nabla_{\mu}f_{Q}\Big( Q_{\nu\alpha\beta} + 2 L_{\beta\alpha\nu}\Big)P^{\mu\alpha\beta}=0, \\
&&E_5+E_8=\frac{1}{2}f_{Q}Q_{\mu}\Big( Q_{\nu\alpha\beta} + 2 L_{\beta\alpha\nu}\Big)P^{\mu\alpha\beta}=0, \\
&&E_4+E_9=f_{Q}\bigg[Q_{\nu\alpha\beta}\Big(\nabla_{\mu}P^{\mu\alpha\beta}\Big) +2L^{\rho}_{\ \ \mu\nu}\nabla_{\alpha}P^{\alpha\mu}_{\ \ \ \ \rho}\bigg] \nonumber \\
&&=f_{Q}\bigg[\Big(Q_{\nu\alpha\beta} +2 L_{\beta\alpha\nu}  \Big)\nabla_{\mu}P^{\mu\alpha\beta} +2L^{\rho}_{\ \ \alpha\nu}Q_{\mu\beta\rho}P^{\mu\alpha\beta}\bigg]\nonumber \\
&&=2f_{Q}L^{\rho}_{\ \ \alpha\nu}Q_{\mu\beta\rho}P^{\mu\alpha\beta}, \\
&& E_3+E_6+E_4+E_9\nonumber \\
&&=f_{Q}\Big( \nabla_{\mu}Q_{\nu\alpha\beta}+2L^{\rho}_{\ \ \alpha\nu}Q_{\mu\beta\rho}+L^{\rho}_{\ \ \mu\nu}Q_{\rho\alpha\beta}\Big)P^{\mu\alpha\beta} \nonumber \\
&&=\frac{1}{2}f_{Q}{\cal D}_{\nu}\Big( Q_{\mu\alpha\beta} P^{\mu\alpha\beta}      \Big)=-\frac{1}{2}f_{Q}\partial_{\nu}Q.
\end{eqnarray}

Finally, we obtain
\begin{eqnarray}
{\cal D}_{\mu}\Big[ f_{T}\big(T^{\mu}_{\ \ \nu}+\Theta^{\mu}_{\ \ \nu}\big)-8\pi T^{\mu}_{\ \ \nu}\Big]+\frac{8\pi}{\sqrt{-g}}\nabla_{\alpha}\nabla_{\mu}H_{\nu}^{\ \ \alpha\mu}\nonumber \\
= \frac{1}{2}\partial_{\nu}f-\frac{1}{2}f_{Q}\partial_{\nu}Q + \frac{1}{\sqrt{-g}} Q_{\mu}\nabla_{\alpha}\Big(f_{Q}\sqrt{-g} P^{\alpha\mu}_{\ \ \ \ \nu}\Big)\nonumber \\
= \frac{1}{2}f_{T}\partial_{\nu}T+ \frac{1}{\sqrt{-g}} Q_{\mu}\nabla_{\alpha}\Big(f_{Q}\sqrt{-g} P^{\alpha\mu}_{\ \ \ \ \nu}\Big).\nonumber \\
\end{eqnarray}

\subsection{Calculation of $ Q=6H^2/N^2$} \label{app5}

Recalling Eq.~(\ref{eq:Qproof}), we have
\begin{eqnarray}
Q=-\frac{1}{4}\Big(-Q_{\alpha\mu\nu}Q^{\alpha \mu \nu}+2 Q_{\alpha\mu\nu}Q^{\mu \alpha  \nu}\nonumber \\
 + Q_{\alpha}Q^{\alpha}-2Q_{\alpha}\tilde{Q}^{\alpha}\Big) .
\end{eqnarray}

By using the relations already presented  in Appendix~\ref{app2}, for the case of the Friedmann-Robertson-Walker metric we obtain
\begin{eqnarray}
&&-Q_{\alpha\mu\nu}Q^{\alpha \mu \nu}=\nabla_{\alpha}g_{\mu\nu}\nabla^{\alpha}g^{\mu\nu}=\frac{4}{N^2}\left( T^2 +3H^2\right), \\
&&Q_{\alpha\mu\nu}Q^{\mu \alpha  \nu}=-\nabla_{\alpha}g_{\mu\nu}\nabla^{\mu}g^{\alpha\nu}=-\frac{4}{N^2}T^2, \\
&&Q_{\alpha}Q^{\alpha}=\left( g_{\rho\mu}\nabla_{\alpha}g^{\rho\mu} \right)\left( g_{\sigma\nu}\nabla^{\alpha}g^{\sigma\nu}\right)=-\frac{4}{N^2}\left( T+3H \right)^2, \nonumber\\
\\
&&Q_{\alpha}\tilde{Q}^{\alpha}=\left( g_{\mu\rho}\nabla_{\alpha}g^{\mu\rho}\right) \left(\nabla_{\beta}g^{\alpha\beta}\right) = -\frac{4}{N^2}\left( T^2 +3HT\right) . \nonumber \\
\end{eqnarray}

Thus, we have
\begin{eqnarray}
Q=-\frac{1}{4}\bigg[\frac{4}{N^2}\left( T^2 +3H^2\right) -\frac{4}{N^2}2T^2 \nonumber \\
-\frac{4}{N^2}\left( T+3H\right)^2+\frac{4}{N^2}\left( 2T^2 +6HT\right) \bigg]=6\frac{H^2}{N^2}.
\end{eqnarray}

\section{Brief review metric-affine gravity theories}

\subsection{Theories with  $F=F(X_{1})$}
\subsubsection{$F(R)$ gravity}
The action of the Myrzakulov $F(R,T,Q, {\cal T})$ gravity or the MG-VIII reads as \cite{7}
\begin{eqnarray}
S=\frac{1}{2\kappa}\int \sqrt{-g}d^{4}x[F(R)+2\kappa L_{m}],
\end{eqnarray}
where $R$ is the curvature  scalar, $T$ is the torsion scalar, $Q$ is the nonmetricity scalar and  ${\cal T}$ is the trace of the energy-momentum tensor (the trace of the stress-energy tensor). The MG-VIII  is for example the unification of $F(R), F(T), F(Q)$ or $F(R,{\cal T}), F(T), F(Q)$ theories.
 The variations of the action (9) with respect to the  metric tensor and the affine connection give the following set of the field equations 
\begin{eqnarray}
	-\frac{1}{2}g_{\mu\nu}F+F_{R}R_{(\mu\nu)}=\kappa T_{\mu\nu},
	\end{eqnarray}
\begin{eqnarray}
	P_{\lambda}^{\;\;\mu\nu}(F_{R})=\kappa \Delta_{\lambda}^{\;\;\mu\nu},
	\end{eqnarray}
	where 
					\begin{eqnarray}
	P_{\lambda}^{\;\;\;\mu\nu}(F_{R}) = -\frac{\nabla_{\lambda}(\sqrt{-g}F_{R}g^{\mu\nu})}{\sqrt{-g}}+\frac{\nabla_{\alpha}(\sqrt{-g}F_{R}g^{\mu\alpha}\delta_{\lambda}^{\nu})}{\sqrt{-g}}+ 	2 F_{R}(S_{\lambda}g^{\mu\nu}-S^{\mu}\delta_{\lambda}^{\nu}-  S_{\lambda}^{\;\;\;\mu\nu}).
	\end{eqnarray}
	
	\subsubsection{$F(T)$ gravity}
The action of the Myrzakulov $F(R,T,Q, {\cal T})$ gravity or the MG-VIII reads as \cite{8}
\begin{eqnarray}
S=\frac{1}{2\kappa}\int \sqrt{-g}d^{4}x[F(T)+2\kappa L_{m}],
\end{eqnarray}
where $R$ is the curvature  scalar, $T$ is the torsion scalar, $Q$ is the nonmetricity scalar and  ${\cal T}$ is the trace of the energy-momentum tensor (the trace of the stress-energy tensor). The MG-VIII  is for example the unification of $F(R), F(T), F(Q)$ or $F(R,{\cal T}), F(T), F(Q)$ theories.
 The variations of the action (9) with respect to the  metric tensor and the affine connection give the following set of the field equations \cite{di1}
\begin{eqnarray}
	-\frac{1}{2}g_{\mu\nu}F+F_{T}\Big(2S_{\nu\alpha\beta}S_{\mu}^{\;\;\;\alpha\beta}-S_{\alpha\beta\mu}S^{\alpha\beta}_{\;\;\;\;\nu}+2S_{\nu\alpha\beta}S_{\mu}^{\;\;\;\beta\alpha}-4S_{\mu}S_{\nu} \Big)=\kappa T_{\mu\nu},
	\end{eqnarray}
\begin{eqnarray}
	2 F_{T}\Big( S^{\mu\nu}_{\;\;\;\;\lambda}-2 S_{\lambda}^{\;\;\;[\mu\nu]}-4 S^{[\mu}\delta^{\nu]}_{\lambda}\Big)
	=\kappa \Delta_{\lambda}^{\;\;\mu\nu}.
	\end{eqnarray}

\subsubsection{$F(Q)$ gravity}
The action of the Myrzakulov $F(R,T,Q, {\cal T})$ gravity or the MG-VIII reads as \cite{9}
\begin{eqnarray}
S=\frac{1}{2\kappa}\int \sqrt{-g}d^{4}x[F(Q)+2\kappa L_{m}],
\end{eqnarray}
where $R$ is the curvature  scalar, $T$ is the torsion scalar, $Q$ is the nonmetricity scalar and  ${\cal T}$ is the trace of the energy-momentum tensor (the trace of the stress-energy tensor). The MG-VIII  is for example the unification of $F(R), F(T), F(Q)$ or $F(R,{\cal T}), F(T), F(Q)$ theories.
 The variations of the action (9) with respect to the  metric tensor and the affine connection give the following set of the field equations \cite{di1}
\begin{eqnarray}
	-\frac{1}{2}g_{\mu\nu}F+F_{Q}L_{(\mu\nu)}
	+\hat{\nabla}_{\lambda}(F_{Q}J^{\lambda}_{\;\;\;(\mu\nu)})+g_{\mu\nu}\hat{\nabla}_{\lambda}(F_{Q}\zeta^{\lambda})=\kappa T_{\mu\nu},
	\end{eqnarray}
\begin{eqnarray}
	F_{Q}\Big( 2 Q^{[\nu\mu]}_{\;\;\;\;\lambda}-Q_{\lambda}^{\;\;\mu\nu}+(q^{\nu}-Q^{\nu})\delta^{\mu}_{\lambda}+Q_{\lambda}g^{\mu\nu}+\frac{1}{2}Q^{\mu}\delta^{\nu}_{\lambda} \Big)=\kappa \Delta_{\lambda}^{\;\;\mu\nu},
	\end{eqnarray}
	where 
		\begin{eqnarray}
		\Omega^{\alpha\mu\nu} = \frac{1}{4}Q^{\alpha\mu\nu}-\frac{1}{2} Q^{\mu\nu\alpha}-\frac{1}{4} g^{\mu\nu}Q^{\alpha}+\frac{1}{2}g^{\alpha\mu}Q^{\nu},
	\end{eqnarray}
		\begin{eqnarray}
	4 L_{\mu\nu}=(Q_{\mu\alpha\beta}-2 Q_{\alpha\beta\mu})Q_{\nu}^{\;\;\;\alpha\beta}+(Q_{\mu}+2q_{\mu})Q_{\nu}
	+(2Q_{\mu\nu\alpha}-Q_{\alpha\mu\nu})Q^{\alpha})-4 \Omega^{\alpha\beta}_{\;\;\;\;\nu}Q_{\alpha\beta\mu}-4 \Omega_{\alpha\mu\beta}Q^{\alpha\beta}_{\;\;\;\;\nu},
	\end{eqnarray}
	\begin{eqnarray}
			J^{\lambda}_{\;\;\;\mu\nu} := \sqrt{-g}\Big( \frac{1}{4} Q^{\lambda}_{\;\;\;\mu\nu}-\frac{1}{2}Q_{\mu\nu}^{\;\;\;\;\lambda}+\Omega^{\lambda}_{\;\;\;\mu\nu}\Big),\quad \zeta^{\lambda}=\sqrt{-g}\Big(\frac{1}{2}q^{\lambda}-\frac{1}{4}Q^{\lambda}\Big).
	\end{eqnarray}

\subsection{Theories with  $F=F(X_{1},X_{2})$}

\subsubsection{$F(R,{\cal T})$ gravity}
The action of the Myrzakulov $F(R,T,Q, {\cal T})$ gravity or the MG-VIII reads as \cite{10}
\begin{eqnarray}
S=\frac{1}{2\kappa}\int \sqrt{-g}d^{4}x[F(R, {\cal T})+2\kappa L_{m}],
\end{eqnarray}
where $R$ is the curvature  scalar, $T$ is the torsion scalar, $Q$ is the nonmetricity scalar and  ${\cal T}$ is the trace of the energy-momentum tensor (the trace of the stress-energy tensor). The MG-VIII  is for example the unification of $F(R), F(T), F(Q)$ or $F(R,{\cal T}), F(T), F(Q)$ theories.
 The variations of the action (9) with respect to the  metric tensor and the affine connection give the following set of the field equations \cite{di1}
\begin{eqnarray}
	-\frac{1}{2}g_{\mu\nu}F+F_{R}R_{(\mu\nu)}+F_{\cal T}(\Theta_{\mu\nu}+T_{\mu\nu})=\kappa T_{\mu\nu},
	\end{eqnarray}
\begin{eqnarray}
	P_{\lambda}^{\;\;\mu\nu}(F_{R})-F_{\cal T}\Theta_{\lambda}^{\;\;\mu\nu}=\kappa \Delta_{\lambda}^{\;\;\mu\nu},
	\end{eqnarray}
	where 
		\begin{eqnarray}
	\hat{\nabla}_{\lambda}:= \frac{1}{\sqrt{-g}}(2S_{\lambda}-\nabla_{\lambda}), 
	\quad \Theta_{\lambda}^{\;\;\mu\nu}:=-\frac{\delta \cal T}{\delta \Gamma^{\lambda}_{\;\;\;\mu\nu}}, \quad 
	\Theta_{\mu\nu}:=g^{\alpha\beta}\frac{\delta T_{\alpha\beta}}{\delta g^{\mu\nu}},
	\end{eqnarray}
			\begin{eqnarray}
	P_{\lambda}^{\;\;\;\mu\nu}(F_{R}) = -\frac{\nabla_{\lambda}(\sqrt{-g}F_{R}g^{\mu\nu})}{\sqrt{-g}}+\frac{\nabla_{\alpha}(\sqrt{-g}F_{R}g^{\mu\alpha}\delta_{\lambda}^{\nu})}{\sqrt{-g}}+ 	2 F_{R}(S_{\lambda}g^{\mu\nu}-S^{\mu}\delta_{\lambda}^{\nu}-  S_{\lambda}^{\;\;\;\mu\nu}).
	\end{eqnarray}
		
\subsubsection{$F(T, {\cal T})$ gravity}
The action of the Myrzakulov $F(R,T,Q, {\cal T})$ gravity or the MG-VIII reads as \cite{11}
\begin{eqnarray}
S=\frac{1}{2\kappa}\int \sqrt{-g}d^{4}x[F(T, {\cal T})+2\kappa L_{m}],
\end{eqnarray}
where $R$ is the curvature  scalar, $T$ is the torsion scalar, $Q$ is the nonmetricity scalar and  ${\cal T}$ is the trace of the energy-momentum tensor (the trace of the stress-energy tensor). The MG-VIII  is for example the unification of $F(R), F(T), F(Q)$ or $F(R,{\cal T}), F(T), F(Q)$ theories.
 The variations of the action (9) with respect to the  metric tensor and the affine connection give the following set of the field equations \cite{di1}
\begin{eqnarray}
	F_{T}\Big(2S_{\nu\alpha\beta}S_{\mu}^{\;\;\;\alpha\beta}-S_{\alpha\beta\mu}S^{\alpha\beta}_{\;\;\;\;\nu}+2S_{\nu\alpha\beta}S_{\mu}^{\;\;\;\beta\alpha}-4S_{\mu}S_{\nu} \Big)-\frac{1}{2}g_{\mu\nu}F
	+F_{\cal T}(\Theta_{\mu\nu}+T_{\mu\nu})=\kappa T_{\mu\nu},
	\end{eqnarray}
\begin{eqnarray}
	2 F_{T}\Big( S^{\mu\nu}_{\;\;\;\;\lambda}-2 S_{\lambda}^{\;\;\;[\mu\nu]}-4 S^{[\mu}\delta^{\nu]}_{\lambda}\Big)-F_{\cal T}\Theta_{\lambda}^{\;\;\mu\nu}=\kappa \Delta_{\lambda}^{\;\;\mu\nu},
	\end{eqnarray}
	where 
		\begin{eqnarray}
	\hat{\nabla}_{\lambda}:= \frac{1}{\sqrt{-g}}(2S_{\lambda}-\nabla_{\lambda}), 
	\quad \Theta_{\lambda}^{\;\;\mu\nu}:=-\frac{\delta \cal T}{\delta \Gamma^{\lambda}_{\;\;\;\mu\nu}},\quad 
	\Theta_{\mu\nu}:=g^{\alpha\beta}\frac{\delta T_{\alpha\beta}}{\delta g^{\mu\nu}}.	\end{eqnarray}

\subsubsection{$F(Q,{\cal T})$ gravity}
The action of the Myrzakulov $F(R,T,Q, {\cal T})$ gravity or the MG-VIII reads as \cite{12}
\begin{eqnarray}
S=\frac{1}{2\kappa}\int \sqrt{-g}d^{4}x[F(Q, {\cal T})+2\kappa L_{m}],
\end{eqnarray}
where $R$ is the curvature  scalar, $T$ is the torsion scalar, $Q$ is the nonmetricity scalar and  ${\cal T}$ is the trace of the energy-momentum tensor (the trace of the stress-energy tensor). The MG-VIII  is for example the unification of $F(R), F(T), F(Q)$ or $F(R,{\cal T}), F(T), F(Q)$ theories.
 The variations of the action (9) with respect to the  metric tensor and the affine connection give the following set of the field equations \cite{di1}
\begin{eqnarray}
	-\frac{1}{2}g_{\mu\nu}F+F_{Q}L_{(\mu\nu)}
	+\hat{\nabla}_{\lambda}(F_{Q}J^{\lambda}_{\;\;\;(\mu\nu)})+g_{\mu\nu}\hat{\nabla}_{\lambda}(F_{Q}\zeta^{\lambda})+F_{\cal T}(\Theta_{\mu\nu}+T_{\mu\nu})=\kappa T_{\mu\nu},
	\end{eqnarray}
\begin{eqnarray}
	F_{Q}\Big( 2 Q^{[\nu\mu]}_{\;\;\;\;\lambda}-Q_{\lambda}^{\;\;\mu\nu}+(q^{\nu}-Q^{\nu})\delta^{\mu}_{\lambda}+Q_{\lambda}g^{\mu\nu}+\frac{1}{2}Q^{\mu}\delta^{\nu}_{\lambda} \Big)-F_{\cal T}\Theta_{\lambda}^{\;\;\mu\nu}=\kappa \Delta_{\lambda}^{\;\;\mu\nu},
	\end{eqnarray}
	where 
			\begin{eqnarray}
	\hat{\nabla}_{\lambda}:= \frac{1}{\sqrt{-g}}(2S_{\lambda}-\nabla_{\lambda}), 
	\Omega^{\alpha\mu\nu} = \frac{1}{4}Q^{\alpha\mu\nu}-\frac{1}{2} Q^{\mu\nu\alpha}-\frac{1}{4} g^{\mu\nu}Q^{\alpha}+\frac{1}{2}g^{\alpha\mu}Q^{\nu},\Theta_{\lambda}^{\;\;\mu\nu}:=-\frac{\delta \cal T}{\delta \Gamma^{\lambda}_{\;\;\;\mu\nu}}.
	\end{eqnarray}
		\begin{eqnarray}
	4 L_{\mu\nu}=(Q_{\mu\alpha\beta}-2 Q_{\alpha\beta\mu})Q_{\nu}^{\;\;\;\alpha\beta}+(Q_{\mu}+2q_{\mu})Q_{\nu}
	+(2Q_{\mu\nu\alpha}-Q_{\alpha\mu\nu})Q^{\alpha})-4 \Omega^{\alpha\beta}_{\;\;\;\;\nu}Q_{\alpha\beta\mu}-4 \Omega_{\alpha\mu\beta}Q^{\alpha\beta}_{\;\;\;\;\nu},
	\end{eqnarray}
	\begin{eqnarray}
	\Theta_{\mu\nu}:=g^{\alpha\beta}\frac{\delta T_{\alpha\beta}}{\delta g^{\mu\nu}},\quad
		J^{\lambda}_{\;\;\;\mu\nu} := \sqrt{-g}\Big( \frac{1}{4} Q^{\lambda}_{\;\;\;\mu\nu}-\frac{1}{2}Q_{\mu\nu}^{\;\;\;\;\lambda}+\Omega^{\lambda}_{\;\;\;\mu\nu}\Big),\quad \zeta^{\lambda}=\sqrt{-g}\Big(\frac{1}{2}q^{\lambda}-\frac{1}{4}Q^{\lambda}\Big).
	\end{eqnarray}

  \subsubsection{MG-I}
 The action of the Myrzakulov  $F(R,T)$  gravity or the MG-I has the following form \cite{1205.5266}
\begin{eqnarray}
S=\frac{1}{2\kappa}\int \sqrt{-g}d^{4}x[F(R,T)+2\kappa L_{m}],
\end{eqnarray}
where $R$ is the curvature scalar, $T$ is the torsion scalar and $L_{m}$ is the matter Lagrangian. This MG-I is some kind generalizations of the well-known $F(R)$ and $F(T)$ gravity theories. If exactly, the MG-I is the unification of the $F(R)$ and $F(T)$ theories.    The variations of the action (9) with respect to the  metric tensor and the affine connection give the following set of the field equations \cite{di1}
\begin{eqnarray}
	-\frac{1}{2}g_{\mu\nu}F+F_{R}R_{(\mu\nu)}+F_{T}\Big(2S_{\nu\alpha\beta}S_{\mu}^{\;\;\;\alpha\beta}-S_{\alpha\beta\mu}S^{\alpha\beta}_{\;\;\;\;\nu}+2S_{\nu\alpha\beta}S_{\mu}^{\;\;\;\beta\alpha}-4S_{\mu}S_{\nu} \Big)=\kappa T_{\mu\nu},
	\end{eqnarray}
\begin{eqnarray}
	P_{\lambda}^{\;\;\mu\nu}(F_{R})+2 F_{T}\Big( S^{\mu\nu}_{\;\;\;\;\lambda}-2 S_{\lambda}^{\;\;\;[\mu\nu]}-4 S^{[\mu}\delta^{\nu]}_{\lambda}\Big)=\kappa \Delta_{\lambda}^{\;\;\mu\nu},
	\end{eqnarray}
	where \cite{di1}
		\begin{eqnarray}
	\hat{\nabla}_{\lambda}:= \frac{1}{\sqrt{-g}}(2S_{\lambda}-\nabla_{\lambda}), 
		\end{eqnarray}
		\begin{eqnarray}
	4 L_{\mu\nu}=(Q_{\mu\alpha\beta}-2 Q_{\alpha\beta\mu})Q_{\nu}^{\;\;\;\alpha\beta}+(Q_{\mu}+2q_{\mu})Q_{\nu}
	+(2Q_{\mu\nu\alpha}-Q_{\alpha\mu\nu})Q^{\alpha})-4 \Omega^{\alpha\beta}_{\;\;\;\;\nu}Q_{\alpha\beta\mu}-4 \Omega_{\alpha\mu\beta}Q^{\alpha\beta}_{\;\;\;\;\nu},
	\end{eqnarray}
				\begin{eqnarray}
	P_{\lambda}^{\;\;\;\mu\nu}(F_{R}) = -\frac{\nabla_{\lambda}(\sqrt{-g}F_{R}g^{\mu\nu})}{\sqrt{-g}}+\frac{\nabla_{\alpha}(\sqrt{-g}F_{R}g^{\mu\alpha}\delta_{\lambda}^{\nu})}{\sqrt{-g}}+ 	2 F_{R}(S_{\lambda}g^{\mu\nu}-S^{\mu}\delta_{\lambda}^{\nu}-  S_{\lambda}^{\;\;\;\mu\nu}).
	\end{eqnarray}

\subsubsection{MG-II}
 The action of the Myrzakulov $F(R,Q)$ gravity or the MG-II reads as \cite{1205.5266}
\begin{eqnarray}
S=\frac{1}{2\kappa}\int \sqrt{-g}d^{4}x[F(R,Q)+2\kappa L_{m}],
\end{eqnarray}
where  $R$ is the curvature scalar and $Q$ is the nonmetricity scalar. The MG-II is the unification of the $F(R)$ and $F(Q)$ theories.    The variations of the action (9) with respect to the  metric tensor and the affine connection give the following set of the field equations \cite{di1}
\begin{eqnarray}
	-\frac{1}{2}g_{\mu\nu}F+F_{R}R_{(\mu\nu)}+F_{Q}L_{(\mu\nu)}	+\hat{\nabla}_{\lambda}(F_{Q}J^{\lambda}_{\;\;\;(\mu\nu)})+g_{\mu\nu}\hat{\nabla}_{\lambda}(F_{Q}\zeta^{\lambda})=\kappa T_{\mu\nu},
	\end{eqnarray}
\begin{eqnarray}
	P_{\lambda}^{\;\;\mu\nu}(F_{R})+
	+F_{Q}\Big( 2 Q^{[\nu\mu]}_{\;\;\;\;\lambda}-Q_{\lambda}^{\;\;\mu\nu}+(q^{\nu}-Q^{\nu})\delta^{\mu}_{\lambda}+Q_{\lambda}g^{\mu\nu}+\frac{1}{2}Q^{\mu}\delta^{\nu}_{\lambda} \Big)=\kappa \Delta_{\lambda}^{\;\;\mu\nu},
	\end{eqnarray}
	where \cite{di1}
		\begin{eqnarray}
	\hat{\nabla}_{\lambda}:= -\frac{1}{\sqrt{-g}}\nabla_{\lambda}, \quad 
	\Omega^{\alpha\mu\nu} = \frac{1}{4}Q^{\alpha\mu\nu}-\frac{1}{2} Q^{\mu\nu\alpha}-\frac{1}{4} g^{\mu\nu}Q^{\alpha}+\frac{1}{2}g^{\alpha\mu}Q^{\nu},
	\end{eqnarray}
		\begin{eqnarray}
	4 L_{\mu\nu}=(Q_{\mu\alpha\beta}-2 Q_{\alpha\beta\mu})Q_{\nu}^{\;\;\;\alpha\beta}+(Q_{\mu}+2q_{\mu})Q_{\nu}
	+(2Q_{\mu\nu\alpha}-Q_{\alpha\mu\nu})Q^{\alpha})-4 \Omega^{\alpha\beta}_{\;\;\;\;\nu}Q_{\alpha\beta\mu}-4 \Omega_{\alpha\mu\beta}Q^{\alpha\beta}_{\;\;\;\;\nu},
	\end{eqnarray}
	\begin{eqnarray}
			J^{\lambda}_{\;\;\;\mu\nu} := \sqrt{-g}\Big( \frac{1}{4} Q^{\lambda}_{\;\;\;\mu\nu}-\frac{1}{2}Q_{\mu\nu}^{\;\;\;\;\lambda}+\Omega^{\lambda}_{\;\;\;\mu\nu}\Big),\quad \zeta^{\lambda}=\sqrt{-g}\Big(\frac{1}{2}q^{\lambda}-\frac{1}{4}Q^{\lambda}\Big),
	\end{eqnarray}
			\begin{eqnarray}
	P_{\lambda}^{\;\;\;\mu\nu}(F_{R}) = -\frac{\nabla_{\lambda}(\sqrt{-g}F_{R}g^{\mu\nu})}{\sqrt{-g}}+\frac{\nabla_{\alpha}(\sqrt{-g}F_{R}g^{\mu\alpha}\delta_{\lambda}^{\nu})}{\sqrt{-g}}.
	\end{eqnarray}

\subsubsection{MG-III}
 The action of the Myrzakulov $F(T,Q)$  gravity or the MG-III reads as \cite{1205.5266}
\begin{eqnarray}
S=\frac{1}{2\kappa}\int \sqrt{-g}d^{4}x[F(T,Q)+2\kappa L_{m}],
\end{eqnarray}
where  $T$ is the torsion  scalar and $Q$ is the nonmetricity scalar. The MG-III is the unification of the $F(T)$ and $F(Q)$ theories.   The variations of the action (9) with respect to the  metric tensor and the affine connection give the following set of the field equations \cite{di1}
\begin{eqnarray}
	-\frac{1}{2}g_{\mu\nu}F+F_{T}\Big(2S_{\nu\alpha\beta}S_{\mu}^{\;\;\;\alpha\beta}-S_{\alpha\beta\mu}S^{\alpha\beta}_{\;\;\;\;\nu}+2S_{\nu\alpha\beta}S_{\mu}^{\;\;\;\beta\alpha}-4S_{\mu}S_{\nu} \Big)+F_{Q}L_{(\mu\nu)}\nonumber \\
	+\hat{\nabla}_{\lambda}(F_{Q}J^{\lambda}_{\;\;\;(\mu\nu)})+g_{\mu\nu}\hat{\nabla}_{\lambda}(F_{Q}\zeta^{\lambda})=\kappa T_{\mu\nu},
	\end{eqnarray}
\begin{eqnarray}
	2 F_{T}\Big( S^{\mu\nu}_{\;\;\;\;\lambda}-2 S_{\lambda}^{\;\;\;[\mu\nu]}-4 S^{[\mu}\delta^{\nu]}_{\lambda}\Big)
	+F_{Q}\Big( 2 Q^{[\nu\mu]}_{\;\;\;\;\lambda}-Q_{\lambda}^{\;\;\mu\nu}+(q^{\nu}-Q^{\nu})\delta^{\mu}_{\lambda}+Q_{\lambda}g^{\mu\nu}+\frac{1}{2}Q^{\mu}\delta^{\nu}_{\lambda} \Big)=\kappa \Delta_{\lambda}^{\;\;\mu\nu},
	\end{eqnarray}
	where \cite{di1}
		\begin{eqnarray}
	\hat{\nabla}_{\lambda}:= \frac{1}{\sqrt{-g}}(2S_{\lambda}-\nabla_{\lambda}), \quad 
	\Omega^{\alpha\mu\nu} = \frac{1}{4}Q^{\alpha\mu\nu}-\frac{1}{2} Q^{\mu\nu\alpha}-\frac{1}{4} g^{\mu\nu}Q^{\alpha}+\frac{1}{2}g^{\alpha\mu}Q^{\nu},
	\end{eqnarray}
		\begin{eqnarray}
	4 L_{\mu\nu}=(Q_{\mu\alpha\beta}-2 Q_{\alpha\beta\mu})Q_{\nu}^{\;\;\;\alpha\beta}+(Q_{\mu}+2q_{\mu})Q_{\nu}
	+(2Q_{\mu\nu\alpha}-Q_{\alpha\mu\nu})Q^{\alpha})-4 \Omega^{\alpha\beta}_{\;\;\;\;\nu}Q_{\alpha\beta\mu}-4 \Omega_{\alpha\mu\beta}Q^{\alpha\beta}_{\;\;\;\;\nu},
	\end{eqnarray}
	\begin{eqnarray}
			J^{\lambda}_{\;\;\;\mu\nu} := \sqrt{-g}\Big( \frac{1}{4} Q^{\lambda}_{\;\;\;\mu\nu}-\frac{1}{2}Q_{\mu\nu}^{\;\;\;\;\lambda}+\Omega^{\lambda}_{\;\;\;\mu\nu}\Big),\quad \zeta^{\lambda}=\sqrt{-g}\Big(\frac{1}{2}q^{\lambda}-\frac{1}{4}Q^{\lambda}\Big).
	\end{eqnarray}

\subsection{Theories with  $F=F(X_{1}, X_{2}, X_{3})$}

\subsubsection{MG-IV}
The action of the Myrzakulov $F(R,T,{\cal T})$ gravity  or the MG-IV has the following form \cite{1205.5266}
\begin{eqnarray}
S=\frac{1}{2\kappa}\int \sqrt{-g}d^{4}x[F(R,T,{\cal T})+2\kappa L_{m}],
\end{eqnarray}
where $R$ is the curvature scalar, $T$ is the torsion scalar and  ${\cal T}$ is the trace of the energy-momentum tensor. The MG-IV is the unification of the $F(R, {\cal T})$ and $F(T)$ theories.  
\begin{eqnarray}
	F_{R}R_{(\mu\nu)}-\frac{1}{2}g_{\mu\nu}F+F_{T}\Big(2S_{\nu\alpha\beta}S_{\mu}^{\;\;\;\alpha\beta}-S_{\alpha\beta\mu}S^{\alpha\beta}_{\;\;\;\;\nu}+2S_{\nu\alpha\beta}S_{\mu}^{\;\;\;\beta\alpha}-4S_{\mu}S_{\nu} \Big)
	+F_{\cal T}(\Theta_{\mu\nu}+T_{\mu\nu})=\kappa T_{\mu\nu},
	\end{eqnarray}
\begin{eqnarray}
	P_{\lambda}^{\;\;\mu\nu}(F_{R})+2 F_{T}\Big( S^{\mu\nu}_{\;\;\;\;\lambda}-2 S_{\lambda}^{\;\;\;[\mu\nu]}-4 S^{[\mu}\delta^{\nu]}_{\lambda}\Big)
	=F_{\cal T}\Theta_{\lambda}^{\;\;\mu\nu}+\kappa \Delta_{\lambda}^{\;\;\mu\nu},
	\end{eqnarray}
	where 
		\begin{eqnarray}
	\hat{\nabla}_{\lambda}:= \frac{1}{\sqrt{-g}}(2S_{\lambda}-\nabla_{\lambda}), 
	\quad \Theta_{\lambda}^{\;\;\mu\nu}:=-\frac{\delta \cal T}{\delta \Gamma^{\lambda}_{\;\;\;\mu\nu}},
	\end{eqnarray}
			\begin{eqnarray}
	\Theta_{\mu\nu}:=g^{\alpha\beta}\frac{\delta T_{\alpha\beta}}{\delta g^{\mu\nu}},
	\end{eqnarray}
			\begin{eqnarray}
	P_{\lambda}^{\;\;\;\mu\nu}(F_{R}) = -\frac{\nabla_{\lambda}(\sqrt{-g}F_{R}g^{\mu\nu})}{\sqrt{-g}}+\frac{\nabla_{\alpha}(\sqrt{-g}F_{R}g^{\mu\alpha}\delta_{\lambda}^{\nu})}{\sqrt{-g}}+ 	2 F_{R}(S_{\lambda}g^{\mu\nu}-S^{\mu}\delta_{\lambda}^{\nu}-  S_{\lambda}^{\;\;\;\mu\nu}). 
	\end{eqnarray}

\subsubsection{MG-V}
The action of the Myrzakulov $F(R,T,Q)$ gravity  or the MG-V is given by \cite{1205.5266}
\begin{eqnarray}
S=\frac{1}{2\kappa}\int \sqrt{-g}d^{4}x[F(R,T,Q)+2\kappa L_{m}],
\end{eqnarray}
where $R$ is the curvature scalar, $T$ is the torsion scalar and  $Q$ is the nonmetricity scalar. The MG-V is the unification of $F(R), F(T), F(Q)$ theories. The variations of the action (9) with respect to the  metric tensor and the affine connection give the following set of the field equations \cite{di1}
\begin{eqnarray}
	-\frac{1}{2}g_{\mu\nu}F+F_{R}R_{(\mu\nu)}+F_{T}\Big(2S_{\nu\alpha\beta}S_{\mu}^{\;\;\;\alpha\beta}-S_{\alpha\beta\mu}S^{\alpha\beta}_{\;\;\;\;\nu}+2S_{\nu\alpha\beta}S_{\mu}^{\;\;\;\beta\alpha}-4S_{\mu}S_{\nu} \Big)+F_{Q}L_{(\mu\nu)}\nonumber \\
	+\hat{\nabla}_{\lambda}(F_{Q}J^{\lambda}_{\;\;\;(\mu\nu)})+g_{\mu\nu}\hat{\nabla}_{\lambda}(F_{Q}\zeta^{\lambda})=\kappa T_{\mu\nu},
	\end{eqnarray}
\begin{eqnarray}
	P_{\lambda}^{\;\;\mu\nu}(F_{R})+2 F_{T}\Big( S^{\mu\nu}_{\;\;\;\;\lambda}-2 S_{\lambda}^{\;\;\;[\mu\nu]}-4 S^{[\mu}\delta^{\nu]}_{\lambda}\Big)\nonumber \\
	+F_{Q}\Big( 2 Q^{[\nu\mu]}_{\;\;\;\;\lambda}-Q_{\lambda}^{\;\;\mu\nu}+(q^{\nu}-Q^{\nu})\delta^{\mu}_{\lambda}+Q_{\lambda}g^{\mu\nu}+\frac{1}{2}Q^{\mu}\delta^{\nu}_{\lambda} \Big)=\kappa \Delta_{\lambda}^{\;\;\mu\nu},
	\end{eqnarray}
	where 
		\begin{eqnarray}
	\hat{\nabla}_{\lambda}:= \frac{1}{\sqrt{-g}}(2S_{\lambda}-\nabla_{\lambda}), 
	\quad \Omega^{\alpha\mu\nu} = \frac{1}{4}Q^{\alpha\mu\nu}-\frac{1}{2} Q^{\mu\nu\alpha}-\frac{1}{4} g^{\mu\nu}Q^{\alpha}+\frac{1}{2}g^{\alpha\mu}Q^{\nu},
	\end{eqnarray}
		\begin{eqnarray}
	4 L_{\mu\nu}=(Q_{\mu\alpha\beta}-2 Q_{\alpha\beta\mu})Q_{\nu}^{\;\;\;\alpha\beta}+(Q_{\mu}+2q_{\mu})Q_{\nu}
	+(2Q_{\mu\nu\alpha}-Q_{\alpha\mu\nu})Q^{\alpha})-4 \Omega^{\alpha\beta}_{\;\;\;\;\nu}Q_{\alpha\beta\mu}-4 \Omega_{\alpha\mu\beta}Q^{\alpha\beta}_{\;\;\;\;\nu},
	\end{eqnarray}
	\begin{eqnarray}
			J^{\lambda}_{\;\;\;\mu\nu} := \sqrt{-g}\Big( \frac{1}{4} Q^{\lambda}_{\;\;\;\mu\nu}-\frac{1}{2}Q_{\mu\nu}^{\;\;\;\;\lambda}+\Omega^{\lambda}_{\;\;\;\mu\nu}\Big),\quad \zeta^{\lambda}=\sqrt{-g}\Big(\frac{1}{2}q^{\lambda}-\frac{1}{4}Q^{\lambda}\Big),
	\end{eqnarray}
			\begin{eqnarray}
	P_{\lambda}^{\;\;\;\mu\nu}(F_{R}) = -\frac{\nabla_{\lambda}(\sqrt{-g}F_{R}g^{\mu\nu})}{\sqrt{-g}}+\frac{\nabla_{\alpha}(\sqrt{-g}F_{R}g^{\mu\alpha}\delta_{\lambda}^{\nu})}{\sqrt{-g}}+ 	2 F_{R}(S_{\lambda}g^{\mu\nu}-S^{\mu}\delta_{\lambda}^{\nu}-  S_{\lambda}^{\;\;\;\mu\nu}).
	\end{eqnarray}

\subsubsection{MG-VI}
 The action of the Myrzakulov $F(R,Q,{\cal T})$  gravity or the MG-VI reads as \cite{1205.5266}
\begin{eqnarray}
S=\frac{1}{2\kappa}\int \sqrt{-g}d^{4}x[F(R,Q, {\cal T})+2\kappa L_{m}],
\end{eqnarray}
where  $R$ is the curvature scalar,  $Q$ is the nonmetricity scalar and  ${\cal T}$ is the trace of the energy-momentum tensor. The MG-VI is the unification of $F(R,{\cal T})$ and $F(Q)$ theories. The variations of the action (9) with respect to the  metric tensor and the affine connection give the following set of the field equations \cite{di1}
\begin{eqnarray}
	F_{R}R_{(\mu\nu)}-\frac{1}{2}g_{\mu\nu}F+F_{Q}L_{(\mu\nu)}
	+\hat{\nabla}_{\lambda}(F_{Q}J^{\lambda}_{\;\;\;(\mu\nu)})+g_{\mu\nu}\hat{\nabla}_{\lambda}(F_{Q}\zeta^{\lambda})+F_{\cal T}(\Theta_{\mu\nu}+T_{\mu\nu})=\kappa T_{\mu\nu},
	\end{eqnarray}
\begin{eqnarray}
	P_{\lambda}^{\;\;\mu\nu}(F_{R})+
	+F_{Q}\Big( 2 Q^{[\nu\mu]}_{\;\;\;\;\lambda}-Q_{\lambda}^{\;\;\mu\nu}+(q^{\nu}-Q^{\nu})\delta^{\mu}_{\lambda}+Q_{\lambda}g^{\mu\nu}+\frac{1}{2}Q^{\mu}\delta^{\nu}_{\lambda} \Big)=F_{\cal T}\Theta_{\lambda}^{\;\;\mu\nu}+\kappa \Delta_{\lambda}^{\;\;\mu\nu},
	\end{eqnarray}
where
		\begin{eqnarray}
	\hat{\nabla}_{\lambda}:= -\frac{1}{\sqrt{-g}}\nabla_{\lambda}, \quad 
	\Omega^{\alpha\mu\nu} = \frac{1}{4}Q^{\alpha\mu\nu}-\frac{1}{2} Q^{\mu\nu\alpha}-\frac{1}{4} g^{\mu\nu}Q^{\alpha}+\frac{1}{2}g^{\alpha\mu}Q^{\nu},\quad \Theta_{\lambda}^{\;\;\mu\nu}:=-\frac{\delta \cal T}{\delta \Gamma^{\lambda}_{\;\;\;\mu\nu}}.
	\end{eqnarray}
		\begin{eqnarray}
	4 L_{\mu\nu}=(Q_{\mu\alpha\beta}-2 Q_{\alpha\beta\mu})Q_{\nu}^{\;\;\;\alpha\beta}+(Q_{\mu}+2q_{\mu})Q_{\nu}
	+(2Q_{\mu\nu\alpha}-Q_{\alpha\mu\nu})Q^{\alpha})-4 \Omega^{\alpha\beta}_{\;\;\;\;\nu}Q_{\alpha\beta\mu}-4 \Omega_{\alpha\mu\beta}Q^{\alpha\beta}_{\;\;\;\;\nu},
	\end{eqnarray}
	\begin{eqnarray}
	\Theta_{\mu\nu}:=g^{\alpha\beta}\frac{\delta T_{\alpha\beta}}{\delta g^{\mu\nu}},\quad
		J^{\lambda}_{\;\;\;\mu\nu} := \sqrt{-g}\Big( \frac{1}{4} Q^{\lambda}_{\;\;\;\mu\nu}-\frac{1}{2}Q_{\mu\nu}^{\;\;\;\;\lambda}+\Omega^{\lambda}_{\;\;\;\mu\nu}\Big),\quad \zeta^{\lambda}=\sqrt{-g}\Big(\frac{1}{2}q^{\lambda}-\frac{1}{4}Q^{\lambda}\Big),
	\end{eqnarray}
			\begin{eqnarray}
	P_{\lambda}^{\;\;\;\mu\nu}(F_{R}) = -\frac{\nabla_{\lambda}(\sqrt{-g}F_{R}g^{\mu\nu})}{\sqrt{-g}}+\frac{\nabla_{\alpha}(\sqrt{-g}F_{R}g^{\mu\alpha}\delta_{\lambda}^{\nu})}{\sqrt{-g}}+ 	2 F_{R}(S_{\lambda}g^{\mu\nu}-S^{\mu}\delta_{\lambda}^{\nu}-  S_{\lambda}^{\;\;\;\mu\nu}). 
	\end{eqnarray}

\subsubsection{MG-VII}
 The action of the Myrzakulov $F(T,Q,{\cal T})$  gravity or the MG-VII reads as \cite{1205.5266}
\begin{eqnarray}
S=\frac{1}{2\kappa}\int \sqrt{-g}d^{4}x[F(T,Q, {\cal T})+2\kappa L_{m}],
\end{eqnarray}
and   $T$ is the torsion  scalar,  $Q$ is the nonmetricity scalar and  ${\cal T}$ is the trace of the energy-momentum tensor.  The variations of the action (9) with respect to the  metric tensor and the affine connection give the following set of the field equations \cite{di1}
\begin{gather}
	-\frac{1}{2}g_{\mu\nu}F+F_{T}\Big(2S_{\nu\alpha\beta}S_{\mu}^{\;\;\;\alpha\beta}-S_{\alpha\beta\mu}S^{\alpha\beta}_{\;\;\;\;\nu}+2S_{\nu\alpha\beta}S_{\mu}^{\;\;\;\beta\alpha}-4S_{\mu}S_{\nu} \Big)+F_{Q}L_{(\mu\nu)}\nonumber \\
	+\hat{\nabla}_{\lambda}(F_{Q}J^{\lambda}_{\;\;\;(\mu\nu)})+g_{\mu\nu}\hat{\nabla}_{\lambda}(F_{Q}\zeta^{\lambda})+F_{\cal T}(\Theta_{\mu\nu}+T_{\mu\nu})=\kappa T_{\mu\nu},
	\end{gather}
\begin{gather}
	2 F_{T}\Big( S^{\mu\nu}_{\;\;\;\;\lambda}-2 S_{\lambda}^{\;\;\;[\mu\nu]}-4 S^{[\mu}\delta^{\nu]}_{\lambda}\Big)\nonumber \\
	+F_{Q}\Big( 2 Q^{[\nu\mu]}_{\;\;\;\;\lambda}-Q_{\lambda}^{\;\;\mu\nu}+(q^{\nu}-Q^{\nu})\delta^{\mu}_{\lambda}+Q_{\lambda}g^{\mu\nu}+\frac{1}{2}Q^{\mu}\delta^{\nu}_{\lambda} \Big)=F_{\cal T}\Theta_{\lambda}^{\;\;\mu\nu}+\kappa \Delta_{\lambda}^{\;\;\mu\nu},
	\end{gather}
	where
		\begin{eqnarray}
	\hat{\nabla}_{\lambda}:= \frac{1}{\sqrt{-g}}(2S_{\lambda}-\nabla_{\lambda}), 
	\Omega^{\alpha\mu\nu} = \frac{1}{4}Q^{\alpha\mu\nu}-\frac{1}{2} Q^{\mu\nu\alpha}-\frac{1}{4} g^{\mu\nu}Q^{\alpha}+\frac{1}{2}g^{\alpha\mu}Q^{\nu},\Theta_{\lambda}^{\;\;\mu\nu}:=-\frac{\delta \cal T}{\delta \Gamma^{\lambda}_{\;\;\;\mu\nu}}.
	\end{eqnarray}
		\begin{eqnarray}
	4 L_{\mu\nu}=(Q_{\mu\alpha\beta}-2 Q_{\alpha\beta\mu})Q_{\nu}^{\;\;\;\alpha\beta}+(Q_{\mu}+2q_{\mu})Q_{\nu}
	+(2Q_{\mu\nu\alpha}-Q_{\alpha\mu\nu})Q^{\alpha})-4 \Omega^{\alpha\beta}_{\;\;\;\;\nu}Q_{\alpha\beta\mu}-4 \Omega_{\alpha\mu\beta}Q^{\alpha\beta}_{\;\;\;\;\nu},
	\end{eqnarray}
	\begin{eqnarray}
	\Theta_{\mu\nu}:=g^{\alpha\beta}\frac{\delta T_{\alpha\beta}}{\delta g^{\mu\nu}},\quad
		J^{\lambda}_{\;\;\;\mu\nu} := \sqrt{-g}\Big( \frac{1}{4} Q^{\lambda}_{\;\;\;\mu\nu}-\frac{1}{2}Q_{\mu\nu}^{\;\;\;\;\lambda}+\Omega^{\lambda}_{\;\;\;\mu\nu}\Big),\quad \zeta^{\lambda}=\sqrt{-g}\Big(\frac{1}{2}q^{\lambda}-\frac{1}{4}Q^{\lambda}\Big),
	\end{eqnarray}
					\begin{eqnarray}
	\Theta_{\lambda}^{\;\;\mu\nu}:=-\frac{\delta \cal T}{\delta \Gamma^{\lambda}_{\;\;\;\mu\nu}}.
	\end{eqnarray}

\subsection{Theories with  $F=F(X_{1}, X_{2}, X_{3}, X_{4})$}

\subsubsection{MG-VIII}
The action of the Myrzakulov $F(R,T,Q, {\cal T})$ gravity or the MG-VIII reads as \cite{1205.5266}
\begin{eqnarray}
S=\frac{1}{2\kappa}\int \sqrt{-g}d^{4}x[F(R,T,Q, {\cal T})+2\kappa L_{m}],
\end{eqnarray}
where $R$ is the curvature  scalar, $T$ is the torsion scalar, $Q$ is the nonmetricity scalar and  ${\cal T}$ is the trace of the energy-momentum tensor (the trace of the stress-energy tensor). The MG-VIII  is for example the unification of $F(R), F(T), F(Q)$ or $F(R,{\cal T}), F(T), F(Q)$ theories.
 The variations of the action (9) with respect to the  metric tensor and the affine connection give the following set of the field equations \cite{di1}
\begin{eqnarray}
	-\frac{1}{2}g_{\mu\nu}F+F_{R}R_{(\mu\nu)}+F_{T}\Big(2S_{\nu\alpha\beta}S_{\mu}^{\;\;\;\alpha\beta}-S_{\alpha\beta\mu}S^{\alpha\beta}_{\;\;\;\;\nu}+2S_{\nu\alpha\beta}S_{\mu}^{\;\;\;\beta\alpha}-4S_{\mu}S_{\nu} \Big)+F_{Q}L_{(\mu\nu)}\nonumber \\
	+\hat{\nabla}_{\lambda}(F_{Q}J^{\lambda}_{\;\;\;(\mu\nu)})+g_{\mu\nu}\hat{\nabla}_{\lambda}(F_{Q}\zeta^{\lambda})+F_{\cal T}(\Theta_{\mu\nu}+T_{\mu\nu})=\kappa T_{\mu\nu},
	\end{eqnarray}
\begin{eqnarray}
	P_{\lambda}^{\;\;\mu\nu}(F_{R})+2 F_{T}\Big( S^{\mu\nu}_{\;\;\;\;\lambda}-2 S_{\lambda}^{\;\;\;[\mu\nu]}-4 S^{[\mu}\delta^{\nu]}_{\lambda}\Big)\nonumber \\
	+F_{Q}\Big( 2 Q^{[\nu\mu]}_{\;\;\;\;\lambda}-Q_{\lambda}^{\;\;\mu\nu}+(q^{\nu}-Q^{\nu})\delta^{\mu}_{\lambda}+Q_{\lambda}g^{\mu\nu}+\frac{1}{2}Q^{\mu}\delta^{\nu}_{\lambda} \Big)=F_{\cal T}\Theta_{\lambda}^{\;\;\mu\nu}+\kappa \Delta_{\lambda}^{\;\;\mu\nu}.
	\end{eqnarray}
	Here \cite{di1}
		\begin{eqnarray}
	\hat{\nabla}_{\lambda}:= \frac{1}{\sqrt{-g}}(2S_{\lambda}-\nabla_{\lambda}), 
	\Omega^{\alpha\mu\nu} = \frac{1}{4}Q^{\alpha\mu\nu}-\frac{1}{2} Q^{\mu\nu\alpha}-\frac{1}{4} g^{\mu\nu}Q^{\alpha}+\frac{1}{2}g^{\alpha\mu}Q^{\nu},\Theta_{\lambda}^{\;\;\mu\nu}:=-\frac{\delta \cal T}{\delta \Gamma^{\lambda}_{\;\;\;\mu\nu}}.
	\end{eqnarray}
		\begin{eqnarray}
	4 L_{\mu\nu}=(Q_{\mu\alpha\beta}-2 Q_{\alpha\beta\mu})Q_{\nu}^{\;\;\;\alpha\beta}+(Q_{\mu}+2q_{\mu})Q_{\nu}
	+(2Q_{\mu\nu\alpha}-Q_{\alpha\mu\nu})Q^{\alpha})-4 \Omega^{\alpha\beta}_{\;\;\;\;\nu}Q_{\alpha\beta\mu}-4 \Omega_{\alpha\mu\beta}Q^{\alpha\beta}_{\;\;\;\;\nu},
	\end{eqnarray}
	\begin{eqnarray}
	\Theta_{\mu\nu}:=g^{\alpha\beta}\frac{\delta T_{\alpha\beta}}{\delta g^{\mu\nu}},\quad
		J^{\lambda}_{\;\;\;\mu\nu} := \sqrt{-g}\Big( \frac{1}{4} Q^{\lambda}_{\;\;\;\mu\nu}-\frac{1}{2}Q_{\mu\nu}^{\;\;\;\;\lambda}+\Omega^{\lambda}_{\;\;\;\mu\nu}\Big),\quad \zeta^{\lambda}=\sqrt{-g}\Big(\frac{1}{2}q^{\lambda}-\frac{1}{4}Q^{\lambda}\Big),
	\end{eqnarray}
			\begin{eqnarray}
	P_{\lambda}^{\;\;\;\mu\nu}(F_{R}) = -\frac{\nabla_{\lambda}(\sqrt{-g}F_{R}g^{\mu\nu})}{\sqrt{-g}}+\frac{\nabla_{\alpha}(\sqrt{-g}F_{R}g^{\mu\alpha}\delta_{\lambda}^{\nu})}{\sqrt{-g}}+ 	2 F_{R}(S_{\lambda}g^{\mu\nu}-S^{\mu}\delta_{\lambda}^{\nu}-  S_{\lambda}^{\;\;\;\mu\nu}).
	\end{eqnarray}

\subsection{Theories with  $F=F(X_{1}, X_{2}, X_{3}, X_{4}, X_{5})$}
Here we present one example of the MAG theories with the five arguments the so-called metric-affine $F(R,T,Q, {\cal T}, {\cal D})$ \cite{di1}.  Its action is given by \cite{di1}
	\begin{eqnarray}
	S[g,\Gamma, \phi]=S_{g}+S_{m}=\frac{1}{2\kappa}\int \sqrt{-g}d^{4}x \left[F(R,T,Q,{\cal T}, {\cal D})+2\kappa \mathcal{L}_{m}\right], \label{2.1}
	\end{eqnarray}
	where  $R$ stands for the Ricci scalar (curvature scalar), $T$ is the torsion scalar, $Q$ is the nonmetricity scalar and  ${\cal T}$ is trace of the energy-momentum tensor of matter Lagrangian $L_{m}$, ${\cal D}$ is the dilaton current scalar. The field equations of this theory have the forms \cite{di1}
	\begin{eqnarray}
	-\frac{1}{2}g_{\mu\nu}F+F_{R}R_{(\mu\nu)}+F_{T}\Big(2S_{\nu\alpha\beta}S_{\mu}^{\;\;\;\alpha\beta}-S_{\alpha\beta\mu}S^{\alpha\beta}_{\;\;\;\;\nu}+2S_{\nu\alpha\beta}S_{\mu}^{\;\;\;\beta\alpha}-4S_{\mu}S_{\nu} \Big)+F_{Q}L_{(\mu\nu)}\nonumber \\
	+\hat{\nabla}_{\lambda}(F_{Q}J^{\lambda}_{\;\;\;(\mu\nu)})+g_{\mu\nu}\hat{\nabla}_{\lambda}(F_{Q}\zeta^{\lambda})+F_{\cal T}(\Theta_{\mu\nu}+T_{\mu\nu})+F_{D}M_{\mu\nu}=\kappa T_{\mu\nu},
	\end{eqnarray}
		\begin{eqnarray}
	P_{\lambda}^{\;\;\mu\nu}(F_{R})+2 F_{T}\Big( S^{\mu\nu}_{\;\;\;\;\lambda}-2 S_{\lambda}^{\;\;\;[\mu\nu]}-4 S^{[\mu}\delta^{\nu]}_{\lambda}\Big)-M_{\lambda}^{\;\;\mu\nu\alpha}\partial_{\alpha}F_{D} \nonumber \\
	+F_{Q}\Big( 2 Q^{[\nu\mu]}_{\;\;\;\;\lambda}-Q_{\lambda}^{\;\;\mu\nu}+(q^{\nu}-Q^{\nu})\delta^{\mu}_{\lambda}+Q_{\lambda}g^{\mu\nu}+\frac{1}{2}Q^{\mu}\delta^{\nu}_{\lambda} \Big)=F_{\cal T}\Theta_{\lambda}^{\;\;\mu\nu}+\kappa \Delta_{\lambda}^{\;\;\mu\nu},
	\end{eqnarray}
	Here  
	\begin{eqnarray}
	t={\cal T}+\frac{1}{2 \sqrt{-g}}\partial_{\nu}(\sqrt{-g}\Delta^{\nu}) \;\;, \;\;\; \Delta^{\nu}:=\Delta_{\mu}^{\;\;\mu\nu}, \quad {\cal D}=\frac{1}{ \sqrt{-g}}\partial_{\nu}(\sqrt{-g}\Delta^{\nu}).
	\end{eqnarray}

	\begin{eqnarray}
	\hat{\nabla}_{\lambda}:= \frac{1}{\sqrt{-g}}(2S_{\lambda}-\nabla_{\lambda}),
	\quad 
	\Omega^{\alpha\mu\nu} = \frac{1}{4}Q^{\alpha\mu\nu}-\frac{1}{2} Q^{\mu\nu\alpha}-\frac{1}{4} g^{\mu\nu}Q^{\alpha}+\frac{1}{2}g^{\alpha\mu}Q^{\nu},
	\end{eqnarray}
	\begin{eqnarray}
	4 L_{\mu\nu}=(Q_{\mu\alpha\beta}-2 Q_{\alpha\beta\mu})Q_{\nu}^{\;\;\;\alpha\beta}+(Q_{\mu}+2q_{\mu})Q_{\nu}
	+(2Q_{\mu\nu\alpha}-Q_{\alpha\mu\nu})Q^{\alpha})\nonumber \\-4 \Omega^{\alpha\beta}_{\;\;\;\;\nu}Q_{\alpha\beta\mu}-4 \Omega_{\alpha\mu\beta}Q^{\alpha\beta}_{\;\;\;\;\nu},
	\end{eqnarray}
	\begin{eqnarray}
	\Theta_{\mu\nu}:=g^{\alpha\beta}\frac{\delta T_{\alpha\beta}}{\delta g^{\mu\nu}},\quad
	M_{\mu\nu}:=\frac{\delta D}{\delta g^{\mu\nu}}, \quad 
	J^{\lambda}_{\;\;\;\mu\nu} := \sqrt{-g}\Big( \frac{1}{4} Q^{\lambda}_{\;\;\;\mu\nu}-\frac{1}{2}Q_{\mu\nu}^{\;\;\;\;\lambda}+\Omega^{\lambda}_{\;\;\;\mu\nu}\Big),
	\end{eqnarray}
	\begin{eqnarray}
	\zeta^{\lambda}=\sqrt{-g}\Big(-\frac{1}{4}Q^{\lambda}+\frac{1}{2}q^{\lambda}\Big), \quad \Theta_{\lambda}^{\;\;\mu\nu}:=-\frac{\delta \cal T}{\delta \Gamma^{\lambda}_{\;\;\;\mu\nu}} \;\;, \;\; M_{\lambda}^{\;\;\mu\nu\alpha}:=\frac{\delta \Delta^{\alpha}}{\delta \Gamma^{\lambda}_{\;\;\;\mu\nu}},
	\end{eqnarray}
		\begin{eqnarray}
	P_{\lambda}^{\;\;\;\mu\nu}(F_{R}) = -\frac{\nabla_{\lambda}(\sqrt{-g}F_{R}g^{\mu\nu})}{\sqrt{-g}}+\frac{\nabla_{\alpha}(\sqrt{-g}F_{R}g^{\mu\alpha}\delta_{\lambda}^{\nu})}{\sqrt{-g}}+ 
	2 F_{R}(S_{\lambda}g^{\mu\nu}-S^{\mu}\delta_{\lambda}^{\nu}-  S_{\lambda}^{\;\;\;\mu\nu}).\end{eqnarray}

	\subsection{Other MAG theories}
	
\subsubsection{Metric-affine $F(R, R_{\mu\nu}R^{\mu\nu})$  gravity}
The action of the metric-affine $F(R, R_{\mu\nu}R^{\mu\nu})$ gravity has the form  (see e.g. \cite{1902.09643} and references therein)
\begin{eqnarray}
S=\frac{1}{2\kappa^{2}}\int \sqrt{-g}d^{4}x[F(R, R_{\mu\nu}R^{\mu\nu})+2\kappa^{2}L_{m}]=S_{g}[g,\Gamma]+S_{m}[g,\Gamma, \psi]. \label{2.1}
\end{eqnarray}
Variations of  the action with respect to the metric tensor $(\delta_{g}S=0)$ and the connection $(\delta_{\Gamma}S=0)$, respectively, give  the following set of two field equations
\begin{eqnarray}
F_{R}R_{(\mu\nu)}-0.5Fg_{\mu\nu}+F_{R_{\mu\nu}R^{\mu\nu}}(R_{\mu\alpha}R_{\nu}^{\,\,\,\alpha}+R_{\alpha\mu}R^{\alpha}_{\,\,\,\nu})&=&\kappa^{2}{\cal T}_{\mu\nu}, \\
\nabla_{\alpha}(\sqrt{-g}B^{\mu\alpha})\delta_{\lambda}^{\nu}-\nabla_{\lambda}(\sqrt{-g}B^{\mu\nu})+2\sqrt{-g}\left[B^{\mu\nu}S_{\lambda}-B^{\mu\alpha}(S_{\lambda\alpha}^{\,\,\, \nu}+S_{\alpha}\delta^{\nu}_{\lambda})\right]&=&\kappa^{2}H_{\lambda}^{\, \, \, \mu\nu},
\end{eqnarray}
where 
\begin{eqnarray}
B^{\mu\nu}=F_{R}g^{\mu\nu}+2F_{R_{\mu\nu}R^{\mu\nu}}R^{\mu\nu}. 
\end{eqnarray}
 
\subsubsection{Metric-affine $F(g_{\mu\nu}, R^{\alpha}_{\,\,\,\beta\gamma\rho})$ gravity}
One of  examples of generalized metric-affine gravity theories  is the metric-affine $F(g_{\mu\nu}, R^{\alpha}_{\,\,\,\beta\gamma\rho})$   gravity. Its action reads as  (see e.g. \cite{1902.09643} and references therein)
\begin{eqnarray}
S[g, \Gamma]=S_{g}+S_{m}=\frac{1}{2\kappa^{2}}\int d^{4}x\sqrt{-g}L_{g}(g_{\mu\nu}, R^{\alpha}_{\,\,\,\beta\gamma\rho})+\int d^{4}x\sqrt{-g}L_{m}(g_{\mu\nu}, R^{\alpha}_{\,\,\,\beta\gamma\rho}, \psi), \label{2.16}
\end{eqnarray}
The two metric-affine gravity theories presented in the previous two subsubsections (2.1.1) and (2.2.2) are particular cases of the more general metric-affine gravity theory given by the action (\ref{2.16}). Varying the action (\ref{2.16}) with respect to the metric tensor and to the affine connection, we come to  the following field  equations \cite{1902.09643}
\begin{eqnarray}
-0.5L_{g}g_{\mu\nu}+\frac{\partial L_{g}}{\partial g^{\mu\nu}}&=&k^{2}{\cal T}_{\mu\nu}, \\
\frac{2}{\sqrt{-g}}\left[(2S_{\alpha}-\nabla_{\alpha})(\sqrt{-g}\Sigma_{\lambda}^{\,\,\, \mu\alpha\nu})-\sqrt{-g}\Sigma_{\lambda}^{\,\,\,\mu\gamma\delta}S_{\gamma\delta}^{\,\, \nu}\right]&=&\kappa^{2}H_{\lambda}^{\,\,\,\mu\nu},
\end{eqnarray}
where  
\begin{eqnarray}
\Sigma_{\lambda}^{\,\,\, \mu\alpha\nu}=\frac{\partial L_{g}}{\partial R^{\lambda}_{\,\,\,\mu\alpha\nu}}.
\end{eqnarray}

\subsubsection{Metric-affine $F(g_{\mu\nu}, R^{\alpha}_{\,\,\,\beta\gamma\rho}, S_{\mu\nu}^{\,\,\,\, \lambda}, Q_{\alpha\mu\nu})$ gravity}
One of most general  examples  of metric-affine gravity theories  is the metric-affine $F(g_{\mu\nu}, R^{\alpha}_{\,\,\,\beta\gamma\rho}, S_{\mu\nu}^{\,\,\,\, \lambda}, Q_{\alpha\mu\nu})$   gravity. Its action is given by (see e.g. \cite{1902.09643} and references therein)
\begin{eqnarray}
S[g, \Gamma]=\frac{1}{2\kappa^{2}}\int d^{4}x\sqrt{-g}L_{g}(g_{\mu\nu}, R^{\alpha}_{\,\,\,\beta\gamma\rho}, S_{\mu\nu}^{\,\,\,\, \lambda}, Q_{\alpha\mu\nu}))+\int d^{4}x\sqrt{-g}L_{m}(g_{\mu\nu}, \Gamma^{\lambda}_{\alpha\beta}, \psi), \label{2.16}
\end{eqnarray}
where
\begin{eqnarray}
S_{g}=\frac{1}{2\kappa^{2}}\int d^{4}x\sqrt{-g}L_{g}(g_{\mu\nu}, R^{\alpha}_{\,\,\,\beta\gamma\rho}, S_{\mu\nu}^{\,\,\,\, \lambda}, Q_{\alpha\mu\nu})), \quad S_{m}=\int d^{4}x\sqrt{-g}L_{m}(g_{\mu\nu}, \Gamma^{\lambda}_{\alpha\beta}, \psi). \label{2.16}
\end{eqnarray}
The two metric-affine gravity theories presented in the previous two subsubsections  are particular cases of the more general metric-affine gravity theory given by the action (3.18). Varying the action (3.18) with respect to the metric tensor and to the affine connection, we come to  the following field  equations \cite{1902.09643}
\begin{eqnarray}
-0.5L_{g}g_{\mu\nu}+\frac{\partial L_{g}}{\partial g^{\mu\nu}}+\frac{1}{\sqrt{-g}}(2S_{\alpha}-\nabla_{\alpha})\sqrt{-g}\frac{\partial L_{g}}{\partial Q_{\alpha}g^{\,\,\,\mu\nu}}&=&k^{2}{\cal T}_{\mu\nu}, \\
-\frac{2\nabla_{\alpha}(\sqrt{-g}\Sigma_{\lambda}^{\,\,\,\mu\alpha\nu})}{\sqrt{-g}}+4\Sigma_{\lambda}^{\,\,\,\mu\alpha\nu}S_{\alpha}-\Sigma_{\lambda}^{\,\,\,\mu\gamma\delta}S_{\gamma\delta}^{\,\,\,\nu}+2W^{\mu\nu}_{\,\,\,\lambda}+V^{\mu\nu}_{\,\,\,\lambda}&=&\kappa^{2}H_{\lambda}^{\,\,\,\mu\nu},
\end{eqnarray}
where  
\begin{eqnarray}
\Sigma_{\lambda}^{\,\,\, \mu\alpha\nu}=\frac{\partial L_{g}}{\partial R^{\lambda}_{\,\,\,\mu\alpha\nu}}, \quad V_{\lambda}^{\,\,\, \mu\nu}=\frac{\partial L_{g}}{\partial S^{\,\,\,\lambda}_{\mu\nu}}, \quad W^{\alpha\mu\nu}=\frac{\partial L_{g}}{\partial Q_{\alpha\mu\nu}}.
\end{eqnarray}

	We consider  some generalized metric - affine spacetime with the  curvature, torsion and nonmetricity. In the previous sections,  we have considered  the  MG-VIII theory.  In this section, we want to collect some other  generalized and/or modified gravity   theories.

 \subsubsection{Einstein-Ricci gravity}
The equations of motion of the Einstein-Ricci gravity  are given by
 \cite{gr-qc/0602054}-\cite{hep-th/0507284}
\begin{eqnarray}
R_{ij}-0.5Rg_{ij}-\kappa T_{ij}+\phi_{ij}&=&0,\\
g_{ij\tau}+2nR_{ij}+f_{ij}&=&0.
\end{eqnarray}
\subsubsection{Einstein-Calabi  gravity}
The equations of motion of the Einstein-Calabi gravity (ECG) read as
 \cite{gr-qc/0602054}-\cite{hep-th/0507284}
\begin{eqnarray}
R_{ij}-0.5Rg_{ij}-kT_{ij}+\phi_{ij}&=&0,\\
g_{ij\tau}-n\frac{\partial^{2}R}{\partial z^{i}\partial z^{j}}+f_{ij}&=&0.
\end{eqnarray}
\subsubsection{Einstein-Cartan gravity}
The action of the Einstein-Cartan gravity (ECG) reads as
\begin{eqnarray}
S=\int L\sqrt{-g}d^{4}x=\frac{1}{16\pi G}\int R(\Gamma, g)\sqrt{-g}d^{4}x+S_{m}.
\end{eqnarray}
The  equations of motion of the ECG are given by
 \cite{gr-qc/0602054}-\cite{hep-th/0507284}
\begin{eqnarray}
G_{ij}+4B_{\beta\mu\, \, }^{[\alpha}B^{\beta] \,}_{\alpha\nu}+2B_{\beta\alpha\mu}B_{\nu}^{\, \beta\alpha}-B_{\mu\beta\alpha}B_{\nu}^{\, \beta\alpha}-0.5g_{ij}(4B^{\, \beta}_{\alpha\, \, [\lambda}B^{\alpha\lambda}_{\,\,\, \beta]}+B_{\alpha\beta\gamma}B^{\alpha\beta\gamma})&=&\kappa T_{ij}, \\ -T^{\lambda}_{\, \mu\nu}+\delta_{\mu}^{\, \lambda}T_{\nu}-\delta_{\nu}^{\, \lambda}T_{\mu}&=&\kappa S^{\lambda}_{\, \mu\nu}, \\
\frac{\partial L}{\partial \phi}+(\nabla_{\lambda}-2T_{\lambda})\frac{\partial L}{\partial \nabla_{\lambda}\phi}&=&0
\end{eqnarray}
where $T_{\mu}=T^{\lambda}_{\mu\lambda}$,  
\begin{eqnarray}
G_{ij}=R_{ij}-0.5Rg_{ij}, \quad T_{ij}=\frac{\delta \sqrt{-g}L_{m}}{\delta g^{ij}}, \quad S^{\lambda}_{\, \mu\nu}=\frac{\delta L_{m}}{\delta T^{\mu\nu}_{\,\, \lambda}}, \quad B^{\lambda}_{\mu\nu}=T^{\lambda}_{\mu\nu} + \delta_{\mu}^{\lambda} T_{\nu} - \delta_{\nu}^{\lambda} T_{\mu}.\end{eqnarray}
\subsubsection{Einstein-Yamabe  gravity}
The equations of motion of the Einstein-Yamabe  gravity  are given by
 \cite{gr-qc/0602054}-\cite{hep-th/0507284}
\begin{eqnarray}
R_{ij}-0.5Rg_{ij}-\kappa T_{ij}+\phi_{ij}&=&0,\\
g_{ij\tau}+2nRg_{ij}+f_{ij}&=&0.
\end{eqnarray}
	

\section{MG-VIII: Myrzakulov $F(R, T, Q, {\cal T})$ gravity}
Let us consider the  general spacetime with the curvature,  torsion and  nonmetricity. In this  spacetime, the action of the  Myrzakulov $F(R, T, Q, {\cal T})$ gravity  (or shortly the MG-VIII gravity) is given by \cite{1205.5266}
\begin{equation}
S=S_{g}+S_{m}=\int \sqrt{-g}\;d^{4}x \left[F(R,T,Q,{\cal T})+L_\mathrm{m}\right]=\int \sqrt{-g}\;d^{4}x F(R,T,Q,{\cal T})+S_{m}, \label{3.1}
\end{equation}
where  $R$ stands for the Ricci scalar (curvature scalar), $T$ is the torsion scalar, $Q$ is the nonmetricity scalar and  ${\cal T}$ is trace of the energy-momentum tensor of matter Lagrangian $L_{m}$.  These fourth scalars are given by
\begin{eqnarray}
R&=& g^{\mu\nu}R_{\mu\nu}\\
T&=&{S_\rho}^{\mu\nu}\,{T^\rho}_{\mu\nu}, \\
Q&=&-g^{\mu\nu}(L^{\alpha}_{\beta\mu}L^{\beta}_{\nu\alpha}-L^{\alpha}_{\beta\alpha}L^{\beta}_{\mu\nu}),\\
{\cal T}&=&g^{\mu\nu}T_{\mu\nu}.
\end{eqnarray}
Here 
\begin{eqnarray}
R_{\mu\nu}&=&R^{\alpha}_{\mu\nu\alpha},\\
{\cal T}_{\mu\nu}&=&-\frac{1}{\sqrt{-g}}\frac{\delta(\sqrt{-g}L_{m})}{\delta g^{\mu\nu}}=L_{m}g_{\mu\nu}-\frac{\delta L_{m}}{\delta g^{\mu\nu}} 
\end{eqnarray} 
are $R^{\beta}_{\mu\nu\alpha}$ is the Riemann curvature tensor and  ${\cal T}_{\mu\nu}$ is the  energy-momentum tensor, respectively.  Note that in the action (\ref{3.1}), we have three independent variables: the metric, the affine connection and the matter fields contained in $S_{m}$. In this case, the energy-momentum tensor and the hypermomentum tensor are given by
\begin{eqnarray}
{\cal T}_{\mu\nu}=-\frac{2}{\sqrt{-g}}\frac{\delta S_{m}}{\delta g^{\mu\nu}}, \quad H_{\lambda}^{\, \, \, \mu\nu}=-\frac{1}{2}\frac{\delta S_{m}}{\delta \Gamma^{\lambda}_{\,\,\, \mu\nu}}. 
\end{eqnarray}
 The variation of the action gives 
\begin{eqnarray}
\delta S=\int[F_{R}\delta R+F_{T}\delta T+F_{Q}\delta Q+F_{{\cal T}}\delta{\cal T}-0.5Fg_{\mu\nu}\delta
g^{\mu\nu}+
\frac{2k}{\sqrt{-g}}\frac{\delta(\sqrt{-g}L_{m})}{
\partial g^{\mu\nu}}\delta g^{\mu\nu}]\sqrt{-g}d^{4}x
.
\end{eqnarray}
Let us find $\delta{\cal T}_{\mu\nu}$. We have
\begin{eqnarray}
\delta{\cal T}_{\mu\nu}&=&L_{m}\delta g_{\mu\nu}+g_{\mu\nu}\frac{\partial L_{m}}{g^{\alpha\beta}}\delta g^{\alpha\beta}-2
\frac{\delta^{2}L_{m}}{\partial g^{\mu\nu}\partial g^{\alpha\beta}}\delta g^{\alpha\beta}=\\
&=&-L_{m}g_{\mu\alpha}g_{\nu\beta}\delta g^{\alpha\beta}+
0.5g_{\mu\nu}(g_{\alpha\beta}L_{m}-
T_{\alpha\beta})\delta g^{\alpha\beta}-2
\frac{\delta^{2}L_{m}}{\partial g^{\mu\nu}\partial g^{\alpha\beta}}\delta g^{\alpha\beta}.
\end{eqnarray} 
This equation gives
\begin{eqnarray}
\frac{\delta{\cal T}_{\mu\nu}}{\delta g^{\alpha\beta}}=
-L_{m}g_{\mu\alpha}g_{\nu\beta}+
0.5g_{\mu\nu}(g_{\alpha\beta}L_{m}-
T_{\alpha\beta})-2
\frac{\delta^{2}L_{m}}{\partial g^{\mu\nu}\partial 
g^{\alpha\beta}}. 
\end{eqnarray}
Note that  the variation of ${\cal T}$  with respect to the metric tensor $g_{\mu\nu}$ is given by 
\begin{eqnarray}
\frac{\delta {\cal T}}{\delta g^{\mu\nu}}=\frac{\delta (g^{\alpha\beta}{\cal T}_{\alpha\beta})}{\delta g^{\mu\nu}}=
{\cal T}_{\mu\nu}+\Theta_{\mu\nu}.
\end{eqnarray}
Hence we obtain 
\begin{eqnarray}
\Theta_{\mu\nu}=g^{\alpha\beta}\frac{\delta {\cal T}_{\alpha\beta}}{\delta g^{\mu\nu}}=-L_{m}g_{\mu\nu}+
2g_{\mu\nu}L_{m}-2{\cal 
T}_{\mu\nu}-2g^{\alpha\beta}
\frac{\delta^{2}L_{m}}{\partial g^{\mu\nu}\partial 
g^{\alpha\beta}}.
\end{eqnarray}
We now ready to write the gravitational  field equation. We have 
\begin{eqnarray}
F_{R}R_{\mu\nu}+\nabla^{\alpha}\nabla_{\alpha}(F_{R}g_{\mu\nu})-
\nabla_{\mu}\nabla_{\nu}F_{R}-0.5Fg_{\mu\nu} + ...  =(\kappa-
 F_{{\cal T}}){\cal T}_{\mu\nu}-F_{{\cal T}}\Theta_{\mu\nu},
\end{eqnarray}
where $\kappa=0.5$. Let us consider the perfect fluid with
\begin{eqnarray}
{\cal T}_{\mu\nu}=(\rho+p)u_{\mu}u_{\nu}+pg_{\mu\nu},
\end{eqnarray}
where $\rho$ and $p$ are the  energy density and matter pressure of the  fluid,
respectively. The $u = (0, 0, 0, 1)$  is the components of the four velocity vector ($u_{\mu}$) in the co-moving
coordinate system which satisfies the conditions $u^{\mu}u_{\mu}=1$ and $u^{\mu}\nabla_{\nu}u_{\mu}=0$.  We choose the
perfect fluid matter as $L_{m}=p$ in the action (\ref{3.1}). Therefore we obtain
\begin{eqnarray}
\Theta_{\mu\nu}=-2{\cal T}_{\mu\nu}+pg_{\mu\nu}.
\end{eqnarray}
Substituting the obtained expressions into the field equations we finally get
\begin{eqnarray}
F_{R}R_{\mu\nu}+g_{\mu\nu}\nabla^{\alpha}\nabla_{\alpha}F_{R}-
\nabla_{\mu}\nabla_{\nu}F_{R}-0.5Fg_{\mu\nu} + ...  =(\kappa-
 F_{{\cal T}}){\cal T}_{\mu\nu}-F_{{\cal T}}pg_{\mu\nu}.
\end{eqnarray}
Now we want rewrite the action of the MG-VIII  with the  lagrangian multipliers as
\begin{eqnarray}
S=\int \sqrt{-g}d^{4}x[F-\lambda_{1}(R-R_{s}-u)-\lambda_{2}(T-T_{s}-v)-\lambda_{3}(Q-Q_{s}-w)- \nonumber \\ \lambda_{4}({\cal T}-{\cal T}_{s}-y)+L_{m}].
\end{eqnarray}
The variations with respect to $R, T, Q, {\cal T}$  of the action give $\lambda_{1} = F_{R}, \lambda_{2}=F_{T}, \lambda_{3}=F_{Q}, \lambda_{4}=F_{{\cal T}}$ respectively. Thus the action of the MG-VIII    takes the form
\begin{eqnarray}
S=\int \sqrt{-g}d^{4}x[F-F_{R}(R-R_{s}-u)-F_{T}(T-T_{s}-v)- \nonumber \\ F_{Q}(Q-Q_{s}-w)-F_{{\cal T}}({\cal T}-{\cal T}_{s}-y)+ L_{m}].
\end{eqnarray}
Let us find the variation  of the curvature  scalar $R$. We  obtain
\begin{eqnarray}
\delta R= \delta(g^{\mu\nu}R_{\mu\nu})=R_{\mu\nu}\delta g^{\mu\nu} + g^{\mu\nu}(\nabla_{\lambda}\delta\Gamma^{\lambda}_{\mu\nu}-\nabla_{\nu}\Gamma^{\lambda}_{\mu\lambda}).
\end{eqnarray}
Note that the variation of the affine connection is given by
\begin{eqnarray}
\delta \Gamma^{\lambda}_{\mu\nu}=0.5g^{\lambda\alpha}(\nabla_{\mu}\delta g_{\nu\alpha}+ \nabla_{\nu}\delta g_{\alpha\mu}-\nabla_{\alpha}\delta g_{\mu\nu}).
\end{eqnarray}
Therefore, for the variation of the curvature  scalar $R$ we obtain
\begin{eqnarray}
\delta R=R_{\mu\nu}\delta g^{\mu\nu}+g_{\mu\nu}\Box \delta g^{\mu\nu}-\nabla_{\mu}\nabla_{\nu}\delta g^{\mu\nu}.
\end{eqnarray}
Using the Palatini formalism 
(see, for example, Refs \cite{1212.6393}, \cite{1101.3864}) and  varying the action with respect to the metric and the affine connection, we obtain the following system of the  two field equations
\begin{eqnarray}
F_{R}R_{\mu\nu}+g_{\mu\nu}\nabla^{\alpha}\nabla_{\alpha}F_{R}-
\nabla_{\mu}\nabla_{\nu}F_{R}-0.5Fg_{\mu\nu} + ...  =(k-
 F_{{\cal T}}){\cal T}_{\mu\nu}-F_{{\cal T}}pg_{\mu\nu}, \\
\nabla_{\rho}\left[\sqrt{-g}(\delta^{\rho}_{\lambda}F_{R}g^{\mu\nu}-0.5\delta^{\mu}_{\lambda}F_{R}g^{\rho\nu}-0.5\delta^{\nu}_{\lambda}F_{R}g^{\mu\rho})\right]+ ... =H_{\lambda}^{\, \, \, \mu\nu}.
\end{eqnarray}
Let us also here present one  important equation. The trace
of the field equation (3.15) of the MG-VIII   becomes
\begin{equation}
F_{R}R-2F+3\Box F_{R}+... =\frac{1}{2}{\cal T}+F_{{\cal T}}{\cal T}-4pF_{{\cal T}}.
\end{equation}
\section{FRW cosmological equations}

For a simplicity, we consider  the flat   FLRW
metric in the following form
\begin{equation}
ds^2= -dt^2+a^2(t)\,\delta_{ij} dx^i dx^j,\label{4.1}
\end{equation}
where  $a (t) $ stands for  the scale factor. If we write down Lagrangian of $F(R,T, Q, {\cal T})$ for this metric and if we assumed that the Universe is filled with matter fields with effective pressure $p$ and energy density $\rho$, we obtain ${\cal T}_{s}=3p-\rho$.  Therefore the trace
of the field equation (3.24)-(3.25) of the MG-VIII  becomes as in (3.26).
In the FRW spacetime, the action of the MG-VIII  reads as 
\begin{equation}
S=\int {\cal L} dt,
\end{equation}
where the point like Lagrangian of the MG-VIII  after an integration  by part takes the form 
\begin{eqnarray}
&&{\cal L}=L+{\bar L}_m=a^3\Big( F-R F_{R}-TF_{T}-QF_{Q}-{\cal T}F_{{\cal T}}\Big)\\
&&\nonumber-6a\dot{a}^{2}\Big(F_{R}+F_{T}-F_{Q})-6F_{Rt}a^{2}\dot{a}+a^{3}[uF_{R}+vF_{T}+wF_{Q}+({\cal T}_{s}+y)F_{{\cal T}}+L_{m}].
\end{eqnarray} 
Here we suppose that $L_m=-\epsilon p(a), \quad (\epsilon=\pm 1)$ and
\begin{eqnarray}
L=a^3\Big( F-R F_{R}-TF_{T}-QF_{Q}-{\cal T}F_{{\cal T}}\Big)
-6a\dot{a}^{2}\Big(F_{R}+F_{T}-F_{Q})-6F_{Rt}a^{2}\dot{a}&=&\\ a^{3}(1)-6a\dot{a}^{2}(2)-6F_{Rt}a^{2}\dot{a}, \nonumber \\
{\bar L}_{m}=a^{3}[uF_{R}+vF_{T}+wF_{Q}+({\cal T}_{s}+y)F_{{\cal T}}+L_{m}],& &
\end{eqnarray}
where
\begin{eqnarray}
(1)&=&F-RF_{R}--TF_{T}-QF_{Q}-{\cal T}F_{{\cal T}},\\
(2)&=&F_{R}+F_{T}-F_{Q}.
\end{eqnarray}
 Here 
 \begin{eqnarray}
 R&=&6(\frac{\ddot {a}}{a}+{\frac 
{{\dot{a}}^{2} }{{a_{{}}}^{2}}})+u,\\
   T&=&-6\,{\frac 
{{\dot{a}}^{2} }{{a_{{}}}^{2}}}+\upsilon,\\
	Q&=&6\,{\frac 
{{\dot{a}}^{2} }{{a_{{}}}^{2}}}+w,\\
	{\cal T}&=&3p-\rho +y,\
 \end{eqnarray}  
where $u, v, w, y$ are some real functions of $a, \dot{a}, \, \,  ... \, \, \, $.
The associated Euler-Lagrange equations  are given by 
 \begin{eqnarray}
\frac{d}{dt}(\frac{\partial{\cal L}}{\partial \dot{q}})-\frac{\partial{\cal L}}{\partial q}=0,
\end{eqnarray}
 where  $q\equiv\{a,R,T, Q,  {\cal T}\}$. Let find the following derivatives
\begin{eqnarray}
\frac{\partial{\cal L}}{\partial a}&=&3a^{2}(1)-6\dot{a}^{2}(2)-12F_{Rt}a\dot{a}+\frac{\partial\bar { L}_{m}}{\partial a},\\
\frac{\partial{\cal L}}{\partial \dot{a}}&=&-12a\dot{a}(2)-6F_{Rt}a^{2}+\frac{\partial\bar{L}_{m}}{\partial \dot{a}},\\
\frac{\partial{\cal L}}{\partial R}&=&a^{3}(1)_{R}-6a\dot{a}^{2}(2)_{R}-6F_{RRt}a^{2}\dot{a}+\frac{\partial\bar { L}_{m}}{\partial R},\\
\frac{\partial\bar{\cal L}}{\partial \dot{R}}&=&-6F_{RR}a^{2}\dot{a}+\frac{\partial\bar{ L}_{m}}{\partial \dot{R}},\\
\frac{\partial{\cal L}}{\partial T}&=&a^{3}(1)_{T}-6a\dot{a}^{2}(2)_{T}-6F_{RTt}a^{2}\dot{a}+\frac{\partial\bar { L}_{m}}{\partial T},\\
\frac{\partial{\cal L}}{\partial \dot{T}}&=&-6F_{RT}a^{2}\dot{a}+\frac{\partial\bar{L}_{m}}{\partial \dot{T}},\\
\frac{\partial{\cal L}}{\partial Q}&=&a^{3}(1)_{Q}-6a\dot{a}^{2}(2)_{Q}-6F_{RQt}a^{2}\dot{a}+\frac{\partial\bar {L}_{m}}{\partial Q},\\
\frac{\partial{\cal L}}{\partial \dot{Q}}&=&-6F_{RQ}a^{2}\dot{a}+\frac{\partial\bar{L}_{m}}{\partial \dot{Q}},\\
\frac{\partial{\cal L}}{\partial {\cal T}}&=&a^{3}(1)_{{\cal T}}-6a\dot{a}^{2}(2)_{{\cal T}}-6F_{R{\cal T}t}a^{2}\dot{a}+\frac{\partial\bar {L}_{m}}{\partial {\cal T}},\\
\frac{\partial{\cal L}}{\partial \dot{{\cal T}}}&=&-6F_{R{\cal T}}a^{2}\dot{a}+\frac{\partial\bar{L}_{m}}{\partial \dot{{\cal T}}}.
\end{eqnarray}
Now we assume that
\begin{eqnarray}
\frac{\partial\bar{L}_{m}}{\partial \dot{R}}=\frac{\partial\bar{L}_{m}}{\partial \dot{T}}=\frac{\partial\bar{L}_{m}}{\partial \dot{Q}}=\frac{\partial\bar{L}_{m}}{\partial \dot{{\cal T}}}=0.
\end{eqnarray}
Thus finally we obtain
\begin{eqnarray}
\frac{\partial{\cal L}}{\partial a}&=&3a^{2}(1)-6\dot{a}^{2}(2)-12F_{Rt}a\dot{a}+\frac{\partial\bar{L}_{m}}{\partial a},\\
\frac{\partial{\cal L}}{\partial \dot{a}}&=&-12a\dot{a}(2)-6F_{Rt}a^{2}+\frac{\partial\bar{L}_{m}}{\partial \dot{a}}, \\
\frac{\partial{\cal L}}{\partial R}&=&a^{3}(1)_{R}-6a\dot{a}^{2}(2)_{R}-6F_{RRt}a^{2}\dot{a}+\frac{\partial\bar{L}_{m}}{\partial R},\\
\frac{\partial\bar{\cal L}}{\partial \dot{R}}&=&-6F_{RR}a^{2}\dot{a},\\
\frac{\partial{\cal L}}{\partial T}&=&a^{3}(1)_{T}-6a\dot{a}^{2}(2)_{T}-6F_{RTt}a^{2}\dot{a}+\frac{\partial\bar{L}_{m}}{\partial T},\\
\frac{\partial{\cal L}}{\partial \dot{T}}&=&-6F_{RT}a^{2}\dot{a},\\
\frac{\partial{\cal L}}{\partial Q}&=&a^{3}(1)_{Q}-6a\dot{a}^{2}(2)_{Q}-6F_{RQt}a^{2}\dot{a}+\frac{\partial\bar{L}_{m}}{\partial Q},\\
\frac{\partial{\cal L}}{\partial \dot{Q}}&=&-6F_{RQ}a^{2}\dot{a},\\
\frac{\partial{\cal L}}{\partial {\cal T}}&=&a^{3}(1)_{{\cal T}}-6a\dot{a}^{2}(2)_{{\cal T}}-6F_{R{\cal T}t}a^{2}\dot{a}+\frac{\partial\bar{L}_{m}}{\partial {\cal T}},\\
\frac{\partial{\cal L}}{\partial \dot{{\cal T}}}&=&-6F_{R{\cal T}}a^{2}\dot{a}.
\end{eqnarray}
As result, we obtain the following five equations
\begin{eqnarray}
6 \left(\dot{a}^2+2a \ddot{a}\right) (2)+3 a^2 (1) +12 a \dot{a} {(2)}_t + 6F_{Rtt} a^2+\frac{\partial {\bar L}_{m}}{\partial  a}-\frac{\partial^{2} {\bar L}_{m}}{\partial t\partial \dot{a}} &=&0 , \\
a^{3}(1)_{R}+6a\dot{a}^{2}[2F_{RR}-(2)_{R}]+6F_{RR}a^{2}\ddot{a}+\frac{\partial\bar{L}_{m}}{\partial R}&=&0,\\
a^{3}(1)_{T}+6a\dot{a}^{2}[2F_{RT}-(2)_{T}]+6F_{RT}a^{2}\ddot{a}+\frac{\partial\bar{L}_{m}}{\partial T}&=&0,\\
a^{3}(1)_{Q}+6a\dot{a}^{2}[2F_{RQ}-(2)_{Q}]+6F_{RQ}a^{2}\ddot{a}+\frac{\partial\bar{L}_{m}}{\partial Q}&=&0,\\
a^{3}(1)_{{\cal T}}+6a\dot{a}^{2}[2F_{R{\cal T}}-(2)_{{\cal T}}]+6F_{R{\cal T}}a^{2}\ddot{a}+\frac{\partial\bar{L}_{m}}{\partial {\cal T}}&=&0. 
\end{eqnarray}
 One more equation we get from the following Hamiltonian constraint 
\begin{equation}
{\cal H}=\dot{a}\frac{\partial {\cal L}}{\partial \dot{a}}+\dot{R}\frac{\partial {\cal L}}{\partial \dot{R}}+\dot{T}\frac{\partial {\cal L}}{\partial \dot{T}}+\dot{Q}\frac{\partial {\cal L}}{\partial \dot{Q}}+\dot{{\cal T}}\frac{\partial {\cal L}}{\partial \dot{{\cal T}}}-{\cal L}=0.
\end{equation}
This constraint gives
\begin{equation}
6 a \dot{a}^2 (2)+6 a^2 \dot{a} [F_{RR}\dot{R} +F_{RT}\dot{T} +F_{RQ}\dot{Q} +F_{R{\cal T}}\dot{{\cal T}}] -\dot{a}\frac{\partial \bar{L}_{m}}{\partial \dot{a}}+\bar{L}_{m}=0.
\end{equation}
Finally we have the following system of the 6 gravitational equations 
\begin{eqnarray}
6 \left(\dot{a}^2+2a \ddot{a}\right) (2)+3 a^2 (1) +12 a \dot{a} {(2)}_t + 6F_{Rtt} a^2+\frac{\partial {\bar L}_{m}}{\partial  a}-\frac{\partial^{2} {\bar L}_{m}}{\partial t\partial \dot{a}} &=&0 , \\
a^{3}(1)_{R}+6a\dot{a}^{2}[2F_{RR}-(2)_{R}]+6F_{RR}a^{2}\ddot{a}+\frac{\partial\bar{L}_{m}}{\partial R}&=&0,\\
a^{3}(1)_{T}+6a\dot{a}^{2}[2F_{RT}-(2)_{T}]+6F_{RT}a^{2}\ddot{a}+\frac{\partial\bar{L}_{m}}{\partial T}&=&0,\\
a^{3}(1)_{Q}+6a\dot{a}^{2}[2F_{RQ}-(2)_{Q}]+6F_{RQ}a^{2}\ddot{a}+\frac{\partial\bar{L}_{m}}{\partial Q}&=&0,\\
a^{3}(1)_{{\cal T}}+6a\dot{a}^{2}[2F_{R{\cal T}}-(2)_{{\cal T}}]+6F_{R{\cal T}}a^{2}\ddot{a}+\frac{\partial\bar{L}_{m}}{\partial {\cal T}}&=&0,\\
6 a \dot{a}^2 (2)+6 a^2 \dot{a} [F_{RR}\dot{R} +F_{RT}\dot{T} +F_{RQ}\dot{Q} +F_{R{\cal T}}\dot{{\cal T}}] -\dot{a}\frac{\partial \bar{L}_{m}}{\partial \dot{a}}+\bar{L}_{m}&=&0. 
\end{eqnarray}

\section{FRW cosmology of  $F=\alpha R+\beta T +\mu Q+\nu{\cal T}$}
To understand the physical and mathematical nature of the Myrzakulov $F(R, T, Q, {\cal T})$ gravity   (that is the MG-VIII), in this section, we consider the following particular model   
\begin{eqnarray}
F(R,T,Q, {\cal T})=\alpha R+\beta T +\mu Q+\nu {\cal T},
\end{eqnarray}
where $\alpha, \beta, \mu, \nu$ are some real constants. Then 
\begin{eqnarray}
(1)=0, \quad (2)=\alpha+ \beta- \mu=\sigma.
\end{eqnarray}
In this particular case,  the Lagrangian (4.3) takes the form
\begin{eqnarray}
{\cal L}=-6\sigma a\dot{a}^{2}+a^{3}[\alpha u+\beta v+\mu w+\nu (y+{\cal T}_{s})+  L_m]=-6\sigma a\dot{a}^{2}+a^{3}B,
\end{eqnarray} 
where 
\begin{equation}
B = \alpha u+\beta v+\mu w+\nu (y+{\cal T}_{s})+  L_m.
\end{equation}
 Let us find the following derivatives:
\begin{eqnarray}
\frac{\partial {\cal L}}{\partial a}&=& -6\sigma\dot{a}^{2}+\{a^{3}[\alpha u+\beta v+\lambda w+\gamma (y+{\cal T}_{s})+  L_m]\}_{a}=-6\sigma\dot{a}^{2}+[a^{3}B]_{a},\\
\frac{\partial {\cal L}}{\partial\dot{a}}&=& -12\sigma a\dot{a}+\{a^{3}[\alpha u_{\dot{a}}+\beta v_{\dot{a}}+\lambda w_{\dot{a}}+\gamma (y_{\dot{a}}+{\cal T}_{s\dot{a}})+  L_{m\dot{a}}]\}=-12\sigma a\dot{a}+a^{3}B_{\dot{a}}, \\
\left(\frac{\partial {\cal L}}{\partial\dot{a}}\right)_{t}&=& -12\sigma (\dot{a}^{2}+a\ddot{a})+[a^{3}B_{\dot{a}}]_{t}. 
 \end{eqnarray}
Hence from the Euler-Lagrange equation 
\begin{eqnarray}
\frac{\partial {\cal L}}{\partial a}-\left(\frac{\partial {\cal L}}{\partial\dot{a}}\right)_{t}=0,
 \end{eqnarray}
we obtain the following first field equation
\begin{eqnarray}
6\sigma(\dot{a}^{2}+2a\ddot{a})+[a^{3}B]_{a}-[a^{3}B_{\dot{a}}]_{t}=0. 
 \end{eqnarray}
From the Hamiltonian constraint 
\begin{eqnarray}
{\cal H}=\dot{a}\frac{\partial {\cal L}}{\partial a}-{\cal L}=0,
 \end{eqnarray}
we get the second field equation
\begin{eqnarray}
  -12\sigma a\dot{a}^{2}+a^{3}\dot{a}B_{\dot{a}}+6\sigma a\dot{a}^{2}-a^{3}B=0
 \end{eqnarray}
or
\begin{eqnarray}
  6\sigma a\dot{a}^{2}-a^{3}\dot{a}B_{\dot{a}}+a^{3}B=0.
 \end{eqnarray}
Finally  we get the following system of the two field equations
\begin{eqnarray}
6\sigma a\dot{a}^{2}-a^{3}\dot{a}B_{\dot{a}}+a^{3}B=0, 
 \\
6\sigma(\dot{a}^{2}+2a\ddot{a})+[a^{3}B]_{a}-[a^{3}B_{\dot{a}}]_{t}=0. 
\end{eqnarray}

We  can rewrite these two equations in the following standard forms
\begin{eqnarray}
3H^{2}&=&\rho,\\
2\dot{H}&=&-(\rho+p),
\end{eqnarray}
where the matter density $\rho$ and the pressure $p$ have the following forms
\begin{eqnarray}
\rho &=& \frac{1}{2 \sigma}[\dot{a}B_{\dot{a}}-B],\label{5.17}\\
p&=& \frac{1}{6\sigma a^2}[(a^{3}B)_{a}-(a^{3}B_{\dot{a}})_{t}]. 
\end{eqnarray}
The EoS have the form
\begin{equation}
\omega=\frac{p}{\rho}=\frac{1}{3a^2}\frac{(a^{3}B)_{a}-(a^{3}B_{\dot{a}})_{t}}{\dot{a}B_{\dot{a}}-B}.
\end{equation}

\section{Cosmological solutions}
As example of the cosmological solutions, let us consider the power-law solution
\begin{equation} 
a=a_0 t^n, 
\end{equation}
where $a_{0}, n$ are some constants. Then
\begin{equation} 
\rho=\frac{3n^2}{t^2}, \quad p=\frac{n(2-3n)}{t^2}, \quad H=\frac{n}{t}, \quad \dot{H}=-\frac{n}{t^{2}}. \label{6.2}
\end{equation}
On the other hand, from (\ref{5.17}) we obtain 
\begin{eqnarray} 
\rho=\frac{1}{2\sigma}[\frac{t}{n-1}B_{t}-B]\label{6.3}
\end{eqnarray}
where we used the following formulas
\begin{eqnarray} 
B_{\dot{a}}=\frac{t^{2-n}}{na_{0}(n-1)}B_{t}, \quad B_{a}=\frac{t^{1-n}}{a_{0}n}B_{t}.
\end{eqnarray}
From (5.18) we get the following expression for the pressure
\begin{eqnarray} 
p=\frac{1}{6\sigma}[3B-\frac{(3+n)t}{n(n-1)}B_{t}-\frac{t^{2}}{n(n-1)}B_{tt}].\label{6.5}
\end{eqnarray}
Now we assume that $B$ has the form
\begin{eqnarray} 
B=\frac{\delta}{t^{2}},
\end{eqnarray}
where $\delta=const$. Eqs. (\ref{6.2}) and (\ref{6.3}) for the density of energy give
\begin{eqnarray} 
\delta=\frac{6\sigma n^{2}(1-n)}{1+n}.\label{6.7}
\end{eqnarray} 
At the same time, the expressions of the pressure (\ref{6.2}) and (\ref{6.5}) give 
\begin{equation}
\delta=\frac{6\sigma (2-3n)n(n-1)}{3n-1}. \label{6.8}
\end{equation}
The last two equations give that $n=0$ that is $a=a_{0}=const$. Thus  the power-law solution is the trivial at least  for our assumptions.

\section{Wheeler-DeWitt equation}

In the Hamiltonian formulation of ordinary classical mechanics the key concept  is the Poisson bracket (PB). In this formalism, the canonical coordinate system consists of canonical position $q_{i}$  and momentum $p_{i}$ variables which  satisfy the following fundamental canonical PB relations
\begin{eqnarray}
\{ q_i , p_j \} = \delta_{ij}.
\end{eqnarray}
Here the PB reads as 
\begin{eqnarray}
\{f,g\} = \sum_{i=1}^N \left(
\frac{\partial f}{\partial q_i} \frac{\partial g}{\partial p_i} - \frac{\partial f}{\partial p_i} \frac{\partial g}{\partial q_i}\right),
\end{eqnarray}
where  $f, g$ are the  phase space functions. Correspondingly, the Hamilton equations have the following forms
\begin{eqnarray}
\dot{q}_i &=& \{ q_i , H \},\\
\dot{p}_i &=& \{ p_i , H \},
\end{eqnarray}
which  can be interpreted as the flow or orbit in phase space generated by $H$. In  quantum case the $q, p$  are promoted to quantum operators $\hat{q}, \hat{p}$ on a  Hilbert space with  the following canonical commutation
\begin{eqnarray}
[\hat{q}, \hat{p}] = i \hbar.
\end{eqnarray}
These operators satisfy the following equations 
\begin{eqnarray}
\hat{q} \psi (q) &=& q \psi (q), \\
\hat{p} \psi (q) &=& -i \hbar \frac{d}{dq} \psi (q).
\end{eqnarray}
Finally we get the following  Schr\"{o}dinger equation
\begin{eqnarray}
i \hbar \frac{\partial}{\partial t} \psi = \hat{H} \psi,
\end{eqnarray}
where $\hat{H}$ is the operator form of the Hamiltonian ${\cal H}$  with the usual replacements 
\begin{eqnarray}
 q \mapsto q, \quad p \mapsto -i \hbar \frac{d}{d q}.
\end{eqnarray}
The momenta conjugate to variable $a$ is given by
\begin{eqnarray}
 \pi_{1}=\frac{\partial L}{\partial \dot{a}}=-12\sigma a\dot{a}+a^{3}B_{\dot{a}}.
\end{eqnarray}
Hence we get
\begin{eqnarray}
 \dot{a}=-\frac{\pi_{1}-a^{3}B_{\dot{a}}}{12\sigma a}.
\end{eqnarray}
Therefore the Hamiltonian takes the form
\begin{eqnarray}
 {\hat H}=\dot{a}\frac{\partial {\cal  L}}{\partial \dot{a}}-{\cal L}=6\sigma a\dot{a}^{2}-a^{3}\dot{a}B_{\dot{a}}+a^{3}B\end{eqnarray}
or
\begin{eqnarray}
 \hat{ H}=\frac{1}{24\sigma a}[\pi_{1}-a^{3}B_{\dot{a}}]^{2}+\frac{a^{3}B_{\dot{a}}(\pi_{1}-a^{3}B_{\dot{a}})}{12\sigma a}+a^{3}B=\frac{1}{24\sigma a}\left(\pi_{1}^{2}-a^{6}B_{\dot{a}}^{2}+24\sigma a^{4}B\right).
\end{eqnarray}
The classical dynamics is governed by the following Hamiltonian equations
\begin{eqnarray}
\dot{a} &=& \{a, {\hat H}\}=\frac{\partial {\hat H}}{\partial \pi_{1}},\\
\dot{\pi}_{1} &=&  \{\pi_{1}, {\hat H}\}=-\frac{\partial {\hat H}}{\partial a}.
\end{eqnarray}
Therefore, we have
\begin{eqnarray}
\dot{a} &=& \frac{\pi_{1}}{12\sigma a},\\
\dot{\pi}_{1} &=& \frac{\left(\pi_{1}^{2}-a^{6}B_{\dot{a}}^{2}+24\sigma a^{4}B\right)}{24\sigma a^{2}}+\frac{\left(a^{6}B_{\dot{a}}^{2}-24\sigma a^{4}B\right)_{a}}{24\sigma a}.
\end{eqnarray}
According to the Dirac quantization approach,
the quantum states of the universe should be annihilated by the operator version of the Hamiltonian, that is
\begin{eqnarray}
 \hat{{\hat H}}\Psi=\left[\frac{1}{24\sigma a}\left(\pi_{1}^{2}-a^{6}B_{\dot{a}}^{2}+24\sigma a^{4}B\right)\right]\Psi=0,
\end{eqnarray}
where $\Psi=\Psi(a)$ is the wave function of the universe. We now use the standard  representation $\pi_{1}\rightarrow -i\partial_{a}$. Then we obtain the Wheeler - DeWitt equation ((WDWE) \cite{1208.3828}-\cite{2006.11935}
\begin{eqnarray}
 \hat{{\hat H}}\Psi=\left[\frac{1}{24\sigma a}\left(-\frac{\partial^{2}}{\partial^{2}a}-a^{6}B_{\dot{a}}^{2}+24\sigma a^{4}B\right)\right]\Psi=0
\end{eqnarray}
or
\begin{eqnarray}
 \left[\frac{1}{24\sigma a}\left(\frac{\partial^{2}}{\partial^{2}a}+a^{6}B_{\dot{a}}^{2}-24\sigma a^{4}B\right)\right]\Psi=0.
\end{eqnarray}
\section{Relation with the soliton theory}
Let us  rewrite the WDWE as
\begin{eqnarray}
L\Psi= -\left[\partial^{2}_{a}-U\right]\Psi=\lambda \Psi,
\end{eqnarray}
where
\begin{eqnarray}
 U=-a^{6}B_{\dot{a}}^{2}+24\sigma a^{4}B.
\end{eqnarray}
Introduce the operator $A$ as
\begin{eqnarray}
A= 4\partial^{3}_{a}-3[U\partial_{a}+\partial_{a} U].
\end{eqnarray}
Then the Lax equation
\begin{eqnarray}
L_{\Lambda}=[L,A] 
\end{eqnarray}
gives the famous Korteweg-de Vries equation
\begin{eqnarray}
U_{\Lambda}+6UU_{a}+U_{aaa}=0.
\end{eqnarray}

\section{Metric-affine MG theories}
In this section, some metric-affine Myrzakulov gravity theories are presented \cite{1205.5266}.
Consider the  metric-affine spacetime with the affine connection $\tilde{\Gamma}^{\lambda}\,_{\mu \nu}$. Then the torsion and nonmetricity tensors are given by
\begin{align}
    T^{\lambda}\,_{\mu \nu}&=2\tilde{\Gamma}^{\lambda}\,_{[\mu \nu]}\,,\\
    Q_{\lambda \mu \nu}&=\tilde{\nabla}_{\lambda}g_{\mu \nu}\,.
\end{align}
The corresponding  covariant derivative of an arbitrary vector $v^{\lambda}$ can be split into a Riemannian contribution and a distortion tensor
\begin{equation}
\tilde{\nabla}_{\mu}v^{\lambda}=\nabla_{\mu}v^{\lambda}+N^{\lambda}\,_{\rho\mu}v^{\rho}\,,
\end{equation}
where 
\begin{equation}
N^{\lambda}\,_{\rho\mu}=K^{\lambda}\,_{\rho\mu}+L^{\lambda}\,_{\rho\mu}.
\end{equation}
Here the contortion and disformation tensors read as
\begin{align}
    K^{\lambda}\,_{\rho\mu}&=\frac{1}{2}(T^{\lambda}\,_{\rho\mu}-T_{\rho}\,^{\lambda}\,_{\mu}-T_{\mu}\,^{\lambda}\,_{\rho})\,,
\end{align}
\begin{align}
    L^{\lambda}\,_{\rho\mu}&=\frac{1}{2}(Q^{\lambda}\,_{\rho\mu}-Q_{\rho}\,^{\lambda}\,_{\mu}-Q_{\mu}\,^{\lambda}\,_{\rho}),
\end{align}
respectively. The commutation of  the covariant derivatives takes the form
\begin{equation}
[\tilde{\nabla}_{\mu},\tilde{\nabla}_{\nu}]\,v^{\lambda}=\tilde{R}^{\lambda}\,_{\rho \mu \nu}\,v^{\rho}+T^{\rho}\,_{\mu \nu}\,\tilde{\nabla}_{\rho}v^{\lambda}\,,
\end{equation}
where
\begin{equation}\label{totalcurvature}
\tilde{R}^{\lambda}\,_{\rho \mu \nu}=\partial_{\mu}\tilde{\Gamma}^{\lambda}\,_{\rho \nu}-\partial_{\nu}\tilde{\Gamma}^{\lambda}\,_{\rho \mu}+\tilde{\Gamma}^{\lambda}\,_{\sigma \mu}\tilde{\Gamma}^{\sigma}\,_{\rho \nu}-\tilde{\Gamma}^{\lambda}\,_{\sigma \nu}\tilde{\Gamma}^{\sigma}\,_{\rho \mu}\,.
\end{equation}
Note that the geometric structure of the metric-affine spacetime is determined by  three tensors: the metric tensor ($g_{\mu\nu}$), the torsion tensor ($T^{\lambda}_{\, \, \, \mu\nu}$) and the nonmetricity tensor ($Q_{\lambda\mu\nu}$). The torsion tensor is the antisymmetric part of the connection and the nonmetricity tensor measures the failure
of the connection to be metric compatible. Note that these three tensors  can be computed
once an affine connection $\tilde{\Gamma}^{\alpha}_{\beta\lambda}$ is given. In this metric-affine  spacetime, let us introduce five  scalars - $R, \, T, \, Q, \, G, \, B$, where $R$ is the metric-affine  curvature scalar, $T$ is the metric-affine torsion scalar,  $Q$ is the metric-affine nonmetricity scalar,    $G$ is the metric-affine Gauss-Bonnet  scalar, $B$ is the boundary term scalar. Below ${\cal T}$ is the trace of the energy-momentum tensor. In the previous sections, we have considered the Myrzakulov gravity-I (MG-I) which has the following action
\begin{eqnarray}
S=\int \sqrt{-g}d^{4}x[F(R,T)+L_{m}],
\end{eqnarray}
where $R$ is the curvature scalar, $T$ is the torsion scalar and $L_{m}$ is the matter Lagrangian. This MG-I is some kind generalization (unification)  of the well-known $F(R)$ and $F(T)$ gravity theories. We now going to present some other examples of   metric-affine Myrzakulov gravity theories, also abbreviated below as MG-N, where N=I, II, III, IV, ... (see, also, Table 1, Table 2  and Table 3).  

  \subsection{MG-I}
 The action of the Myrzakulov    gravity - I (MG-I) has the following form 
\begin{eqnarray}
S=\frac{1}{2\kappa^{2}}\int \sqrt{-g}d^{4}x[F(R,T)+2\kappa^{2}L_{m}],
\end{eqnarray}
where $R$ is the curvature scalar, $T$ is the torsion scalar and $L_{m}$ is the matter Lagrangian. This MG-I is some kind generalizations of the well-known $F(R)$ and $F(T)$ gravity theories. If exactly, the MG-I is the unification of the $F(R)$ and $F(T)$ theories.    
\subsection{MG-II}
 The action of the Myrzakulov  gravity - II (MG-II) reads as 
\begin{eqnarray}
S=\frac{1}{2\kappa^{2}}\int \sqrt{-g}d^{4}x[F(R,Q)+2\kappa^{2}L_{m}],
\end{eqnarray}
where  $R$ is the curvature scalar and $Q$ is the nonmetricity scalar of the   metric-affine spacetime.
\subsection{MG-III}
 The action of the Myrzakulov   gravity - III (MG-III) reads as 
\begin{eqnarray}
S=\frac{1}{2\kappa^{2}}\int \sqrt{-g}d^{4}x[F(T,Q)+2\kappa^{2}L_{m}],
\end{eqnarray}
where  $T$ is the torsion  scalar and $Q$ is the nonmetricity scalar of the   metric-affine spacetime.
\subsection{MG-IV}
The action of the Myrzakulov  gravity - IV (MG-IV) has the following form
\begin{eqnarray}
S=\frac{1}{2\kappa^{2}}\int \sqrt{-g}d^{4}x[F(R,T,{\cal T})+2\kappa^{2}L_{m}],
\end{eqnarray}
where $R$ is the curvature scalar, $T$ is the torsion scalar and  ${\cal T}$ is the trace of the energy-momentum tensor.

\subsection{MG-V}
The action of the Myrzakulov gravity - V (MG-V) is given by
\begin{eqnarray}
S=\frac{1}{2\kappa^{2}}\int \sqrt{-g}d^{4}x[F(R,T,Q)+2\kappa^{2}L_{m}],
\end{eqnarray}
where $R$ is the curvature scalar, $T$ is the torsion scalar and  $Q$ is the nonmetricity scalar of the   metric-affine spacetime. 
\subsection{MG-VI}
 The action of the Myrzakulov   gravity - VI (MG-VI) reads as 
\begin{eqnarray}
S=\frac{1}{2\kappa^{2}}\int \sqrt{-g}d^{4}x[F(R,Q, {\cal T})+2\kappa^{2}L_{m}],
\end{eqnarray}
where  $R$ is the curvature scalar,  $Q$ is the nonmetricity scalar and  ${\cal T}$ is the trace of the energy-momentum tensor
 of our generalized spacetime.
\subsection{MG-VII}
 The action of the Myrzakulov   gravity - VII (MG-VII) reads as 
\begin{eqnarray}
S=\frac{1}{2\kappa^{2}}\int \sqrt{-g}d^{4}x[F(T,Q, {\cal T})+2\kappa^{2}L_{m}],
\end{eqnarray}
and   $T$ is the torsion  scalar,  $Q$ is the nonmetricity scalar and  ${\cal T}$ is the trace of the energy-momentum tensor
 of the   metric-affine spacetime.

\subsection{MG-VIII}
The action of the Myrzakulov gravity - VIII (MG-VIII) reads as
\begin{eqnarray}
S=\frac{1}{2\kappa^{2}}\int \sqrt{-g}d^{4}x[F(R,T,Q, {\cal T})+2\kappa^{2}L_{m}],
\end{eqnarray}
where $R$ is the curvature  scalar, $T$ is the torsion scalar, $Q$ is the nonmetricity scalar and  ${\cal T}$ is the trace of the energy-momentum tensor (the trace of the stress-energy tensor) of the   metric-affine spacetime.

\newpage

\begin{table}
\caption{Metric-affine Myrzakulov gravity theories}
\begin{tabular}{|c|c|c|}
\hline 
N & Name & Action \\ 
\hline 
1 &Myrzakulov Gravity - I (MG-I) & $S=\frac{1}{2k^2}\int d^4x \sqrt{-g}\left[F(R,T)+2k^{2}L_m\right]$ \\ 
\hline 
2 &Myrzakulov Gravity - II (MG-II) & $S=\frac{1}{2k^2}\int d^4x \sqrt{-g}\left[F(R,Q)+2k^{2}L_m\right]$\\ 
\hline 
3 &Myrzakulov Gravity - III (MG-III) & $S=\frac{1}{2k^2}\int d^4x \sqrt{-g}\left[F(T,Q)+2k^{2}L_m\right]$ \\ 
\hline 
4 & Myrzakulov Gravity - IV (MG-IV) & $S=\frac{1}{2k^2}\int d^4x \sqrt{-g}\left[F(R,T,{\cal T})+2k^{2}L_m\right]$ \\ 
\hline 
5 & Myrzakulov Gravity - V (MG-V) & $S=\frac{1}{2k^2}\int d^4x \sqrt{-g}\left[F(R,T,Q)+2k^{2}L_m\right]$ \\ 
\hline 
6 & Myrzakulov Gravity - VI (MG-VI) & $S=\frac{1}{2k^2}\int d^4x \sqrt{-g}\left[F(R,Q,{\cal T})+2k^{2}L_m\right]$ \\ 
\hline 
7 & Myrzakulov Gravity - VII (MG-VII) & $S=\frac{1}{2k^2}\int d^4x \sqrt{-g}\left[F(T,Q,{\cal T})+2k^{2}L_m\right]$ \\ 
\hline 
8 & Myrzakulov Gravity - VIII (MG-VIII) & $S=\frac{1}{2k^2}\int d^4x \sqrt{-g}\left[F(R,T,Q,{\cal T})+2k^{2}L_m\right]$ \\ 
\hline  
\end{tabular} 
\end{table}

 \section{Metric-affine MG theories with the Gauss-Bonnet scalars}

The metric-affine MG theories with the Gauss-Bonnet scalars ($G$)  were proposed in \cite{1205.5266}. For our convenience, let us present these models (see e.g. Table 2).
  \subsection{MG-IX}
 The action of the Myrzakulov   gravity - IX (MG-IX) has the following form 
\begin{eqnarray}
S=\frac{1}{2\kappa^{2}}\int \sqrt{-g}d^{4}x[F(R,T,G)+2\kappa^{2}L_{m}],
\end{eqnarray}
where $R$ is the curvature scalar, $T$ is the torsion scalar, $G$ is the metric-affine Gauss-Bonnet scalar  of the   metric-affine spacetime.     
\subsection{MG-X}
 The action of the Myrzakulov gravity - X (MG-X) reads as 
\begin{eqnarray}
S=\frac{1}{2\kappa^{2}}\int \sqrt{-g}d^{4}x[F(R,Q,G)+2\kappa^{2}L_{m}],
\end{eqnarray}
where  $R$ is the curvature scalar,  $Q$ is the nonmetricity scalar, $G$ is the metric-affine Gauss-Bonnet scalar of the metric-affine spacetime.
\subsection{MG-XI}
 The action of the Myrzakulov gravity - XI (MG-XI) reads as 
\begin{eqnarray}
S=\frac{1}{2\kappa^{2}}\int \sqrt{-g}d^{4}x[F(T,Q,G)+2\kappa^{2}L_{m}],
\end{eqnarray}
where  $T$ is the metric-affine torsion  scalar,  $Q$ is the metric-affine  nonmetricity scalar and $G$ is the metric-affine Gauss-Bonnet scalar of our metric-affine  spacetime.
\subsection{MG-XII}
The action of the Myrzakulov  gravity - XII (MG-XII) has the following form
\begin{eqnarray}
S=\frac{1}{2\kappa^{2}}\int \sqrt{-g}d^{4}x[F(R,T,G, {\cal T})+2\kappa^{2}L_{m}],
\end{eqnarray}
where $R$ is the metric-affine curvature scalar, $T$ is the metric-affine torsion scalar, $G$ is the metric-affine Gauss-Bonnet scalar and  ${\cal T}$ is the trace of the energy-momentum tensor.

\subsection{MG-XIII}
The action of the Myrzakulov  gravity - XIII (MG-XIII) is given by
\begin{eqnarray}
S=\frac{1}{2\kappa^{2}}\int \sqrt{-g}d^{4}x[F(R,T,Q,G)+2\kappa^{2}L_{m}],
\end{eqnarray}
where $R$ is the curvature scalar, $T$ is the torsion scalar,   $Q$ is the nonmetricity scalar and $G$ is the metric-affine Gauss-Bonnet scalar of the   metric-affine spacetime.
\subsection{MG-XIV}
 The action of the Myrzakulov   gravity - XIV (MG-XIV) reads as 
\begin{eqnarray}
S=\frac{1}{2\kappa^{2}}\int \sqrt{-g}d^{4}x[F(R,Q, G, {\cal T})+2\kappa^{2}L_{m}],
\end{eqnarray}
where  $R$ is the metric-affine curvature scalar,  $Q$ is the metric-affine nonmetricity scalar, $G$ is the metric-affine Gauss-Bonnet scalar and  ${\cal T}$ is the trace of the energy-momentum tensor
 of the  metric-affine  spacetime.
\subsection{MG-XV}
 The action of the Myrzakulov   gravity - XV (MG-XV) reads as 
\begin{eqnarray}
S=\frac{1}{2\kappa^{2}}\int \sqrt{-g}d^{4}x[F(T,Q,G,  {\cal T})+2\kappa^{2}L_{m}],
\end{eqnarray}
and   $T$ is the metric-affine torsion  scalar,  $Q$ is the metric-affine nonmetricity scalar, $G$ is the metric-affine Gauss-Bonnet scalar and  ${\cal T}$ is the trace of the energy-momentum tensor
 of our metric-affine  spacetime. 

\subsection{MG-XVI}
The action of the Myrzakulov gravity - XVI (MG-XVI) reads as
\begin{eqnarray}
S=\frac{1}{2\kappa^{2}}\int \sqrt{-g}d^{4}x[F(R,T,Q, G, {\cal T})+2\kappa^{2}L_{m}],
\end{eqnarray}
where $R$ is the metric-affine curvature  scalar, $T$ is the metric-affine torsion scalar, $Q$ is the metric-affine nonmetricity scalar, $G$ is the metric-affine Gauss-Bonnet scalar and  ${\cal T}$ is the trace of the energy-momentum tensor  of the  metric-affine  spacetime.
 \subsection{MG-XVII}
The action of the Myrzakulov gravity - XVII (MG-XVII) reads as
\begin{eqnarray}
S=\frac{1}{2\kappa^{2}}\int \sqrt{-g}d^{4}x[F(Q, G)+2\kappa^{2}L_{m}],
\end{eqnarray}
where  $Q$ is the metric-affine nonmetricity scalar and $G$ is the metric-affine Gauss-Bonnet scalar  of the metric-affine  spacetime.

\subsection{MG-XVIII}
The action of the Myrzakulov gravity - XVIII (MG-XVIII) reads as
\begin{eqnarray}
S=\frac{1}{2\kappa^{2}}\int \sqrt{-g}d^{4}x[F(R,T,G)+2\kappa^{2}L_{m}],
\end{eqnarray}
where $R$ is the metric-affine curvature  scalar, $T$ is the metric-affine torsion scalar and  $G$ is the metric-affine Gauss-Bonnet scalar of the metric-affine  spacetime.
\subsection{MG-XIX}
The action of the Myrzakulov gravity - XIX (MG-XIX) reads as
\begin{eqnarray}
S=\frac{1}{2\kappa^{2}}\int \sqrt{-g}d^{4}x[F(T,G, {\cal T})+2\kappa^{2}L_{m}],
\end{eqnarray}
where  $T$ is the metric-affine torsion scalar,  $G$ is the metric-affine Gauss-Bonnet scalar and  ${\cal T}$ is the trace of the energy-momentum tensor  of the  metric-affine  spacetime.

\newpage

\begin{table}
\caption{Metric-affine Myrzakulov gravity theories with Gauss-Bonnet scalars}
\begin{tabular}{|c|c|c|}
\hline 
N & Name & Action \\ 
\hline  
9 & Myrzakulov Gravity - IX (MG-IX) & $S=\frac{1}{2k^2}\int d^4x \sqrt{-g}\left[F(R,T,G)+2k^{2}L_m\right]$ \\ 
\hline 
10 &Myrzakulov Gravity - X (MG-X) & $S=\frac{1}{2k^2}\int d^4x \sqrt{-g}\left[F(R,Q,G)+2k^{2}L_m\right]$\\ 
\hline 
11 &Myrzakulov Gravity - XI (MG-XI) & $S=\frac{1}{2k^2}\int d^4x \sqrt{-g}\left[F(T,Q,G)+2k^{2}L_m\right]$ \\ 
\hline 
12 & Myrzakulov Gravity - XII (MG-XII) & $S=\frac{1}{2k^2}\int d^4x \sqrt{-g}\left[F(R,T,G,{\cal T})+2k^{2}L_m\right]$ \\ 
\hline 
13 & Myrzakulov Gravity - XIII (MG-XIII) & $S=\frac{1}{2k^2}\int d^4x \sqrt{-g}\left[F(R,T,Q,G)+2k^{2}L_m\right]$ \\ 
\hline 
14 & Myrzakulov Gravity - XIV (MG-XIV) & $S=\frac{1}{2k^2}\int d^4x \sqrt{-g}\left[F(R,Q,G,{\cal T})+2k^{2}L_m\right]$ \\ 
\hline 
15 & Myrzakulov Gravity - XV (MG-XV) & $S=\frac{1}{2k^2}\int d^4x \sqrt{-g}\left[F(T,Q,G,{\cal T})+2k^{2}L_m\right]$ \\ 
\hline 
16 & Myrzakulov Gravity - XVI (MG-XVI) & $S=\frac{1}{2k^2}\int d^4x \sqrt{-g}\left[F(R,T,Q,G,{\cal T})+2k^{2}L_m\right]$ \\ 
\hline  
17 & Myrzakulov Gravity - XVII (MG-XVII) & $S=\frac{1}{2k^2}\int d^4x \sqrt{-g}\left[F(Q,G)+2k^{2}L_m\right]$ \\ 
\hline 
18 & Myrzakulov Gravity - XVIII (MG-XVIII) & $S=\frac{1}{2k^2}\int d^4x \sqrt{-g}\left[F(R,T,G)+2k^{2}L_m\right]$ \\ 
\hline  
19 & Myrzakulov Gravity - XIX (MG-XIX) & {$S=\frac{1}{2k^2}\int d^4x \sqrt{-g}\left[F(T,G,{\cal T})+2k^{2}L_m\right]$} \\ 
\hline 
\end{tabular} 
\end{table}

 \section{Metric-affine MG theories with  boundary  term scalars}
In this section, we would like to present some metric-affine MG theories with the boundary term scalars ($B$). Note that these MG theories with the boundary term scalars were proposed in 
\cite{1205.5266} (see e.g. Table 3). 

  \subsection{MG-XX}
 The action of the Myrzakulov    gravity - XX (MG-XX) has the following form 
\begin{eqnarray}
S=\frac{1}{2\kappa^{2}}\int \sqrt{-g}d^{4}x[F(R,T,B)+2\kappa^{2}L_{m}],
\end{eqnarray}
where $R$ is the curvature scalar, $T$ is the torsion scalar, $B$ is  the boundary term scalar and $L_{m}$ is the matter Lagrangian. This MG-I is some kind generalizations of the well-known $F(R)$ and $F(T)$ gravity theories. If exactly, the MG-I is the unification of the $F(R)$ and $F(T)$ theories.    
\subsection{MG-XXI}
 The action of the Myrzakulov  gravity - XXI (MG-XXI) reads as 
\begin{eqnarray}
S=\frac{1}{2\kappa^{2}}\int \sqrt{-g}d^{4}x[F(R,Q,B)+2\kappa^{2}L_{m}],
\end{eqnarray}
where  $R$ is the curvature scalar, $B$ is  the boundary term scalar and $Q$ is the nonmetricity scalar of the   metric-affine spacetime.
\subsection{MG-XXII}
 The action of the Myrzakulov   gravity - XXII (MG-XXII) reads as 
\begin{eqnarray}
S=\frac{1}{2\kappa^{2}}\int \sqrt{-g}d^{4}x[F(T,Q, B)+2\kappa^{2}L_{m}],
\end{eqnarray}
where  $T$ is the torsion  scalar, $B$ is  the boundary term scalar and $Q$ is the nonmetricity scalar of the   metric-affine spacetime.
\subsection{MG-XXIII}
The action of the Myrzakulov  gravity - XXIII (MG-XXIII) has the following form
\begin{eqnarray}
S=\frac{1}{2\kappa^{2}}\int \sqrt{-g}d^{4}x[F(R,T,B, {\cal T})+2\kappa^{2}L_{m}],
\end{eqnarray}
where $R$ is the curvature scalar, $T$ is the torsion scalar, $B$ is  the boundary term scalar and  ${\cal T}$ is the trace of the energy-momentum tensor.

\subsection{MG-XXIV}
The action of the Myrzakulov gravity - XXIV (MG-XXIV) is given by
\begin{eqnarray}
S=\frac{1}{2\kappa^{2}}\int \sqrt{-g}d^{4}x[F(R,T,Q, B)+2\kappa^{2}L_{m}],
\end{eqnarray}
where $R$ is the curvature scalar, $T$ is the torsion scalar, $B$ is  the boundary term scalar and  $Q$ is the nonmetricity scalar of the   metric-affine spacetime. 
\subsection{MG-XXV}
 The action of the Myrzakulov   gravity - XXV (MG-XXV) reads as 
\begin{eqnarray}
S=\frac{1}{2\kappa^{2}}\int \sqrt{-g}d^{4}x[F(R,Q, B,  {\cal T})+2\kappa^{2}L_{m}],
\end{eqnarray}
where  $R$ is the curvature scalar,  $Q$ is the nonmetricity scalar, $B$ is  the boundary term scalar and  ${\cal T}$ is the trace of the energy-momentum tensor
 of our generalized spacetime.
\subsection{MG-XXVI}
 The action of the Myrzakulov   gravity - XXVI (MG-XXVI) reads as 
\begin{eqnarray}
S=\frac{1}{2\kappa^{2}}\int \sqrt{-g}d^{4}x[F(T,Q, B, {\cal T})+2\kappa^{2}L_{m}],
\end{eqnarray}
and   $T$ is the torsion  scalar,  $Q$ is the nonmetricity scalar, $B$ is  the boundary term scalar  and  ${\cal T}$ is the trace of the energy-momentum tensor
 of the   metric-affine spacetime.

\subsection{MG-XXVII}
The action of the Myrzakulov gravity - XXVII (MG-XXVII) reads as
\begin{eqnarray}
S=\frac{1}{2\kappa^{2}}\int \sqrt{-g}d^{4}x[F(R,T,Q, B,  {\cal T})+2\kappa^{2}L_{m}],
\end{eqnarray}
where $R$ is the curvature  scalar, $T$ is the torsion scalar, $Q$ is the nonmetricity scalar, $B$ is  the boundary term scalar and  ${\cal T}$ is the trace of the energy-momentum tensor (the trace of the stress-energy tensor) of the   metric-affine spacetime.
 
  \subsection{MG-XXVIII}
 The action of the Myrzakulov   gravity - XXVIII (MG-XXVIII) has the following form 
\begin{eqnarray}
S=\frac{1}{2\kappa^{2}}\int \sqrt{-g}d^{4}x[F(R,T,G, B)+2\kappa^{2}L_{m}],
\end{eqnarray}
where $R$ is the curvature scalar, $T$ is the torsion scalar, $B$ is  the boundary term scalar, $G$ is the metric-affine Gauss-Bonnet scalar  of the   metric-affine spacetime.     
\subsection{MG-XXIX}
 The action of the Myrzakulov gravity - XXIX (MG-XXIX) reads as 
\begin{eqnarray}
S=\frac{1}{2\kappa^{2}}\int \sqrt{-g}d^{4}x[F(R,Q,G, B)+2\kappa^{2}L_{m}],
\end{eqnarray}
where  $R$ is the curvature scalar,  $Q$ is the nonmetricity scalar, $B$ is  the boundary term scalar,  $G$ is the metric-affine Gauss-Bonnet scalar of the metric-affine spacetime.
\subsection{MG-XXX}
 The action of the Myrzakulov gravity - XXX (MG-XXX) reads as 
\begin{eqnarray}
S=\frac{1}{2\kappa^{2}}\int \sqrt{-g}d^{4}x[F(T,Q,G, B)+2\kappa^{2}L_{m}],
\end{eqnarray}
where  $T$ is the metric-affine torsion  scalar,  $Q$ is the metric-affine  nonmetricity scalar, $B$ is  the boundary term scalar and $G$ is the metric-affine Gauss-Bonnet scalar of our metric-affine  spacetime.
\subsection{MG-XXXI}
The action of the Myrzakulov  gravity - XXXI (MG-XXXI) has the following form
\begin{eqnarray}
S=\frac{1}{2\kappa^{2}}\int \sqrt{-g}d^{4}x[F(R,T,G, B, {\cal T})+2\kappa^{2}L_{m}],
\end{eqnarray}
where $R$ is the metric-affine curvature scalar, $T$ is the metric-affine torsion scalar, $G$ is the metric-affine Gauss-Bonnet scalar, $B$ is  the boundary term scalar and  ${\cal T}$ is the trace of the energy-momentum tensor.

\subsection{MG-XXXII}
The action of the Myrzakulov  gravity - XXXII (MG-XXXII) is given by
\begin{eqnarray}
S=\frac{1}{2\kappa^{2}}\int \sqrt{-g}d^{4}x[F(R,T,Q,G, B)+2\kappa^{2}L_{m}],
\end{eqnarray}
where $R$ is the curvature scalar, $T$ is the torsion scalar,   $Q$ is the nonmetricity scalar, $B$ is  the boundary term scalar and $G$ is the metric-affine Gauss-Bonnet scalar of the   metric-affine spacetime.
\subsection{MG-XXXIII}
 The action of the Myrzakulov   gravity - XXXIII (MG-XXXIII) reads as 
\begin{eqnarray}
S=\frac{1}{2\kappa^{2}}\int \sqrt{-g}d^{4}x[F(R,Q, G, B,  {\cal T})+2\kappa^{2}L_{m}],
\end{eqnarray}
where  $R$ is the metric-affine curvature scalar,  $Q$ is the metric-affine nonmetricity scalar, $G$ is the metric-affine Gauss-Bonnet scalar, $B$ is  the boundary term scalar and  ${\cal T}$ is the trace of the energy-momentum tensor
 of the  metric-affine  spacetime.
\subsection{MG-XXXIV}
 The action of the Myrzakulov   gravity - XXXIV (MG-XXXIV) reads as 
\begin{eqnarray}
S=\frac{1}{2\kappa^{2}}\int \sqrt{-g}d^{4}x[F(T,Q,G, B,  {\cal T})+2\kappa^{2}L_{m}],
\end{eqnarray}
and   $T$ is the metric-affine torsion  scalar,  $Q$ is the metric-affine nonmetricity scalar, $G$ is the metric-affine Gauss-Bonnet scalar, $B$ is  the boundary term scalar and  ${\cal T}$ is the trace of the energy-momentum tensor
 of our metric-affine  spacetime. 

\subsection{MG-XXXV}
The action of the Myrzakulov gravity - XXXV (MG-XXXV) reads as
\begin{eqnarray}
S=\frac{1}{2\kappa^{2}}\int \sqrt{-g}d^{4}x[F(R,T,Q, G, B,  {\cal T})+2\kappa^{2}L_{m}],
\end{eqnarray}
where $R$ is the metric-affine curvature  scalar, $T$ is the metric-affine torsion scalar, $Q$ is the metric-affine nonmetricity scalar, $G$ is the metric-affine Gauss-Bonnet scalar, $B$ is  the boundary term scalar and  ${\cal T}$ is the trace of the energy-momentum tensor  of the  metric-affine  spacetime.
 \subsection{MG-XXXVI}
The action of the Myrzakulov gravity - XXXVI (MG-XXXVI) reads as
\begin{eqnarray}
S=\frac{1}{2\kappa^{2}}\int \sqrt{-g}d^{4}x[F(Q, G, B)+2\kappa^{2}L_{m}],
\end{eqnarray}
where  $Q$ is the metric-affine nonmetricity scalar, $B$ is  the boundary term scalar and $G$ is the metric-affine Gauss-Bonnet scalar  of the metric-affine  spacetime.

\subsection{MG-XXXVII}
The action of the Myrzakulov gravity - XXXVII (MG-XXXVII) reads as
\begin{eqnarray}
S=\frac{1}{2\kappa^{2}}\int \sqrt{-g}d^{4}x[F(R,T,G, B)+2\kappa^{2}L_{m}],
\end{eqnarray}
where $R$ is the metric-affine curvature  scalar, $T$ is the metric-affine torsion scalar, $B$ is  the boundary term scalar and  $G$ is the metric-affine Gauss-Bonnet scalar of the metric-affine  spacetime.
\subsection{MG-XXXVIII}
The action of the Myrzakulov gravity - XXXVIII (MG-XXXVIII) reads as
\begin{eqnarray}
S=\frac{1}{2\kappa^{2}}\int \sqrt{-g}d^{4}x[F(T,G, B, {\cal T})+2\kappa^{2}L_{m}],
\end{eqnarray}
where  $T$ is the metric-affine torsion scalar,  $G$ is the metric-affine Gauss-Bonnet scalar, $B$ is  the boundary term scalar and  ${\cal T}$ is the trace of the energy-momentum tensor  of the  metric-affine  spacetime.

\newpage

\begin{table}
\caption{Metric-affine MG theories  with  boundary term scalars}
\begin{tabular}{|c|c|c|}
\hline 
N & Name & Action \\ 
\hline 
1 &Myrzakulov Gravity - XX (MG-XX) & $S=\frac{1}{2k^2}\int d^4x \sqrt{-g}\left[F(R,T,B)+2k^{2}L_m\right]$ \\ 
\hline 
2 &Myrzakulov Gravity - XXI (MG-XXI) & $S=\frac{1}{2k^2}\int d^4x \sqrt{-g}\left[F(R,Q,B)+2k^{2}L_m\right]$\\ 
\hline 
3 &Myrzakulov Gravity - XXII (MG-XXII) & $S=\frac{1}{2k^2}\int d^4x \sqrt{-g}\left[F(T,Q,B)+2k^{2}L_m\right]$ \\ 
\hline 
4 & Myrzakulov Gravity - XXIII (MG-XXIII) & $S=\frac{1}{2k^2}\int d^4x \sqrt{-g}\left[F(R,T,B,{\cal T})+2k^{2}L_m\right]$ \\ 
\hline 
5 & Myrzakulov Gravity - XXIV (MG-XXIV) & $S=\frac{1}{2k^2}\int d^4x \sqrt{-g}\left[F(R,T,Q,B)+2k^{2}L_m\right]$ \\ 
\hline 
6 & Myrzakulov Gravity - XXV (MG-XXV) & $S=\frac{1}{2k^2}\int d^4x \sqrt{-g}\left[F(R,Q,B,{\cal T})+2k^{2}L_m\right]$ \\ 
\hline 
7 & Myrzakulov Gravity - XXVI (MG-XXVI) & $S=\frac{1}{2k^2}\int d^4x \sqrt{-g}\left[F(T,Q,B, {\cal T})+2k^{2}L_m\right]$ \\ 
\hline 
8 & Myrzakulov Gravity - XXVII (MG-XXVII) & $S=\frac{1}{2k^2}\int d^4x \sqrt{-g}\left[F(R,T,Q,B,{\cal T})+2k^{2}L_m\right]$ \\ 
\hline  
9 & Myrzakulov Gravity - XXVIII (MG-XXVIII) & $S=\frac{1}{2k^2}\int d^4x \sqrt{-g}\left[F(R,T,G,B)+2k^{2}L_m\right]$ \\ 
\hline 
10 &Myrzakulov Gravity - XXIX (MG-XXIX) & $S=\frac{1}{2k^2}\int d^4x \sqrt{-g}\left[F(R,Q,G,B)+2k^{2}L_m\right]$\\ 
\hline 
11 &Myrzakulov Gravity - XXX (MG-XXX) & $S=\frac{1}{2k^2}\int d^4x \sqrt{-g}\left[F(T,Q,G,B)+2k^{2}L_m\right]$ \\ 
\hline 
12 & Myrzakulov Gravity - XXXI (MG-XXXI) & $S=\frac{1}{2k^2}\int d^4x \sqrt{-g}\left[F(R,T,G,B,{\cal T})+2k^{2}L_m\right]$ \\ 
\hline 
13 & Myrzakulov Gravity - XXXII (MG-XXXII) & $S=\frac{1}{2k^2}\int d^4x \sqrt{-g}\left[F(R,T,Q,G,B)+2k^{2}L_m\right]$ \\ 
\hline 
14 & Myrzakulov Gravity - XXXIII (MG-XXXIII) & $S=\frac{1}{2k^2}\int d^4x \sqrt{-g}\left[F(R,Q,G,B,{\cal T})+2k^{2}L_m\right]$ \\ 
\hline 
15 & Myrzakulov Gravity - XXXIV (MG-XXXXIV) & $S=\frac{1}{2k^2}\int d^4x \sqrt{-g}\left[F(T,Q,G,B,{\cal T})+2k^{2}L_m\right]$ \\ 
\hline 
16 & Myrzakulov Gravity - XXXV (MG-XXXV) & $S=\frac{1}{2k^2}\int d^4x \sqrt{-g}\left[F(R,T,Q,G,B,{\cal T})+2k^{2}L_m\right]$ \\ 
\hline  
17 & Myrzakulov Gravity - XXXVI (MG-XXXVI) & $S=\frac{1}{2k^2}\int d^4x \sqrt{-g}\left[F(Q,G,B)+2k^{2}L_m\right]$ \\ 
\hline 
18 & Myrzakulov Gravity - XXXVII (MG-XXXVII) & $S=\frac{1}{2k^2}\int d^4x \sqrt{-g}\left[F(R,T,G,B)+2k^{2}L_m\right]$ \\ 
\hline  
19 & Myrzakulov Gravity - XXXVIII (MG-XXXVIII) & {$S=\frac{1}{2k^2}\int d^4x \sqrt{-g}\left[F(T,G,B,{\cal T})+2k^{2}L_m\right]$} \\ 
\hline 
\end{tabular} 
\end{table}

\section{Cosmology in metric-affine MG theories}
Consider the FRW universe.  The flat FRW spacetime is described by the metric
 \begin{equation}\label{1.1}
 ds^2=-dt^2+a^2(t)(dx^2+dy^2+dz^2),
 \end{equation}
 where $a=a(t)$ is the scale factor.
Let $R$, $T$, $Q$ are the Ricci, torsion,  nonmetricity  scalars. For the FRW metric they have the forms: i)  $R=R_{0}$, where $T=Q=0$;  ii) $T=T_{0}$, where $R=Q=0$; iii) $Q=Q_{0}$, where $R=T=0$. For the FRW metric, they have the forms: 
\begin{eqnarray}
 R_{0}&=&6(\dot{H}+2H^2),\\
T_{0}&=&-6H^2,\\
 Q_{0}&=& 6H^2,
   \end{eqnarray}  
 where $H=(\ln a)_{t}$  is the Hubble parameter. In the  metric-affine spacetime case, we assume that the Ricci, torsion and nonmetricity scalars take  the forms 
   \begin{eqnarray}
 R&=&6(\dot{H}+2H^2)+u,\\
   T&=&-6H^2+v,\\
	Q&=&6H^{2}+w.
 \end{eqnarray}
 Similarly, we can write the boundary term scalar ($B$) and the GB scalar ($G$)  as \cite{1205.5266}
\begin{eqnarray}
 G&=&G_{0}+p,\\
   B&=&B_{0}+f,
 \end{eqnarray}
where $u, v, w, p, f$ are some real functions of $t, a, \dot{a}, \ddot{a}$. 
\section{Spherically symmetric and black hole solutions in metric-affine MG theories}
Let us we now  present our idea to study,  for example, the black hole solutions of metric-affine MG theories. For this aim, we consider the following static and spherically symmetric metric \cite{1205.5266}
\begin{eqnarray}
ds^{2}= A(r)dt^{2}-B(r)dr^{2}-C(r) (d\theta^{2}+\sin^{2}\theta d\phi^{2}),
\end{eqnarray}
where $A(r)$, $B(r)$  and $C(r)$ are real functions of the radial coordinate $r$. Consider two connections: the Levi-Civita connection and the Weitzenb{\"{o}}ck connection.
First, let us consider the  Levi-Civita connection case. In this case, the nonmetricity  and torsion scalars are  equal to zero that is $T_{0}=Q_{0}=0$. Then   the corresponding Ricci scalar has the form
\begin{eqnarray}
R_{0}=\frac{A^{\prime\prime}}{AB}+2\frac{C^{\prime\prime}}{BC}+\frac{A^{\prime}C^{\prime}}{ABC}-\frac{A^{\prime 2}}{2A^{2}B}-\frac{C^{\prime 2}}{2BC^{2}}-
\frac{A^{\prime}B^{\prime}}{2AB^{2}}
-\frac{B^{\prime}C^{\prime}}{B^{2}C}-\frac{2}{C}.
\end{eqnarray}
Here and below primes denote differentiation with respect to the radial coordinate $r$. Let us we now  consider  the Weitzenb{\"{o}}ck connection case. In this case,  the Ricci scalar and nonmetricity scalar are  equal to zero that is $R_{0}=Q_{0}=0$ and  the
 torsion scalar is given by
\begin{eqnarray}
T_{0}=\frac{C^{\prime}(2A^{\prime}C+AC^{\prime}}{2ABC^{2}}.
\end{eqnarray}
Similarly, we can calculate  the nonmetricity scalar $Q_{0}$. 		For the metric (11.1), it has the form
\begin{eqnarray}
Q_{0}=-\frac{C^{\prime}(2A^{\prime}C+AC^{\prime}}{2ABC^{2}},
\end{eqnarray}
where $R_{0}=T_{0}=0$. 
The geometry of the  MG theories is the metric-affine spacetime. For that reason, now let us consider the more general case, namely, the metric-affine spacetime. For this metric-affine spacetime, we have the  metric-affine connection. In this metric-affine connection case, the Ricci scalar,
the torsion scalar
and the nonmetricity scalar take the forms
\begin{eqnarray}
R&=&R_{0}+u, \\
T&=&T_{0}+v, \\
Q&=&Q_{0}+w.
\end{eqnarray}
Here the metric-affine contributions are given by the following functions \cite{1205.5266}
\begin{eqnarray}
u&=&u(A,B,C,A^{\prime},B^{\prime},C^{\prime},A^{\prime\prime},B^{\prime\prime},C^{\prime\prime}), \\
v&=&v(A,B,C,A^{\prime},B^{\prime},C^{\prime},A^{\prime\prime},B^{\prime\prime},C^{\prime\prime}), \\
w&=&w(A,B,C,A^{\prime},B^{\prime},C^{\prime},A^{\prime\prime},B^{\prime\prime},C^{\prime\prime}).
\end{eqnarray}
They are some real functions of the metric tensor components $g_{ij}$  (11.1). 
\section{Metric-affine MG theories  with  boundary term scalars}
Next, we very briefly mention the main moments of metric-affine MG theories  with the boundary term scalars \cite{1205.5266}. According our idea, we assume that  the boundary term scalar has the form \cite{1205.5266}
\begin{eqnarray}
B=B_{0}+f.
\end{eqnarray}
Similarly, we can write the GB scalar for the metric-affine spacetime as
\begin{eqnarray}
G=G_{0}+p.
\end{eqnarray}
In the last two equations, $p$ and $f$ are metric-affine contributions and some functions of $A,B,C$ and their derivatives.

\section{Conclusion}
As we mentioned in the introduction, GR has several generalizations like $F(R), F(T)$ and so on.  Among these  generalizations of GR, the metric-affine gravity theories   have a nice feature by extending to admit not only curvature but both torsion and nonmetricity. This means  the MAG  is described by a pseudo - Riemannian geometry. The  geometrical structure of the  MAG can  be studied once
a metric tensor and a connection are given. In this way, we can calculate   the affine connection for the underlying theory. 
In this paper, we have considered the  so-called generalized Myrzakulov gravity or MG-VIII  which can be considered as  the particular case of the MAG. To simplify the problem,  we consider the FRW spacetime case in detail. For this case the point-like Lagrangian and Hamiltonian  of the theory is derived. Using this Lagrangian and the Euler-Lagrangian equation, the gravitational equations  of the MG-VIII  is presented. For simplicity,  the particular case of the MG-VIII  when $F=\alpha R+\beta T +\mu Q+\nu {\cal T}$ is investigated. For this particular case, the gravitational equations is considered in detail. For the  quantum case, the corresponding  Wheeler - DeWitt equation  is presented.  The relation with the soliton theory is shortly discussed. These results show that  altogether one can say that some ingredients of the MG-VIII  are present and work as expected, but some other aspects remain to be properly understood.  These aspects of the MG-VIII  certainly worth further investigation (see e.i. refs. \cite{2011.012488}-\cite{1412.1471}).

\section*{Acknowledgments}
The work was  supported by the Ministry of Education and Science of the Republic of Kazakhstan, Grant  AP09058240.

 \end{document}